\documentclass[12pt]{article}
\usepackage{amsmath,amsfonts,amssymb,graphicx,fullpage}
\usepackage[tight]{subfigure}
\usepackage{ifpdf}
\ifpdf
  \usepackage{color}
  \definecolor{darkblue}{rgb}{0.3,0.3,0.6}
  \usepackage[pdftitle={Millions of Standard Models on Z6'?},pdfauthor={Florian Gmeiner and Gabriele Honecker},pdfsubject={String theory},pdfkeywords={string phenomenology, compactification, orientifold},colorlinks=true,linkcolor=black,urlcolor=darkblue,filecolor=darkblue,citecolor=darkblue,menucolor=darkblue,breaklinks=true,bookmarks=true,anchorcolor=black,unicode=true]{hyperref}
\fi
\usepackage[sort,compress]{cite}

\def\muc{\multicolumn}

\def\bZ{\mathbb{Z}}
\def\unity{1\!\!{\rm I}}
\def\ov{\overline}
\def\N{\mathbf{N}}
\def\Sym{\mathbf{Sym}}
\def\Anti{\mathbf{Anti}}
\def\Adj{\mathbf{Adj}}
\def\cc{c.c.}
\def\ov{\overline}
\def\1{{\bf 1}}
\def\2{{\bf 2}}
\def\3{{\bf 3}}
\def\4{{\bf 4}}
\def\6{{\bf 6}}
\def\OR{\Omega\mathcal{R}}
\def\half{\frac{1}{2}}

\newcommand{\captionfonts}{\small}
\makeatletter
\long\def\@makecaption#1#2{%
  \vskip\abovecaptionskip
  \sbox\@tempboxa{{\captionfonts #1: #2}}%
  \ifdim \wd\@tempboxa >\hsize
    {\captionfonts #1: #2\par}
  \else
    \hbox to\hsize{\hfil\box\@tempboxa\hfil}%
  \fi
  \vskip\belowcaptionskip}
\makeatother

\makeatletter
\let\ORIGINALlatex@openbib@code=\@openbib@code
\renewcommand{\@openbib@code}{\ORIGINALlatex@openbib@code\setlength{\itemsep}{1ex plus.5ex minus.5ex}\setlength{\parsep}{0pt}}
\makeatother

\def\incfig#1#2{\includegraphics[width=#2\linewidth]{#1}}
\def\fig#1#2{\begin{figure}[tb]\centering\incfig{fig_#1}{0.75}\caption{#2}\label{fig:#1}\end{figure}}
\def\wfig#1#2{\begin{figure}[tb]\centering\incfig{fig_#1}{0.99}\caption{#2}\label{fig:#1}\end{figure}}
\def\twofig#1#2#3#4{\begin{figure}[tb]\centering\subfigure[\label{fig:#1}]{\incfig{fig_#1}{0.49}}\hfill\subfigure[\label{fig:#2}]{\incfig{fig_#2}{0.49}}\caption{#3}\label{#4}\end{figure}}
\def\twofigvar#1#2#3#4#5#6{\begin{figure}[tb]\centering\hspace{15mm}\subfigure[\label{fig:#1}]{\incfig{fig_#1}{#2}}\hfill\subfigure[\label{fig:#3}]{\incfig{fig_#3}{#4}}\hspace*{10mm}\caption{#5}\label{#6}\end{figure}}

\begin{document}
\begin{center}
\begin{flushright}
{\small CERN-PH-TH/2008-123\\NIKHEF/2008-004\\June 18, 2008}
\end{flushright}

\vspace{30mm}
{\Large\bf Millions of Standard Models on $\bZ_6'$?}

\vspace{15mm}
{\large Florian Gmeiner$^1$ and Gabriele Honecker$^2$}

\vspace{5mm}
{~$^1$\it NIKHEF, Kruislaan 409, 1098 SJ Amsterdam, The Netherlands}\\[1.5ex]
{~$^2$\it PH-TH Division, CERN, 1211 Geneva 23, Switzerland}

\vspace{20mm}{\bf Abstract}\\[2ex]\parbox{140mm}{
Using previous results on the statistics of intersecting D6-brane models on
$T^6/\bZ_6'$ we analyse solutions that resemble the gauge group and matter
content of the supersymmetric standard model.
In particular the structure of the hidden sector, the gauge coupling
constants and chiral exotic matter content are computed and classified
for all possible standard model--like configurations on this background.
It turns out that the number of chiral exotics, Higgses and values of 
gauge couplings are strongly correlated.
Some examples with the chiral spectrum of the supersymmetric standard model
plus vector-like states and a massless $U(1)_{B-L}$ are discussed in detail.}
\end{center}
\vfill
\begin{flushleft}\small\tt
fgmeiner@nikhef.nl\\
Gabriele.Honecker@cern.ch
\end{flushleft}

\thispagestyle{empty}
\clearpage

\tableofcontents
\newpage
\setlength{\parskip}{1em plus1ex minus.5ex}
\section{Introduction}\label{sec:intro}
Understanding the structure of the large amount of string theory compactifications,
also known as the landscape, is one of the big challenges for string phenomenology.
In order to make contact with physics at accessible energy scales, it is necessary
to explore all feasible ways to obtain a standard model--like theory, a task that
is complicated by the huge amount of possibilities.
A deeper understanding of patterns in the space of solutions could lead not only to
a more refined string model building, but if we are lucky to direct predictions based on
a statistical analysis of correlations in the landscape.

In recent years, a large number of string theory vacua has been found by using a
statistical approach, see e.g.~\cite{le06} for heterotic orbifolds,~\cite{di06} for 
free fermionic models,~\cite{adks06} for rational conformal field theories 
and~\cite{bghlw04,gbhlw05,gm05,dota06,gmst06,gm06,gls07,gmho07} for intersecting D-branes in orbifold backgrounds.
For the latter examples, CFT techniques have recently been applied to compute non-perturbative Yukawa couplings
and Majorana mass terms, see e.g.~\cite{bcw06,ibur06}, 
which can in particular lead to non-vanishing results for instantons on rigid cycles.
However, to our knowledge, no fully-fledged intersecting brane model with standard model 
spectrum on rigid cycles has been found so far.\footnote{See e.g.~\cite{bcms05,cmn06,clmn07,sfyz08} for
model searches on $T^6/\bZ_2 \times \bZ_2$ with torsion and rigid cycles.}

In this article, we focus on properties and correlations of the intersecting D6-brane
models on $T^6/\bZ_6'$ found in~\cite{gmho07} (for the set-up see also~\cite{balo06,balo07,balo08})
with supersymmetric standard model--like spectra. In contrast to the smooth heterotic $E_8 \times E_8$ 
compactifications of, e.g.,~\cite{bodo08}, intersecting D6-brane models in $T^6/\bZ_{2N}$ backgrounds 
are built on {\it fractional} but {\it non-rigid} branes and 
do always have some amount of non-chiral matter states, in particular chiral multiplets in the adjoint 
representation. We use a combined method of three-cycle intersection numbers and Chan-Paton matrices 
to determine the full massless spectrum needed to compute the one-loop beta function coefficients of the
gauge couplings. Due to the large number of ${\cal O}(10^{23})$ supersymmetric RR tadpole solutions,
our focus in the present article is on a statistical evaluation of supersymmetric standard model--like string spectra,
in particular the subset of ${\cal O}(10^{7})$ models without chiral exotic matter. While the 
bigger set of around ${\cal O}(10^{15})$ three generation models displays a large variety of 
abundances of chiral exotics and ``chiral" Higgs candidates, hidden sectors  as well as diverse ratios of gauge couplings,
it turns out that these quantities are strongly correlated. Our analysis is performed for one out of
two possible background geometries for which we prove the equivalence of bulk solutions.
Some examples of spectra with exactly three quark-lepton generations plus vector-like states and a 
massless $B-L$ symmetry are discussed in detail. Selection rules for perturbative Yukawa couplings are 
considered as well as those mass terms which can arise through parallel displacement of branes.

The methods to analyse the ensemble of solutions to the supersymmetry and RR tadpole equations for
models on a particular orbifold background have been developed in an earlier publication~\cite{gmho07}
and are based on the general idea to classify large amounts of models in the landscape.
While this approach has shown to be useful to gain insights into the structure of the space of
solutions, there are some caveats not to be overlooked.
In particular, one has to be very careful which solutions to the equations one wants to count as
individual models and which are just different realisations of the same low energy theory that
might be related by a symmetry. Moreover, by using a statistical approach there exists the possibility
that one finds correlations between properties of the solutions that are artefacts of the
method by which the solutions are generated~\cite{dile06}. We do not have to worry about the second issue
in the case at hand, since we construct all possible solutions explicitly (although we do not
analyse every single solution in full detail) and thereby can be sure that no unwanted bias is
introduced. With respect to the first issue, we are providing a careful analysis of the geometric set-up,
but since not all solutions could be checked for the complete non-chiral matter content, there
remains an uncertainty about the true value of completely unrelated models.

The paper is organised as follows. In section~\ref{sec:set-up} we review the geometric set-up and 
consistency conditions for intersecting D-brane models in the $T^6/\bZ_6'$ orientifold background.
In section~\ref{sec:CMOSS} we determine the complete massless spectrum using a combined 
approach of intersection numbers and Chan-Paton matrices with special emphasis on the cases where 
branes are parallel along some direction, clarifying in particular the cases with symmetric
and antisymmetric  
representations. In section~\ref{Sec:GC}, tree level gauge couplings and the one-loop running due to 
massless string modes are discussed, and in section~\ref{sec:SM-constraints} our constraints on
standard model--like spectra are given. In section~\ref{sec:Statistics}, the results of the 
statistical analysis are presented with some standard model--like examples without chiral exotics
discussed in detail in section~\ref{App:Realisations}.
Our conclusions are given in section~\ref{Sec:Concl}, and finally technical details are collected in 
appendices~\ref{App:CPlabels} to~\ref{App_bulkrelation}, and the existence of trinification spectra 
is debated in appendix~\ref{App_Trinification}.

\section{Set-up}\label{sec:set-up}

The set-up of the $T^6/\bZ_6'$ orientifold with fractional D6-branes
has been discussed in detail in~\cite{gmho07}.\footnote{Our choice of a basis of three-cycles differs 
from the one in~\cite{balo06,balo07,balo08}.} 
We briefly review the main
features here.
 The $\bZ_6'$ orbifold generator acts as
\begin{equation}
\theta: z^k \rightarrow e^{2 \pi i v_k} z^k
\quad
{\rm with}
\quad 
\vec{v}=\frac{1}{6}(1,2,-3) 
\end{equation}
on the complex coordinates $z^k$ $(k=1,2,3)$ of a factorised six-torus $T^2_1 \times T^2_2 \times T^2_3$.
\footnote{We are dealing with factorisable tori only, for orientifold models on non-factorisable tori see~\cite{ftz07,kot07,sfyz08}.}
The geometric part ${\cal R}$ of the orientifold action $\OR$, 
\begin{equation}
{\cal R}: z^k \rightarrow \ov{z}^k,
\end{equation}
enforces the $SU(3) \times SU(3) \times SU(2)^2$ lattice to have one of two possible orientations {\bf A} or {\bf B}
 per $SU(3)$ and {\bf a}  or {\bf b} per $SU(2)^2$ lattice with respect to the ${\rm Re}(z^k)$ directions.
The {\bf a} and {\bf b} torus are conveniently parameterised by $b=0,1/2$, respectively.
 The $T^6/\bZ_6'$ orbifold has 24 three-cycles with four,  $\rho_i (i=1\ldots 4)$, stemming from the underlying torus, eight, $\delta_j, \tilde{\delta}_j (j=1\ldots 4)$, associated to the $\bZ_2$ sub-symmetry and twelve more pertaining to the $\bZ_3$ sub-symmetry. The former two kinds $(\rho_i,\delta_j,\tilde{\delta}_j)$ form a twelve dimensional sub-lattice from which fractional cycles 
\begin{equation}\label{Eq:Def-frac}\begin{aligned}
 \Pi^{\rm frac} &= \frac{1}{2} \Pi^{\rm bulk} +\frac{1}{2}\Pi^{\rm ex} \\
 &= \frac{1}{2}\left( \sum_{i=1}^4 \tilde{a}_i \rho_i  +\sum_{j=1}^4 \left( d_j \delta_j +e_j \tilde{\delta}_j  \right)   \right)
\end{aligned}
\end{equation}
are built with coefficients $\tilde{a}_i, d_j,e_j \in \bZ$ which 
are composed of toroidal one-cycle wrapping numbers $(n_i,m_i)$  discussed in more detail in 
appendix~\ref{App_bulkrelation}.\footnote{For the sake of brevity we have set $(\tilde{a}_1,\tilde{a}_2,\tilde{a}_3,\tilde{a}_4) \equiv (P,Q,U,V)$ in the notation 
of~\cite{gmho07}.}

The RR tadpole cancellation conditions can be written in the general form
\begin{equation}\label{Eq:RRtcc}
\sum_a N_a \vec{X}_a = \vec{L}
\end{equation}
with $\vec{L}=(L_1,L_2,0,0,0,0)^T$ and the bulk entries $X_1,X_2$ listed in table~\ref{Tab:Def-XYL}
descending from the toroidal cycles. The contributions from exceptional cycles, $X_3 \ldots X_6$,
 are given in appendix~\ref{App_bulkrelation} in equations~\eqref{Eq:App-X3X6}
and~\eqref{Eq:App-X4X5}.

\begin{table}[ht]
\begin{center}
\begin{equation*}
\begin{array}{|c||c|c|c|}
\hline
{\rm lattice}
& \begin{array}{c} X_1\\ X_2 \end{array} 
& \begin{array}{c} Y_1 \\ Y_2 \end{array}
& \begin{array}{c} L_1 \\ L_2 \end{array}
\\\hline\hline
{\bf AA} 
& \begin{array}{c} 2\tilde{a}_1 + \tilde{a}_2 \\ -(\tilde{a}_4 + b \, \tilde{a}_2)
\end{array}
& \begin{array}{c} 2 \tilde{a}_3 + \tilde{a}_4 + b(2\tilde{a}_1+\tilde{a}_2) \\ \tilde{a}_2
\end{array}
& \begin{array}{c} 8 \\ 8 \, (1-b) 
\end{array}
\\\hline
\!\!\!\left.\!\! \begin{array}{c} {\bf AB} \\ {\bf BA} \end{array}\!\!\right\}\!\!\! 
&  \!\!\!\begin{array}{c} \tilde{a}_1 + \tilde{a}_2 \\ \tilde{a}_3 - \tilde{a}_b + b (\tilde{a}_1-\tilde{a}_2)
\end{array}\!\!\!
& \begin{array}{c} \tilde{a}_3 + \tilde{a}_4 + b (\tilde{a}_1 + \tilde{a}_2) \\ \tilde{a}_2-\tilde{a}_1
\end{array}
& \begin{array}{c} 8 \\ k \, 8 \, (1-b) 
\end{array}
\\\hline
{\bf BB} 
&  \begin{array}{c} \tilde{a}_1 + 2 \tilde{a}_2 \\ \tilde{a}_3 + b \tilde{a}_4
\end{array}
& \begin{array}{c} \tilde{a}_3 + 2 \tilde{a}_4 + b (\tilde{a}_1 + 2 \tilde{a}_2) \\ -\tilde{a}_1
\end{array}
& \begin{array}{c} 24 \\ 8 \, (1-b) 
\end{array}
\\\hline
\end{array}
\end{equation*}
\end{center}
\caption{Bulk coefficients of RR tadpole cancellation and supersymmetry conditions. For shortness, we have 
set 
$k=3$ for the {\bf AB} lattice on $T^2_1 \times T^2_2$ and $k=1$ for {\bf BA}.
$b=0,1/2$ parameterises the {\bf a} and {\bf b} shape on $T^2_3$.}
\label{Tab:Def-XYL}
\end{table}

The supersymmetry condition for the toroidal part of a three-cycle $\Pi^{\rm frac}$ can be cast into the form
\begin{equation}\label{Eq:SUSY}
\vec{Y} \cdot \vec{F}(U) = 0,
\quad
\vec{X} \cdot \vec{U} > 0,
\end{equation}
with the bulk coefficients given in table~\ref{Tab:Def-XYL} and 
\begin{equation}
\begin{aligned}
\left(\begin{array}{c} U_1\\ U_2 \end{array}\right) &\sim \left(\begin{array}{c} 1 \\ c_1
\end{array}\right)
\quad {\rm with} \quad c_1 =
\left\{\begin{array}{cc} 2 \varrho & {\bf AA},{\bf BB} \\ \frac{2\varrho}{3} & {\bf AB}, {\bf BA}
\end{array}\right.
,\\
\left(\begin{array}{c} F_1(U)\\ F_2(U) \end{array}\right) &\sim \left(\begin{array}{c} 1 \\ c_2
\end{array}\right)
\quad {\rm with} \quad c_2 =
\left\{\begin{array}{cc} \frac{3}{2 \varrho} & {\bf AA}, {\bf BB} \\ \frac{1}{2 \varrho} & {\bf AB}, {\bf BA}
\end{array}\right.
,
\end{aligned}
\end{equation}
up to overall normalisation. The complex structure modulus $\varrho = \sqrt{3} R_2 / 2 R_1$ parameterises the 
ratio of radii $R_1,R_2$ along the real and imaginary direction on $T^2_3$, and the constants $c_i$ are 
given in dependence of the lattice orientations on $T_1^2 \times T_2^2$.

The RR tadpole cancellation conditions on exceptional cycles have no contribution from O6-planes,
 $(L_3 \ldots L_6) = (0\ldots 0)$, 
and fractional cycles are supersymmetric if they are composed of 
 supersymmetric toroidal cycles and only those exceptional cycles which wrap the four $\bZ_2$ fixed points on 
$T^2_1 \times T^2_3$  traversed 
 by the toroidal cycle with three independent signs of the four potential contributions corresponding 
to a $\bZ_2$ eigenvalue and two discrete Wilson lines on $T^2_1 \times T^2_3$. More details
are given in appendix~\ref{App_bulkrelation} and~\cite{gmho07}.

It turns out that the solutions to the {\it bulk} RR tadpole and supersymmetry conditions 
are independent of the choice of the $T^2_1$ lattice as we prove in appendix~\ref{App_bulkrelation}.

\section{Complete massless spectrum}\label{sec:CMOSS}

In this section, we review the computation of the complete massless, i.e. chiral and
non-chiral, open string  spectrum
in terms of three-cycle intersection numbers and compare the results with those arising from 
the computation of Chan-Paton 
matrices. The latter method confirms the formulae inferred in~\cite{gmho07}.

In the present discussion, special emphasis is placed on states living on branes which are parallel 
along some two-torus $T^2_m$ of any $T^6/\bZ_{2N}$ background. 
The generic formulae listed in~\cite{gmho07} for those cases 
are found to be correct except for the  case in equation~\eqref{eq:a-|OR||T2} 
where orientifold image branes $a$ and $(\theta^k a')$
are parallel to each other for some $k\in \{0,\ldots,N-1\}$
but orthogonal to the $\OR\theta^{-k}$ and $\OR\theta^{-k+N}$ invariant O6-planes 
on the two-torus  where the $\bZ_2$ action is trivial.
There is furthermore a so far unobserved sign subtlety in~\eqref{eq:a||T3-S+W}
for brane $a$ on top of  $(\theta^k a')$ but displaced from the origin on $T^2_3$ in case of a Wilson line.

For concreteness, the notation is adapted to the $T^6/\bZ_6'$ background with $\vec{v}=(1/6,1/3,-1/2)$,
the $\bZ_2$ invariant two-torus being $T^2_2$,
but the line of reasoning is valid for any other $T^6/\bZ_{2N}$ orbifold upon 
suitable permutation of two-tori.
The complete matter spectrum will be needed in order to determine the one-loop running of gauge couplings in
section~\ref{Sec:GC}. One should, however, keep in mind that ${\cal N}=2$ supersymmetric
sectors might become massive, for example by a parallel displacement of branes according to
equation~\eqref{Eq:masses}.

For completeness, we also give the closed string spectrum for the $T^6/\bZ_6'$ orbifold.

\subsection{Closed string spectrum}

The untwisted sector contains the supergravity and universal dilaton-axion multiplet.
The numbers $n_C$ and $n_V$ of non-universal chiral and vector multiplets depends on the background.
For the $T^6/\bZ_6'$ orbifold on {\bf AA} or {\bf BA} type lattices on $T^2_1 \times T^2_2$, 
we obtained $(n_C,n_V)= (38-6b, 8+6b)$ in~\cite{gmho07} and for the {\bf AB} and {\bf BB} 
orientations
$(n_C,n_V)= (46-10b,10b)$ where $b=0,\half$ parameterises the {\bf a} or {\bf b} type shape of $T^2_3$,
respectively. Vectors arise always from RR states whereas scalars stem from both NS-NS and RR states.
The closed string spectrum preserves ${\cal N}=1$ supersymmetry with the fermionic superpartners
arising from R-NS and NS-R sectors.

In the Green Schwarz mechanism, an open string $U(1)$ gauge field wrapped around the three-cycle $\Pi$
with orientifold image cycle $\Pi'$
becomes massive by absorbing one linear combination of RR scalars dual to the two-form $\int_{\Pi-\Pi'} C_5$ 
where $C_5$ is the ten dimensional RR 5-form. Due to supersymmetry, at the same time the  NS-NS state 
parameterising a linear combination of complex structure moduli and pertaining to the same chiral multiplet is frozen. 
Since these closed string sector states carry no gauge 
representation, the Green Schwarz mechanism in intersecting D-branes on orbifolds {\it does not induce} any 
Fayet-Iliopoulos term provided  the three-cycles wrapped by the branes fulfil the three-cycle calibration condition 
with toroidal part given in equation~\eqref{Eq:SUSY}, for more details
see e.g. the discussion  in~\cite{cim02,bbkl02}.
This is in contrast to heterotic orbifolds where the anomalous $U(1)$ factor has a non-vanishing 
Fayet-Iliopoulos term and a {\it charged} scalar forms the transverse degrees of freedom of the massive vector boson.

\subsection{Massless open spectrum from three-cycle intersection numbers}\label{sec:ospec}

\subsubsection{For branes at three non-vanishing angles}

In~\cite{gmho07}, we showed that the intersection number or net chirality
$\chi^{ab}\equiv\Pi_a^{\rm frac} \circ \Pi_b^{\rm frac}$ between fractional 
three-cycles $\Pi_a^{\rm frac}$ and $\Pi_b^{\rm frac}$ and the resulting total amount  $\varphi^{ab}$
of bifundamental matter in the $(\N_a,\ov{\N}_b)$ representation
can be expressed in terms of a sum of toroidal and $\bZ_2$ invariant intersection numbers,
\begin{equation}\label{Eq:Def_Inters}
\begin{aligned}
I_{ab} &
\equiv\prod_{i=1}^3 I_{ab}^{(i)}= \prod_{i=1}^3 (n^a_i m^b_i - m^a_i n^b_i) ,
\\[1ex]
I_{ab}^{\bZ_2} &
\equiv \sum_{x_a^k x_b^k} (-1)^{\tau_{x_a^1x_a^3} +\tau_{x_b^1x_b^3}} 
\delta_{x_a^1 x_b^1} \delta_{x_a^3 x_b^3}  I_{ab}^{(2)} ,
\end{aligned}
\end{equation} 
among brane $a$ and all orbifold images $(\theta^k b)$ for $k=0\ldots N-1$ on $T^6/\bZ_{2N}$,
\begin{equation}\label{ChiFractional}
\begin{aligned}
\chi^{ab} \equiv  \chi_L^{ab} - \chi_R^{ab} &=
-  \sum_{k=0}^{N-1} \frac{ I_{a (\theta^k  b)}
+ I^{\bZ_2}_{a (\theta^k b)}}{2} ,
\\[1ex]
\varphi^{ab} \equiv \chi_L^{ab} + \chi_R^{ab} &=
  \sum_{k=0}^{N-1} \left| \frac{ I_{a (\theta^k b)} + I^{\bZ_2}_{a (\theta^k b)}}{2}  \right|,
\end{aligned}
\end{equation}
the formula for the chiral plus non-chiral bifundamental matter states $\varphi^{ab}$ being valid only 
in case of three non-vanishing angles. Here, $x_a^i$ labels the $\bZ_2$ fixed points traversed by the toroidal
 cycle with wrapping number $(n_i^a,m_i^a)$ along $T^2_i$ and $\tau_{x_a^1x_a^3} \in \{0,1\}$ are combinations of 
Wilson lines and a $\bZ_2$ eigenvalue.
The complete matter spectrum for this case is  given in table~\ref{NonChiralSpectrum}.
\begin{table}[ht]
  \begin{center}
    \begin{equation*}
      \begin{array}{|c|c|} \hline
        \multicolumn{2}{|c|}{\rule[-3mm]{0mm}{8mm}
\text{\bf Chiral and non-chiral massless matter on } T^6/(\bZ_{2N} \times \OR)  }\\ \hline\hline
\text{rep.} & \text{total number} \;\; \varphi
\\\hline\hline
({\bf Adj}_a) & 1 +\frac{1}{4} \sum_{k=1}^{N-1} \left| I_{a(\theta^k a)} +I_{a(\theta^k a)}^{\bZ_2}\right|
\\
({\bf Anti}_a) &  \frac{1}{4} \sum_{k=0}^{N-1}\left|I_{a(\theta^k a')}
    +I_{a(\theta^k a')}^{\bZ_2} + I_a^{\OR\theta^{-k}} + I_a^{\OR\theta^{-k+N} }
 \right|
\\
({\bf Sym}_a) &  \frac{1}{4} \sum_{k=0}^{N-1}\left|I_{a(\theta^k a')}
    +I_{a(\theta^k a')}^{\bZ_2} - I_a^{\OR\theta^{-k}} - I_a^{\OR\theta^{-k+N}}  \right|
\\
({\bf N}_a,\overline{\bf N}_b) & \frac{1}{2} \sum_{k=0}^{N-1}\left| I_{a(\theta^k b)} +I_{a(\theta^k b)}^{\bZ_2} \right|
\\
({\bf N}_a,{\bf N}_b) & \frac{1}{2}  \sum_{k=0}^{N-1}\left| I_{a(\theta^k b')} +I_{a(\theta^k b')}^{\bZ_2} \right|
\\ \hline
     \end{array}
    \end{equation*}
  \end{center}
\caption{Chiral plus non-chiral matter states $\varphi$ in $T^6/(\bZ_{2N} \times \OR)$ models for generic
non-vanishing angles. 
For vanishing angles, the formulae are modified as discussed in sections~\protect\ref{Sss:||inters} to~\protect\ref{Sss:||T3}.}
\label{NonChiralSpectrum}
\end{table}

Supersymmetric brane configurations are also possible with either one or three vanishing angles.
The net-chirality $\chi^{ab}$ is still computed by means of three-cycle intersection numbers~\eqref{Eq:Def_Inters}, 
but a zero can either 
correspond to no massless state or ${\cal N}=2$ supersymmetric non-chiral matter pairs.
If the branes are parallel along the $\bZ_2$ invariant two-torus $T^2_2$ (and also parallel to 
the relevant O6-planes for $b=(\theta^k a')$), the number of non-chiral  pairs is simply computed from the
intersection numbers on $T^2_1 \times T^2_3$ in a six-dimensional set-up. In~\cite{gmho07},
we were only able to list the remaining cases with branes parallel
along $T^2_1$ or $T^2_3$ by silently employing the method of  Chan-Paton matrices illustrated in detail 
in section~\ref{Subsec:CPmatrices}.
Furthermore, we did not consider before the case with brane $a$ and its orientifold image
$ (\theta^k a')$ parallel along $T^2_2$ but orthogonal
to the $\OR\theta^{-k}$ and $\OR\theta^{-k+N}$ invariant  O6-planes given below in~\eqref{eq:a-|OR||T2}.
There is also a sign subtlety for branes parallel on, e.g.,  $T^2_3$ when carrying a Wilson line
and displaced from the origin on the same two-torus, see equation~\eqref{eq:a||T3-S+W}. 

The formulae for branes parallel along some two-torus 
are modified as follows, with a detailed discussion of the derivation
to follow below in section~\ref{Subsec:CPmatrices}:

\subsubsection{For branes parallel along all three tori}\label{Sss:||inters}

\begin{itemize}
\item
The $aa$ sector provides a vector multiplet of $U(N_a)$ and one chiral multiplet in the adjoint 
representation.
\item
Branes $a_1$, $a_2$ with identical position, opposite $\bZ_2$ eigenvalues and no relative Wilson line, i.e. $\tau_{a_1}^0 = \tau_{a_2}^0 +1 \; {\rm mod} \; 2$ and $\tau_{a_1}^i = \tau_{a_2}^i$ ($i=1,3$), contribute
$2 \times [ ({\bf N^1}_a,\ov{\bf N^2}_a) + \cc]$, or in the special case of orientifold
image branes, i.e. $a_2 =(\theta^k a_1')$ for some $k \in \{0,\ldots,N-1\}$, 
the bifundamental representations are replaced by $ 2 \times [\Anti_a + \cc]$.  
\item
Branes $a$ and $b$ parallel on $T^2_i$ but either spatially separated, $\Delta\sigma^i_{ab} \neq 0$ ($i=1,3$),\footnote{
In slight abuse of notation, displacements from the origin 
along one-cycles $\pi_k$ are labelled by $\sigma_k \in \{0,1/2\}$, but 
$\Delta\sigma^i$ denotes a relative distance of branes along $T^2_i$.}
 or with a relative Wilson line, $\Delta \tau^i_{ab} \neq 0$ ($i=1,3$), do not contribute to the massless spectrum
due to a shift in the mass formula for a state with given momentum and winding numbers $p_i,q_i \in \bZ$
along $T^2_i$,
\begin{equation}\label{Eq:masses}
\left( {\rm mass}  \right)^2 \sim 
\left(\left( \frac{1}{r_{\parallel}^{(i)}}\right)^2 
\left(p_i + \frac{\Delta\tau^i}{2} \right)^2 + 
\left( \frac{r_{\perp}^{(i)}}{\alpha^{\prime}}\right)^2 \left(q_i + \Delta\sigma^i \right)^2 \right),
\end{equation}
where $r_{\parallel}^{(i)}$ is the length of the one-cycle on $T^2_i$
and $r_{\perp}^{(i)}$ the distance of two copies of the same one-cycle on $T^2_i$, 
both in appropriate units, and the relative Wilson lines and spatial displacements
are parameterised by $\Delta\tau^i \in \{0,1\}$ and $\Delta\sigma^i \in \{0,1/2\}$ for
$i=1,3$, whereas  $\Delta\tau^2$ and $\Delta\sigma^2$ can vary continuously.
\end{itemize}

\subsubsection[]{For branes parallel along the $\bZ_2$ invariant $T^2_2$}\label{Sss:||inv}

\begin{itemize}
\item
The bifundamental states for $a$ and $(\theta^k b)$ parallel on $T^2_2$ for a given $k$
and at angles on $T^2_1 \times T^2_3$
are grouped in ${\cal N}=2$ supersymmetric 
non-chiral pairs counted by twice the intersection number on 
$T^2_1 \times T^2_3$. The corresponding term for the given $k$ in table~\ref{NonChiralSpectrum} is replaced by
\begin{equation}
\begin{aligned}
\varphi^{ab,\parallel T^2_2}
&\rightarrow \phantom{\frac{1}{2}}
\left|  I_{a(\theta^k b)}^{(1 \cdot 3)} +I_{a(\theta^k b)}^{\bZ_2,(1 \cdot 3)} \right|  ,
\\
\varphi^{\Adj_a, \parallel  T^2_2}  
&\rightarrow  \frac{1}{2} \left| I_{a(\theta^k a)}^{(1 \cdot 3)} 
+I_{a(\theta^k a)}^{\bZ_2,(1 \cdot 3)}\right|  ,
\end{aligned}
\end{equation}
where the upper index $(1\!\cdot\!3)$ indicates that intersection numbers are computed
only on $T^2_1 \times T^2_3$, e.g. $ I_{ab}^{(1 \cdot 3)} = \prod_{i=1,3} (n^a_i m^b_i - m^a_i n^b_i) $.
This case applies e.g. to the adjoint representations in a $T^6/\bZ_4$ background and is found to agree with the 
$({\bf 6}_i,{\bf 6}_{i+2})$ spectra of~\cite{bgk00,bcs04}.
The modification  for $a$ parallel to some orientifold image brane $(\theta^k b')$ is obvious. 
\item
If brane $a$ and its orientifold image $(\theta^k a')$ are {\it parallel} to the
$\OR\theta^{-k}$ and $\OR\theta^{-k+N}$ invariant O6-planes on $T^2_2$ for some $k$, the corresponding 
sector stays ${\cal N}=2$ supersymmetric,
and the  matter states for the given $k$ are counted by
\begin{equation}
\left.\begin{array}{c} 
\varphi^{\Anti_a,\parallel T^2_2\parallel \OR\theta^{-k(+N)}}  
 \\ \varphi^{\Sym_a,\parallel T^2_2\parallel \OR\theta^{-k(+N)}}  
\end{array} \right\}  
 \rightarrow
\frac{1 }{2}\left| I_{a(\theta^k a')}^{(1\cdot 3)} + I_{a(\theta^k a')}^{\bZ_2,(1\cdot 3)}
 \pm \left( I_{a}^{\OR\theta^{-k},(1\cdot 3)} + I_{a}^{\OR\theta^{-k+N},(1\cdot
3)} \right)\right|,
\end{equation}
the upper sign being valid for antisymmetric representations and the lower one for symmetric ones.
\item
If brane $a$ and its orientifold image $(\theta^k a')$ are parallel among themselves 
but {\it orthogonal} to the
$\OR\theta^{-k}$ and $\OR\theta^{-k+N}$ invariant O6-planes on $T^2_2$ for some given $k$, the O6-planes break 
  ${\cal N}=2$ down to ${\cal N}=1$ supersymmetry, as can be seen also from the fact that 
the net-chirality of symmetric and antisymmetric states is in general opposite and non-vanishing
for the given $k$, 
\begin{equation}\label{Eq:chirality||T2-|OR}
\left.\begin{array}{c} \chi^{\Anti_a, \parallel T^2_2\perp \OR\theta^{-k(+N)}} 
\\ \chi^{\Sym_a, \parallel T^2_2\perp \OR\theta^{-k(+N)}}  
\end{array} \right\} 
\rightarrow
\mp \frac{1}{4} \left( I_a^{\OR\theta^{-k}} + I_a^{\OR\theta^{-k+N}}  \right).
\end{equation}
There is an equal amount of antisymmetric and symmetric representations (up to complex conjugation)
for the given $k$, 
\begin{equation}\label{eq:a-|OR||T2}
\left.\begin{array}{c} \varphi^{\Anti_a,\parallel T^2_2\perp \OR\theta^{-k(+N)}} 
\\ \varphi^{\Sym_a,\parallel T^2_2\perp \OR\theta^{-k(+N)}} 
\end{array} \right\} 
\rightarrow
\frac{1 }{2}\left| I_{a(\theta^k a')}^{(1\cdot 3)} + I_{a(\theta^k a')}^{\bZ_2,(1\cdot 3)}\right|.
\end{equation}
This case had not been discussed in~\cite{gmho07} and is presented here for the first time.
\end{itemize}

\subsubsection[]{For branes parallel along $T^2_3$ where $\bZ_2$ acts non-trivially}\label{Sss:||T3}

The case with branes parallel along $T^2_1$ arises by permutation of indices and is not listed here in detail. 
\begin{itemize}
\item
The sector of branes  $a$ and $(\theta^k b)$ parallel to each other on $T^2_3$ for some $k$ 
is in general ${\cal N}=1$ supersymmetric
with net-chirality
\begin{equation}\label{Eq:T3-netchiral}
\chi^{ab,\parallel T^2_3} 
 \rightarrow -\frac{1}{2}I_{a(\theta^k b)}^{\bZ_2} ,
\end{equation}
which turns out to be zero for a non-vanishing relative distance or Wilson line,
\begin{equation}
(\Delta\sigma^3_{a(\theta^k b)},\Delta\tau^3_{a(\theta^k b)}) \neq (0,0).
\end{equation}
The total counting of massless bifundamental states for branes $a$ and $(\theta^k b)$ on top of each other on $T^2_3$ is
\begin{equation}
\begin{aligned}
\varphi^{ab,\parallel T^2_3} 
&\rightarrow  
\left\{\begin{array}{cc}
\left| I_{a(\theta^k b)}^{(1 \cdot 2)}   \right| & (\Delta\sigma^3_{a(\theta^k b)},\Delta\tau^3_{a(\theta^k b)}) = (0,0)
\\
0 & (\Delta\sigma^3_{a(\theta^k b)},\Delta\tau^3_{a(\theta^k b)}) \neq (0,0)
\end{array}\right.,
\\[1ex]
\varphi^{\Adj_a, \parallel T^2_3} 
&\rightarrow  \frac{1}{2}\left| I_{a(\theta^k a)}^{(1 \cdot 2)}   \right|
\quad {\rm for }\quad k \neq 0.
\end{aligned}
\end{equation}
This formula in particular applies to the counting of adjoint representation in a $T^6/\bZ_6'$ 
background where $\varphi^{\Adj} = 1 + \prod_{i=1}^2 |n_i^2 + n_i m_i + m_i^2|$.
\item
The O6-planes invariant under $\OR\theta^{-k}$ and $\OR\theta^{-k+N}$ are orthogonal to each other on $T^2_3$, and 
branes $a$ lie on one of these if they are on top of $(\theta^k a')$ for some $k$.
The sector is ${\cal N}=1$ supersymmetric with net-chirality
\begin{equation}\label{Eq:T2-netchiral}
\begin{aligned}
\left.\begin{array}{c} \chi^{\Anti_a, \parallel T^2_3} 
\\ \chi^{\Sym_a,\parallel T^2_3} 
\end{array} \right\} 
&\rightarrow
- \frac{1}{4} \left( I_{a(\theta^k a')}^{\bZ_2} \pm  I_a^{\OR\theta^{-x}} \right)
\\ 
&\text{with the exponent} \quad  x=
\left\{\begin{array}{cc} 
k & a \perp \OR\theta^{-k} \; {\rm on} \; T^2_3
\\
k+N & a \perp \OR\theta^{-k+N} \; {\rm on} \; T^2_3
\end{array}\right.
.
\end{aligned}
\end{equation}
The total number of antisymmetric and symmetric states is counted by 
\begin{equation}\label{eq:a||T3-S+W}
\begin{aligned}
\left.\begin{array}{c} \varphi^{\Anti_a,\parallel T^2_3} 
\\ \varphi^{\Sym_a,\parallel T^2_3} 
\end{array} \right\} 
&\rightarrow
\frac{1 }{2}\left| I_{a(\theta^k a')}^{(1 \cdot 2)} \pm \tilde{c} \, I_a^{\OR\theta^{-y}}  \right|
\\ 
&{\rm with} \quad  y=
\left\{\begin{array}{cc} 
k & a \parallel \OR\theta^{-k} \; {\rm on} \; T^2_3
\\
k+N & a \parallel \OR\theta^{-k+N} \; {\rm on} \; T^2_3
\end{array}\right.
.
\end{aligned}
\end{equation}
The constant $\tilde{c}= e^{2 \pi i \tau^3 \Delta\sigma^3}$ takes into account that for branes which are 
displaced from the origin on $T^2_3$, the $\bZ_2$ invariant intersection points are exchanged under
the orientifold projection, and a sign factor arises if the brane carries a discrete Wilson line on $T^2_3$.
This permutation of $\bZ_2$ fixed points only occurs on tilted tori such as the {\bf b} type torus on $T^2_3$
or the $SU(3)$ invariant ones {\bf A} and {\bf B} on $T^2_1$, but not on the {\bf a} type $T^2_3$.   
\\
The sign factor $\tilde{c}$ had not been observed in~\cite{gmho07}, but
is necessary in section~\ref{App:Realisations} in order to avoid
a mismatch in the  chiral and non-chiral counting of antisymmetric states, $|\chi^{\Anti_b, \parallel T^2_1}|
\leq \varphi^{\Anti_b,\parallel T^2_1}$, on stack $b$  defined in tables~\ref{tab:sm_bulk_branes} and~\ref{tab:sm_ex_branes}. 
\end{itemize}

In the  following section we verify explicitly the correctness of these formulae (in the absence 
of Wilson lines) using Chan-Paton labels. The antisymmetric and symmetric representations on orientifold
invariant D6-branes with $SO(2N)$ or $Sp(2N)$ gauge groups require a different treatment displayed in appendix~\ref{App:SO-Sp}.

\subsection{Massless spectrum from Chan-Paton matrices}\label{Subsec:CPmatrices}

The $\gamma$-matrices associated to the $\bZ_2 \equiv \theta^N$ and $\OR\theta^{-k}$ action on the Chan-Paton matrices 
$\lambda_{a(\theta^k b)}$ and $\lambda_{a(\theta^k a')}$  in a $T^6/\bZ_{2N}$ background
can be chosen in 
agreement with~\cite{blkFB02}, where  {\it bulk} D7-branes on $T^4/\bZ_2$ had been analysed,
\begin{equation}\label{Eq:Gamma-Matrix-Reps}
\begin{aligned}
&\gamma_{\bZ_2} = \left(\begin{array}{cc}
\unity & 0 \\ 0 & -\unity
\end{array}\right),
\quad 
\gamma_{\OR\theta^{-k}} = \left(\begin{array}{cc}
0 & \unity \\ \unity & 0
\end{array}\right) ,
\\&\gamma_{\OR\theta^{-k+N}} \equiv \gamma_{\OR\theta^{-k}} \cdot \gamma_{\bZ_2} 
= \left(\begin{array}{cc}
0 & -\unity \\ \unity & 0
\end{array}\right)
.
\end{aligned}
\end{equation}
The gauge group for a $2N_a \times 2N_a$ matrix $\lambda_{aa}$ derived in~\cite{blkFB02} is  
$U(N_a) \times U(N_a)$ for a generic brane $a \neq (\theta^k a')$ not on top of the O7-planes and 
$U(N_a)$ if $a =(\theta^k a')$ for some $k$ .
This set-up can be viewed as the special case of {\it  fractional} D7-branes $a_i$ ($i=1,2$) 
and their orbifold images $(\theta^k a_i)$ wrapping  fractional cycles $\Pi^{\rm frac}_{a_i}$ such that 
\begin{equation}
\Pi^{\rm bulk}_a = \Pi^{\rm frac}_{a_1} + \Pi^{\rm frac}_{a_2} \quad {\rm and} \quad N_a=N_a^1=N_a^2 .
\end{equation}
Allowing for different stack sizes $N_a^i$, $N_b^j \geq 0$  for $i,j=1,2$, the Chan-Paton labels decompose as 
\begin{equation}\label{Eq:CP-fractional}
\begin{aligned}
\lambda_{a(\theta^k b)}  &
\simeq \left(\begin{array}{cc}
({\bf N^1}_a,\ov{\bf N^1}_b) & ({\bf N^1}_a,\ov{\bf N^2}_b) \\
({\bf N^2}_a,\ov{\bf N^1}_b) & ({\bf N^2}_a,\ov{\bf N^2}_b)
\end{array}\right), 
\\[1ex]
\lambda_{a(\theta^k a')}  &
\simeq \left(\begin{array}{cc}
({\bf N^1}_a,{\bf N^2}_a) & \Anti^{\bf 1}_a + \Sym^{\bf 1}_a \\
 \Anti^{\bf 2}_a + \Sym^{\bf 2}_a & ({\bf N^2}_a,{\bf N^1}_a)
\end{array}\right),
\end{aligned}
\end{equation}
where for the $a(\theta^k a')$ states it was used that the orientifold projection
exchanges $\bZ_2$ eigenvalues $\tau^0_a \rightarrow \tau^0_{a'} =
\tau^0_a +1$ and acts as 
complex conjugation on the representations, i.e.
${\bf N^1}_{a'} = \ov{\bf N^2}_a$ and ${\bf N^2}_{a'} = \ov{\bf N^1}_a$.

The line of argument carries directly over to fractional D6-branes and O6-planes in a $T^6/\bZ_{2N}$ background
with a single $\bZ_2$ sub-symmetry. The cases with $T^6/\bZ_{2N} \times \bZ_{2M}$ backgrounds 
require a second $\gamma_{\bZ_2}$ matrix for the other $\bZ_2$ sub-symmetry and will be discussed elsewhere~\cite{WIP}.

In order to determine the massless spectrum, one further needs to know the invariance properties
of the brane intersections in order to determine which projection condition on the Chan-Paton
label applies, and finally the number of massless states, their chiralities and the associated eigenvalues 
$(a_{\bZ_2}, a_{\OR\theta^{-k}},a_{\OR\theta^{-k+N}})$ under the 
orbifold and orientifold projections with $a_{\OR\theta^{-k+N}} = a_{\bZ_2} \cdot a_{\OR\theta^{-k}}$ for
consistency.

Let us start by listing the massless multiplets for branes at relative angles 
$\pi(\phi_1,\phi_2,\phi_3)$ which comply with the supersymmetry constraint 
$\sum_{i=1}^3 \phi_i = 0 \; {\rm mod} \; 2$:
\begin{itemize}
\item
On parallel branes with $\pi(\phi_1,\phi_2,\phi_3)=(0,0,0)$ there exist
a $\bZ_2$ even vector and chiral multiplet plus  two further $\bZ_2$ odd chiral multiplets.
\item
At intersections of branes $a$ and $(\theta^k b)$ with  all $\phi_i \neq 0$, 
the $a(\theta^k b)$ sector provides one bosonic and fermionic degree of freedom. Together with the 
$(\theta^k b)a$ sector, these group into a $\bZ_2$ even chiral multiplet.
\item
Branes parallel along the $\bZ_2$ invariant two-torus $T^2_2$ at angles $\pi(\phi,0,-\phi)$ 
provide two $\bZ_2$ odd multiplets of opposite chirality.
\item
Branes parallel along a two-torus where $\bZ_2$ acts non-trivially, i.e. $\pi(\phi,-\phi,0)$ or
$\pi(0,\phi,-\phi)$, provide one $\bZ_2$ even and one $\bZ_2$ odd  multiplet
with opposite chiralities.
\end{itemize}
More details on the NS and R states are given in appendix~\ref{A:MasslessStates}, see 
in particular table~\ref{Tab:Massless-States}.

The chirality of the massless multiplets in each sector can in principle directly be read off from the massless
R state in  table~\ref{Tab:Massless-States}. We will in the following, however, ignore relative chiralities
among different sectors since they are easily recovered from the three-cycle intersection numbers in equations~\eqref{Eq:Def_Inters},
\eqref{eq:a-|OR||T2}, \eqref{Eq:T3-netchiral} and \eqref{Eq:T2-netchiral}.
Opposite chiralities within a given sector are on the other hand taken into account.

\subsubsection{Branes parallel on all tori}

The case for parallel branes has been discussed in detail in section~\ref{Sss:||inters}.
In the absence of Wilson lines, the representations pertaining to $\bZ_2$ even states
are read off from the diagonal entries in $\lambda_{a(\theta^k b)}$, those of the $\bZ_2$ odd states
from the off-diagonals in~\eqref{Eq:CP-fractional}.

\subsubsection{Branes at non-trivial angles}\label{Sss:Angles}

The matter states on branes with non-vanishing intersection numbers on all tori can be computed wholly
from the intersection numbers. We verify here that using the method of Chan-Paton labels, 
one recovers the spectrum in table~\ref{NonChiralSpectrum}. The present computation carries over
to the case with a vanishing angle as discussed below in sections~\ref{Sss:CP||T2} and~\ref{Sss:CP||T3}.

For  two branes $a$ and $(\theta^k b)$ at non-trivial angles,
the intersections can be either $\bZ_2$ invariant with their abundance
counted by $x_{\bZ_2} \equiv |I_{a(\theta^k b)}^{\bZ_2}|$ or form pairs under $\bZ_2$ which are counted by
\begin{equation}
x_{\bZ_2-{\rm pairs}}\equiv \frac{|I_{a(\theta^k b)}| -|I_{a(\theta^k b)}^{\bZ_2}| }{2}.
\end{equation}
At this point, we focus on branes without relative Wilson lines, $\tau^i_a=\tau^i_b$ for $i=1,3$,
but include arbitrary $\bZ_2$ eigenvalues $\tau^0_a,\tau^0_b \in \{0,1\}$. 
We will briefly comment on the case with relative Wilson lines below.
The following representations occur:
\begin{itemize}
\item
At intersection points which are exchanged by the $\bZ_2$ symmetry, there is no projection on the 
Chan-Paton matrix. Independently of any relative $\bZ_2$ eigenvalue $i,j=1,2$, the contribution to the massless spectrum is
\begin{equation}
x_{\bZ_2-{\rm pairs}} \times 
({\bf N}_a^i,\overline{\bf N}_b^j).
\end{equation} 
\item
At $\bZ_2$ invariant intersections, the projection on the Chan-Paton label 
leads to
\begin{equation}
a_{\bZ_2} = 1: 
x_{\bZ_2}  \times ({\bf N}_a^i,\overline{\bf N}_b^i),
\quad {\rm or} 
\quad
a_{\bZ_2} = -1: 
x_{\bZ_2}  \times ({\bf N}_a^i,\overline{\bf N}_b^j)_{i \neq j}.
\end{equation} 
\item 
Adding up the contributions from $\bZ_2$ invariant intersection points and pairs which are exchanged
while taking care of relative signs among $I_{a(\theta^k b)}$ and $ I_{a(\theta^k b)}^{\bZ_2}$ leads to
\begin{equation}\label{Eq:ab-mult}
\frac{|I_{a(\theta^k b)} + I_{a(\theta^k b)}^{\bZ_2}| }{2} \times ({\bf N}_a^i,\overline{\bf N}_b^j)
\quad
{\rm for} 
\quad
i,j=1,2
,
\end{equation} 
which is clearly in agreement with the multiplicity derived from intersection numbers alone in 
table~\ref{NonChiralSpectrum}. 
\end{itemize}
Our derivation of massless matter representations 
is tailor-made for branes without relative Wilson lines. However, from the construction
and by comparison with the result in table~\ref{NonChiralSpectrum},
one sees clearly that a relative Wilson line will provide a flip of the  sign $a_{\bZ_2}$ in the 
projection of the Chan-Paton matrix at some of the $\bZ_2$ invariant intersection points. 
The counting of massless bifundamental representations~\eqref{Eq:ab-mult} remains valid, but 
$x_{\bZ_2} = |I_{a(\theta^k b)}^{\bZ_2}|$ looses the geometric interpretation of counting simply $\bZ_2$ 
invariant intersection points. Instead, it counts intersections with relative signs.

For the computation of  symmetric and antisymmetric representations 
the orbits of intersection numbers have to be divided differently into 
their invariance properties and abundances as follows:
\begin{itemize}
\item
points fixed under both $\OR\theta^{-k}$ and $\bZ_2$:
\begin{equation}
y_{\bZ_2 + \OR\theta^{-k}} \equiv |I_a^{\bZ_2 + \OR\theta^{-k}}|,
\end{equation}
\item
orbits of points fixed under $\bZ_2$ but forming pairs under 
$\OR\theta^{-k}$:
\begin{equation}
y_{\bZ_2} \equiv \frac{|I_{a(\theta^k a')}^{\bZ_2}| - |I_a^{\bZ_2 + \OR\theta^{-k}}|}{2},
\end{equation}
\item
orbits of points fixed under $\OR\theta^{-k}$ or  $\OR\theta^{-k+N}$ but forming 
pairs under $\bZ_2$:
\begin{equation}
\begin{aligned}
y_{\OR\theta^{-k}}&
\equiv \frac{|I_{a}^{\OR\theta^{-k}}| - |I_a^{\bZ_2 + \OR\theta^{-k}}|}{2}\quad\mbox{and}\\
y_{\OR\theta^{-k+N}}&
\equiv \frac{|I_{a}^{\OR\theta^{-k+N}}| - |I_a^{\bZ_2 + \OR\theta^{-k}}|}{2},
\end{aligned}
\end{equation}
\item
orbits of intersection points which are not fixed under $\bZ_2$, $\OR\theta^{-k}$ or  $\OR\theta^{-k+N}$:
 \begin{equation}
y_0 \equiv
\frac{1}{4} \, \bigl\{ %
|I_{a(\theta^k a')}|
-|I_{a(\theta^k a')}^{\bZ_2}|
-|I_{a}^{\OR\theta^{-k}}|
-|I_{a}^{\OR\theta^{-k+N}}|
+ 2 \, |I_a^{\bZ_2 + \OR\theta^{-k}}|
\bigr\}.
\end{equation}
\end{itemize}
The quantity $|I_a^{\bZ_2 + \OR\theta^{-k}}|$ can be determined on a case-by-case basis 
for a given brane configuration and $T^6/\bZ_{2N}$ background.
As a simple check, one can verify that these orbits add up to the total
number of intersections,
\begin{equation}
|I_{a(\theta^k a')}| = 4 \, y_0 + 2 \, \left(y_{\bZ_2} + y_{\OR\theta^{-k}} + y_{\OR\theta^{-k+N}}  \right) + y_{\bZ_2 + \OR\theta^{-k}} .
\end{equation}
We do not give the details here since this number does not appear in the final result of 
counting chiral and non-chiral matter representations. 

Depending on the various signs in the
orbifold or orientifold projections, the states are as given in table~\ref{Tab:CP-AntiSym}.
\begin{table}[ht]
  \begin{center}
    \begin{equation*}
      \begin{array}{|c|c|c|} \hline
        \multicolumn{3}{|c|}{\rule[-3mm]{0mm}{8mm}
\text{\bf Matter at $a(\theta^k a')+(\theta^k a')a$ intersections on }
 T^6/\bZ_{2N}   }\\ \hline\hline
{\rm mult.} & (a_{\bZ_2},a_{\OR\theta^{-k}},a_{\OR\theta^{-k+N}}) & {\rm rep.}
\\\hline\hline
y_0 & (*,*,*) & \Anti_a^i+ \Sym_a^i
\\\hline
y_{\bZ_2} & (+,*,*) &  -
\\
& (-,*,*) & \Anti_a^i+ \Sym_a^i
\\\hline
y_{\OR\theta^{-k}} & (* ,+ ,*) & \Sym_a^i
\\
& (*,-,*) & \Anti_a^i
\\\hline 
y_{\OR\theta^{-k+N}} & (*,*,+) & \Anti_a^i
\\
& (*,*,-) & \Sym_a^i
\\\hline 
y_{\bZ_2+\OR\theta^{-k}} & (+,\pm,\pm) & -
\\
& (-,+,-) & \Sym_a^i
\\
& (-,-,+) & \Anti_a^i
\\ \hline
     \end{array}
    \end{equation*}
  \end{center}
\caption{Counting of symmetric and antisymmetric representations at 
$a(\theta^k a')+(\theta^k a')a$ intersections in a $T^6/(\bZ_{2N} \times \OR)$ background.
A star * denotes no projection condition. Only two signs are independent, 
$a_{\OR\theta^{-k+N}}=a_{\bZ_2} \cdot a_{\OR\theta^{-k}}$. }
\label{Tab:CP-AntiSym}
\end{table}
The formula in table~\ref{NonChiralSpectrum} is recovered by
adding all matter states up for a given choice of $(a_{\bZ_2},a_{\OR\theta^{-k}},a_{\OR\theta^{-k+N}})$.

\subsubsection[]{Branes parallel along the $\bZ_2$ invariant torus}\label{Sss:CP||T2}

For branes parallel along $T^2_m$ and at angles on $T^2_n \times T^2_p$ with
$\{m,n,p\}$ cyclic permutations of $\{1,2,3\}$, a similar counting of orbits 
of intersection points can be performed as in 
section~\ref{Sss:Angles}. 
In the following, we will label these orbits of points on $T^2_n \times T^2_p$ for branes parallel along $T^2_m$ 
by an upper index $(m)$. The abundance of $\bZ_2$ invariant 
intersection points on $T^2_1 \times T^2_3$ for branes parallel along $T^2_2$ 
is, e.g., denoted by 
$x_{\bZ_2}^{(2)} =|I_{ab}^{\bZ_2,(1 \cdot 3)}| 
= | \sum_{x_a^k x_b^k} (-1)^{\tau_{x_a^1x_a^3} +\tau_{x_b^1x_b^3}} \delta_{x_a^1 x_b^1} \delta_{x_a^3 x_b^3}|$.\footnote{The geometric interpretation is again strictly only valid for vanishing relative Wilson lines $\tau^i_a=\tau^i_b$ ($i=1,3$)
as in the case with
three non-trivial angles.}

The two massless multiplets for branes parallel along $T^2_2$ have the same (negative) $\bZ_2$ eigenvalue as listed
in table~\ref{Tab:Massless-States} with opposite chirality thereby forming an ${\cal N}=2$ hyper multiplet. 
This statement remains true for branes $a$ and $(\theta^k a')$ parallel along $T^2_2$ for some $k$ and also parallel to the
$\OR\theta^{-k}$ and $\OR\theta^{-k+N}$ invariant O6-planes there. In these two cases, the spectrum is computed simply
from intersection numbers on $T^2_1 \times T^2_3$ leading to the states listed in table~\ref{Tab:CP-parallel-T2}.
\begin{table}[ht]
  \begin{center}
    \begin{equation*}
      \begin{array}{|c|c|c|} \hline
        \multicolumn{3}{|c|}{\rule[-3mm]{0mm}{8mm}
\text{\bf ${\cal N}=2$ Matter for 
branes} \parallel \text{ on } T^2_2 \text{ and } \parallel \OR\theta^{-k} }\\ \hline\hline
{\rm mult.} & (a_{\bZ_2},a_{\OR\theta^{-k}},a_{\OR\theta^{-k+N}}) & {\rm rep.}
\\\hline\hline
x_{\bZ_2-{\rm pairs}}^{(2)} & (*,*,*) &  ({\bf N}_a^i,\overline{\bf N}_b^j) + \cc
\\\hline
x_{\bZ_2}^{(2)} & (+,*,*) & ({\bf N}_a^i,\overline{\bf N}_b^i)+ \cc
\\
& (-,*,*) & ({\bf N}_a^i,\overline{\bf N}_b^j)_{i \neq j}+ \cc
\\\hline\hline
y_0^{(2)} & (*,*,*) & \Anti_a^i + \Sym_a^i + \cc
\\\hline
y_{\bZ_2}^{(2)} & (+,*,*) &  -
\\
& (-,*,*) & \Anti_a^i+ \Sym_a^i + \cc
\\\hline
y_{\OR\theta^{-k}}^{(2)} & (* ,+ ,*) & \Sym_a^i+ \cc
\\
& (*,-,*) & \Anti_a^i+ \cc
\\\hline 
y_{\OR\theta^{-k+N}}^{(2)} & (*,*,+) & \Anti_a^i + \cc
\\
& (*,*,-) & \Sym_a^i + \cc
\\\hline 
y_{\bZ_2+\OR\theta^{-k}}^{(2)} & (+,\pm,\pm) & -
\\
& (-,+,-) & \Sym_a^i+ \cc
\\
& (-,-,+) & \Anti_a^i+ \cc
\\ \hline
     \end{array}
    \end{equation*}
  \end{center}
\caption{${\cal N}=2$ supersymmetric sectors occurring for branes parallel along the $\bZ_2$ invariant $T^2_2$. 
The symmetric and antisymmetric representations only fit into ${\cal N}=2$ multiplets if they are also parallel
to the $\OR\theta^{-k}$ and $\OR\theta^{-k+N}$ invariant O6-planes. For branes orthogonal to these O6-planes see 
table~\protect\ref{Tab:CP-perp-T2}.}
\label{Tab:CP-parallel-T2}
\end{table}

The situation changes if brane $a$ and its orientifold image $(\theta^k a')$ are parallel along $T^2_2$ for some $k$, 
but orthogonal to the $\OR\theta^{-k}$ and $\OR\theta^{-k+N}$ invariant O6-planes. In this case, the orientifold
projection breaks half of the supersymmetry by assigning $a_{\OR\theta^{-k}}=1$ to one 
${\cal N}=1$ chiral multiplet and
$a_{\OR\theta^{-k}}=-1$ to the one with opposite chirality. The resulting matter spectrum is listed in 
table~\ref{Tab:CP-perp-T2}.
\begin{table}[ht]
  \begin{center}
    \begin{equation*}
      \begin{array}{|c|c|c|} \hline
        \multicolumn{3}{|c|}{\rule[-3mm]{0mm}{8mm}
\text{\bf ${\cal N}=1$ Matter for 
image branes } \parallel
\text{ on } T^2_2 \text{ and } \perp \OR\theta^{-k} }\\ \hline\hline
{\rm mult.} & (a_{\bZ_2},a_{\OR\theta^{-k}},a_{\OR\theta^{-k+N}}) & {\rm rep.}
\\\hline\hline
y_0^{(2)} & (*,*,*) & \Anti_a^i+ \Sym_a^i + \cc
\\\hline
y_{\bZ_2}^{(2)} & (+,*,*) &  -
\\
& (-,*,*) &  \Anti_a^i+ \Sym_a^i + \cc
\\\hline
y_{\OR\theta^{-k}}^{(2)} & (* ,+ ,*) &  \ov{\Anti}_a^i+ \Sym_a^i
\\
& (*,-,*) & \Anti_a^i+ \ov{\Sym}_a^i
\\\hline 
y_{\OR\theta^{-k+N}}^{(2)} & (*,*,+) & \Anti_a^i+ \ov{\Sym}_a^i
\\
& (*,*,-) & \ov{\Anti}_a^i+ \Sym_a^i
\\\hline 
y_{\bZ_2+\OR\theta^{-k}}^{(2)} & (+,\pm,\pm) & -
\\
& (-,+,-) & \ov{\Anti}_a^i+ \Sym_a^i
\\
& (-,-,+) & \Anti_a^i+ \ov{\Sym}_a^i
\\ \hline
     \end{array}
    \end{equation*}
  \end{center}
\caption{${\cal N}=1$ supersymmetric sectors occurring for branes $a$ and $(\theta^k a')$ parallel along the 
$\bZ_2$ invariant $T^2_2$ but perpendicular to the $\OR\theta^{-k}$ and $\OR\theta^{-k+N}$ invariant O6-planes. }
\label{Tab:CP-perp-T2}
\end{table}
This part of the spectrum has net-chirality $\pm \frac{1}{2} 
\left(\left|I_a^{\OR\theta^{-k},(1\cdot 3)}\right| - \left|I_a^{\OR\theta^{-k+N},(1\cdot 3)}\right|\right)$
for antisymmetric and symmetric representations
with the sign inside the parenthesis corresponding to the fact that we have identified the multiplets
as being $\bZ_2$ odd. This result fits nicely with~\eqref{Eq:chirality||T2-|OR} since the contribution from $T^2_2$ is  
\begin{equation}
N_{O6}^{\OR\theta^{-x},(2)} |I_a^{\OR\theta^{-x},(2)}|=2
\end{equation}
for any lattice and any exponent $x$.\footnote{Throughout the  article we use the sloppy notation $I_a^{\OR\theta^{-x}}$
instead of $N_{O6}^{\OR\theta^{-x}} I_a^{\OR\theta^{-x}}$ since for a $\bZ_3$ or $\bZ_6$ invariant two-torus $T^2_m$, 
the number of $\OR\theta^{-x}$ invariant planes is 
$N_{O6}^{\OR\theta^{-x},(m)}=1$. However, on a $\bZ_2$ invariant torus such as $T^2_3$ for $T^6/\bZ_6'$ we have 
$N_{O6}^{\OR\theta^{-x},(m)}=2 \, (1-b)$, and for a  $\bZ_4$ invariant $T^2_m$, $N_{O6}^{\OR\theta^{-x},(m)}=1$ 
for $x=1,3$ on the {\bf A} type lattice and   $N_{O6}^{\OR\theta^{-x},(m)}=2$ for $x=0,2$. For the {\bf B} orientation
on a $\bZ_4$ invariant torus, the values of $N_{O6}^{\OR\theta^{-x},(m)}$ for $x$ even and odd are exchanged
compared to the {\bf A} torus.}

\subsubsection[]{Branes parallel along a torus where $\bZ_2$ acts}\label{Sss:CP||T3}

The present case applies to branes parallel on $T^2_1$ or $T^2_3$ in the $T^6/\bZ_6'$ background. 
For concreteness, we focus on the latter.
As shown in table~\ref{Tab:Massless-States}, the two massless multiplets have opposite chirality and opposite
$\bZ_2$ eigenvalue, and according to the intersection numbers in section~\ref{Sss:||T3}, net-chiralities arise 
in this sector. 
\begin{table}[ht]
  \begin{center}
    \begin{equation*}
      \begin{array}{|c|c|c|} \hline
        \multicolumn{3}{|c|}{\rule[-3mm]{0mm}{8mm}
\text{\bf ${\cal N}=1$ Matter for 
branes parallel on } T^2_3  }\\ \hline\hline
{\rm mult.} & (a_{\OR\theta^{-k}}) & {\rm rep.}
\\\hline\hline
x_{\bZ_2-{\rm pairs}}^{(3)} & (*) & ({\bf N}_a^i,\ov{\bf N}_b^j) + \cc
\\\hline
x_{\bZ_2}^{(3)} & (*) &  ({\bf N}_a^i,\ov{\bf N}_b^i) \quad {\rm or} \quad (\ov{\bf N}_a^i,{\bf N}_b^j)_{i\neq j}
\\\hline\hline
y_0^{(3)} & (*) & \Anti_a^i + \Sym_a^i + \cc
\\\hline
y_{\bZ_2}^{(3)} & (*) &  \ov{\Anti}_a^i + \ov{\Sym}_a^i
\\\hline
y_{\OR\theta^{-k}}^{(3)} & (+) & \Sym_a^i + \cc
\\
& (-) & \Anti_a^i + \cc
\\\hline
y_{\OR\theta^{-k+N}}^{(3)} & (+) & \Sym_a^i + \ov{\Anti}_a^i
\\
& (-) & \Anti_a^i + \ov{\Sym}_a^i
\\\hline
y_{\bZ_2+\OR\theta^{-k}}^{(3)} & (+) & \ov{\Sym}_a^i
\\
& (-) & \ov{\Anti}_a^i
\\ \hline
     \end{array}
    \end{equation*}
  \end{center}
\caption{${\cal N}=1$ supersymmetric sectors occurring for branes $a$ and $(\theta^k b)$ or 
$(\theta^k a')$ parallel along the two-torus $T^2_3$ where $\bZ_2$ acts non-trivially. The symmetric and antisymmetric 
representations are 
listed for the case of identical $\OR\theta^{-k}$ eigenvalue for the two multiplets of opposite chirality and $\bZ_2$ 
eigenvalue. 
}
\label{Tab:CP-par-T3}
\end{table}
In the absence of Wilson lines, the spectrum is given in table~\ref{Tab:CP-par-T3}, where we assumed that
both massless multiplets have the same $\OR\theta^{-k}$ transformation. 
If instead the $\OR\theta^{-k+N}$ eigenvalue is identical, the representations for $y_{\OR\theta^{-k}}^{(3)}$
and $y_{\OR\theta^{-k+N}}^{(3)}$ have to be exchanged while replacing $a_{\OR\theta^{-k}} \leftrightarrow - a_{\OR\theta^{-k+N}}$
and taking complex conjugates except for the $y_{\bZ_2}^{(3)}$ entry which is independent of the orientifold projection.
The generalisation to including Wilson lines along the two-torus with $\bZ_2$ action and a non-vanishing angle 
is implemented as before by adjusting the interpretation of $x^{(3)}_{\bZ_2-{\rm pairs}}$ as a sum 
of $\bZ_2$ invariant intersection points including relative signs.
A relative Wilson line along the parallel direction provides a mass, see equation~\eqref{Eq:masses}.

Bifundamental matter at intersections with orientifold invariant branes is computed in the same way, 
the $\OR\theta^{-k}$ action on Chan-Paton labels, however, differs and is discussed in appendix~\ref{App:SO-Sp}.

\section{Gauge couplings}\label{Sec:GC}

The gauge coupling constant $g_a$  for a gauge factor $G_a$ at the energy scale $\mu<M_{\rm string} $
is at one loop given by
\begin{equation}
\frac{8 \pi^2}{g_a^2(\mu)} = \frac{8 \pi^2}{g_{a,{\rm string}}^2}
+\frac{b_a}{2} \ln\left(\frac{M_{\rm string}^2}{\mu^2}  \right) + \frac{\Delta_a}{2}.
\end{equation}
The three contributions on the right hand side are 
\begin{itemize}
\item
The tree level gauge coupling $g_{a,{\rm string}}$ and fine structure constant $\alpha_{a,{\rm string}}$
(see e.g.~\cite{bls03})
\begin{equation}\label{Eq:alpha}
\frac{1}{\alpha_{a,{\rm string}}} =
\frac{4 \pi }{g^2_{a,{\rm string}}} = \frac{M_{\rm Planck}}{2 \sqrt{2}\kappa_a M_{\rm string}}\frac{V_a}{\sqrt{V_6}},
\end{equation}
where $V_6$ is the six dimensional compact volume, $V_a$ the volume of the three-cycle wrapped by brane $a$ 
and $\kappa_a = 1$ for $SU(N_a)$ gauge groups. $\kappa_a=2$ applies to $Sp(2N_a)$ and $SO(2N_a)$ gauge groups.
The dilaton dependence has been eliminated from~\eqref{Eq:alpha} by inserting 
the gravitational and string scales, $M_{\rm Planck}$ and $M_{\rm string}$, and $\alpha^{-1}_{a,{\rm string}}$ depends
only on the complex structure moduli. A {\it universal} one-loop correction is included in~\eqref{Eq:alpha}
when the redefinition of the dilaton and complex structure moduli at one-loop level is inserted~\cite{abls07b}.  
At the orbifold point, all exceptional cycles have zero volume, and the volume of a fractional cycle
is simply given by its toroidal part,
\begin{equation}
\begin{aligned}
V_a =& c \, L_1^a \cdot L_2^a \cdot L_3^a 
\\
=& c \left(r_1 r_2 R_1\right) \left[\,
\left(1 + b^2 \frac{R_2^2}{R_1^2} \right) \left(\tilde{a}_1^2 + \tilde{a}_1\tilde{a}_2 + \tilde{a}_2^2  \right)\right.
+ 2 b \, \frac{R_2^2}{R_1^2}\left( \tilde{a}_1 \tilde{a}_3 + \frac{\tilde{a}_1 \tilde{a}_4+ \tilde{a}_2 \tilde{a}_3}{2} + \tilde{a}_2 \tilde{a}_4 \right)
\\
& \quad\quad\quad\quad\quad
+ \left.\frac{R_2^2}{R_1^2}\left(\tilde{a}_3^2 + \tilde{a}_3 \tilde{a}_4 + \tilde{a}_4^2 \right)
\,\right]^{1/2}
,
\end{aligned}
\end{equation}
where $r_i$ are the radii on $T^2_i$ for $i=1,2$ and $R_1,R_2$ those on $T^2_3$ and $\frac{\tilde{a}_1 \tilde{a}_4+ \tilde{a}_2 \tilde{a}_3}{2} =
\tilde{a}_1 \tilde{a}_4= \tilde{a}_2 \tilde{a}_3$ with the bulk wrapping numbers $\tilde{a}_i$ introduced in section~\ref{sec:set-up}. 
The constant $c$ takes care of the different normalisation of a fractional cycle compared 
to a toroidal one.\\
For the massless linear combination $U(1)_X = \sum_i x_i U(1)_i$ with $U(1)_i \subset U(N_i)$,
the fine structure constant is given by~\footnote{The factor $2N_i$ is due to the different prefactors 
of the canonical four dimensional kinetic terms for Abelian and non-Abelian gauge fields, $-\int_{\mathbb{R}^{1,3}} \left( \frac{1}{4 g^2_{U(1)}} F_{U(1)} \wedge 
\star  F_{U(1)} + \frac{1}{2 g^2_{SU(N)}} {\rm tr} \left( F_{SU(N)} \wedge \star F_{SU(N)} \right)\right)$,
with the quadratic Casimir of the fundamental representation of $SU(N)$ normalised to 1/2 and the fact that 
${\rm tr} \left( F_{U(N)} \wedge \star F_{U(N)} \right)={\rm tr} \left( F_{SU(N)} \wedge \star F_{SU(N)} \right)
+ N \,  F_{U(1)_{\rm diag}} \wedge \star  F_{U(1)_{\rm diag}}$  as noted e.g. in~\cite{giiq02}.  }
\begin{equation}
\frac{1}{\alpha_X} =  \sum_i 2 \, N_i \, x_i^2 \frac{1}{\alpha_i}. 
\end{equation}
\item
The running of the gauge coupling at one-loop due to massless string modes charged under $G_a=SU(N_a)$
encoded in the beta function coefficient $b_a$ with
\begin{equation}
b_{SU(N_a)} = 
-  N_a \left( 3 - \varphi^{\Adj_a}\right) +\sum_{b\neq a} \frac{N_b}{2} \left( \varphi^{ab} + \varphi^{ab'}\right)  
+ \frac{N_a-2 }{2} \, \varphi^{\Anti_a} + \frac{N_a+2}{2} \, \varphi^{\Sym_a}
.
\end{equation} 
For $U(1)_a$ gauge groups inside $U(N_a)$ factors, the coefficient is
\begin{equation}
b_{U(1)_a} =  
N_a \Bigl(\sum_{b\neq a}  N_b \left( \varphi^{ab} + \varphi^{ab'}\right) 
+ 2  \, (N_a +1) \, \varphi^{\Sym_a} + 2 \, (N_a -1) \, \varphi^{\Anti_a} \Bigr)
,
\end{equation} 
with the beta function coefficient for a massless $U(1)_X= \sum_i x_i U(1)_i$ factor given by
\begin{equation}
b_X = \sum_i  x_i^2 \, b_i + 2 \, \sum_{i < j} N_i N_j x_i x_j \left(-\varphi^{ij} + \varphi^{ij'}\right).
\end{equation} 
\item
The one-loop gauge threshold correction $\Delta_a$ due to charged massive string modes.
This correction has been computed for bulk D6-branes in~\cite{lust03,abls07a,absy06} 
and for rigid D6-branes in~\cite{abls07b,blsc07}.
For the fractional non-rigid D6-branes employed in this article, to our knowledge no explicit
result has been obtained so far.
Since the threshold corrections are expected to be tiny at energy scales $\mu$ well below the string scale, 
we are postponing their discussion to future work~\cite{WIP} and
neglecting them at this point in our statistical analysis of vacua.
\end{itemize}

\section{Standard model constraints}\label{sec:SM-constraints}

In~\cite{gmho07}, we considered the most general way to obtain $n$ standard model generations from
at most four different D6-branes with initial gauge group $U(3)_a \times U(2) / Sp(2)_b \times U(1)_c \times U(1)_d$
 and right-handed quarks and leptons realised either as bifundamental or antisymmetric representations. We 
furthermore imposed supersymmetry and RR tadpole cancellation on all possible brane configurations.
It turned out that only one single option of the four possibilities led to $n$ generation standard model--like spectra. 
The prolific configuration is displayed in 
table~\ref{Tab:SM-Confs}.\footnote{To be exhaustive, all four possible combinations $Q_Y = \frac{1}{6} Q_a 
\pm \frac{1}{2} Q_c \pm \frac{1}{2} Q_d$ corresponding to permutations of orientifold image branes
$c \leftrightarrow c'$ and $d\leftrightarrow d'$ 
should be considered. A check of random samples showed that all four choices have the same statistical behaviour.
The standard model configurations with $Q_Y$ as defined in table~\ref{Tab:SM-Confs} is therefore complete up to
some statistical factor of the order of ${\cal O}(4)$.
}
\begin{table}[ht]
\begin{center}
\begin{equation*}
\begin{array}{|c|c|}
\hline
\multicolumn{2}{|c|}{\rule[-3mm]{0mm}{8mm}
 U(3)_a \times U(2)_b \times U(1)_c \times U(1)_d
}
\\\hline\hline
{\rm particle} & n
\\\hline\hline
Q_L & \chi^{ab} + \chi^{ab'}
\\
u_R & \chi^{a'c}+\chi^{a'd}
\\
d_R & \chi^{a'c'}+\chi^{a'd'} + \chi^{\Anti_a}
\\\hline
L & \chi^{bc} + \chi^{bd} +\chi^{b'c} + \chi^{b'd}
\\
e_R & \chi^{cd'} + \chi^{\Sym_c} + \chi^{\Sym_d}
\\\hline
\multicolumn{2}{|c|}{\rule[-3mm]{0mm}{7mm}
Q_Y =  \frac{1}{6} Q_a + \frac{1}{2} Q_c + \frac{1}{2} Q_d
}
\\\hline
\end{array}
\end{equation*}
\end{center}
\caption{The fertile standard model--like configuration with $n$ generations.
The number of right-handed neutrinos $\nu_R$ as well as the number of ``chiral" Higgs 
candidates $(H_u,H_d)$ are left as free parameters. In the last line, the 
hyper charge $Q_Y$ is given as a linear combination of the original $U(1)$ factors.}
\label{Tab:SM-Confs}
\end{table}

The number of right-handed neutrinos $\nu_R$ and ``chiral" Higgs particles~\footnote{Only 
right-handed neutrinos and Higgs particles stemming from non-vanishing intersection numbers are 
expected to contribute to the Yukawa couplings in the superpotential.
Pairs of Higgs particles $(H_u,H_d)$ are ``chiral'' in the sense of having different charges under
massive $U(1)$ symmetries which impose selection rules on {\it perturbative} couplings, 
but non-chiral w.r.t. the standard model gauge group.
}
are kept as free parameters. The standard model constraints in table~\ref{Tab:SM-Confs} imply that 
vector like pairs w.r.t. the standard model, $(\1,\2)_{\bf 1/2} +(\1,\2)_{\bf -1/2}$, 
can occur. In the absence of any other gauge group, these representations
have an interpretation of Higgs pairs $H_u+H_d$, and as discussed in section~\ref{App:Realisations},
in the presence of of a $B-L$ symmetry, they might either form Higgs pairs or 
lepton-anti-lepton pairs. In the statistical approach in section~\ref{sec:Statistics},
possible  $B-L$ symmetries are not explored, but only the number $h$ 
of vector like pairs w.r.t. the standard model gauge group in the  $(\1,\2)_{\bf 1/2} +(\1,\2)_{\bf -1/2}$
representation which stem from non-vanishing intersection numbers are
computed as
\begin{equation}\label{eq:chh}
h = \frac{1}{2} \left(|\chi^{bc}| + |\chi^{bd}| + |\chi^{b'c}| + |\chi^{b'd}|
    - |\chi^{bc}+\chi^{bd}+\chi^{b'c}+\chi^{b'd}|\right).
\end{equation}

The hyper charge $Q_Y$ remains massless after the generalised Green Schwarz mechanism 
provided that its effective three-cycle
\begin{equation}\label{Eq:Q_Y-cycle}
\Pi_Y = \frac{1}{2} \left(  \Pi_a + \Pi_c + \Pi_d \right)
\end{equation}
is $\OR$ invariant, i.e. $\Pi_Y = \Pi_Y'$, which is equivalent to 
$\vec{Y}_a + \vec{Y}_c + \vec{Y}_d= 0$ with the toroidal contributions $Y_1,Y_2$ defined in table~\ref{Tab:Def-XYL} 
and corresponding entries $(Y_3 \ldots Y_6)$ for the exceptional parts given in appendix~\ref{App_bulkrelation} in
equation~\eqref{Eq:App-X3X6}.

While individual gauge couplings depend on the string, Planck and compactification scale as well as some numerical 
factor, their ratios are independent of these quantities. The quotient of strong and electro-weak fine structure constants,
\begin{equation}
\frac{\alpha_s}{\alpha_w}=\frac{V_{U(2)_b}}{V_{U(3)_a}},
\end{equation}
is one characteristic quantity, the weak mixing angle $\theta_w$ with
\begin{equation}\label{Eq:sin2theta}
{\rm sin}^2 \theta_w = \frac{\alpha_Y}{\alpha_Y+\alpha_w}
\end{equation} 
another one. We compute both parameters at tree level in section~\ref{sec:Statistics}.
 
In~\cite{bls03}, it was further argued that if there is some underlying Pati-Salam
symmetry, the fine structure constants are related by
\begin{equation}\label{Eq:PS-fine-relations}
\frac{1}{\alpha_Y} =  \frac{2}{3} \frac{1}{\alpha_s} + \frac{1}{\alpha_w},
\end{equation} 
and in case of an $SU(5)$ GUT even more constrained by
\begin{equation}\label{Eq:SU5-relations}
\frac{1}{\alpha_s} = \frac{1}{\alpha_w} = \frac{3}{5}\frac{1}{\alpha_Y},
\end{equation} 
which includes the previous case.
The statistical evaluation in section~\ref{Subsec:GC-SM} does not point to the presence of any of these GUT relations
in the $T^6/\bZ_6'$ background.

In~\cite{gmho07}, also $SU(5)$ and Pati-Salam configurations have been considered with the result that on the $T^6/\bZ_6'$ 
orbifold 
$SU(5)$ models occur only with an even number of generations and some chiral states in the {\bf 15} 
representation of $SU(5)$ whereas a Pati-Salam group arises only for odd numbers of generations and 
generically with a large number of chiral exotics.~\footnote{
The search for Pati-Salam configurations in~\cite{gmho07}
was, however, restricted to a very specific intersection pattern and consequently there might be 
undetected models with less chiral exotics or $Sp(2)$ gauge factors.
}
For the sake of completeness of GUT searches, we present in appendix~\ref{App_Trinification}
the solutions for trinification models which turn out to give results only for two generations and 
a bunch of chiral exotics.

The statistical analysis in the remainder of this article focuses on supersymmetric standard model--like spectra 
with the hyper charge assignment as in table~\ref{Tab:SM-Confs}.

\section{Statistics}\label{sec:Statistics}
In this section we present the results of a statistical survey of standard
model-like solutions. The results have been obtained by constructing all models
explicitly, such that we do not have to worry about intrinsic problems of
samples or unwanted correlations~\cite{dile06}.

\subsection{Relation between solutions on different geometries}
In~\cite{gmho07}, we found that for a given shape of the $T^2_3$ lattice (or a given $b \in \{0,1/2\}$), 
the number of solutions to the bulk supersymmetry and RR tadpole cancellation conditions~\eqref{Eq:SUSY}
and~\eqref{Eq:RRtcc} with bulk entries listed in table~\ref{Tab:Def-XYL} is identical for 
the {\bf AB} and {\bf BB} geometries on $T^2_1 \times T^2_2$, similarly for {\bf AA} and {\bf BA}.
The frequency of the former is by a factor $\mathcal{O}(10)$ higher than the latter, and bulk
models with $b=1/2$ are suppressed by a factor of 10 and 6, respectively, compared to the $b=0$ ones.
In appendix~\ref{App_bulkrelation}, we prove analytically that the bulk solutions on these pairs of lattices 
are related by a redefinition of the complex structure modulus $\varrho \rightarrow \frac{3}{4 \varrho}$
in combination with a rotation $\pi(1/3,0,-1/2)$ of the bulk cycle wrapping numbers. This change of parameters carries
over to the toroidal part of the full solutions with fractional cycles and standard model--like properties 
to be discussed below, where,
however the suppression of {\bf AA} and {\bf BA} geometries is enhanced to a factor of ${\cal O}(10^8 - 10^9)$  for 
$b=0$ compared to {\bf AB} and {\bf BB}, and the suppression factors for $b=1/2$ are ${\cal O}(10^{-4})$ and
${\cal O}(10^{-6} -10^{-7})$, respectively.
 
\subsection{Counting standard model--like solutions}
We found $4.43 \times 10^{15}$ supersymmetric standard model--like solutions with three generations and
massless hyper charge $Q_Y$ in~\cite{gmho07}. One or two generations and a massless $Q_Y$
occurs for $3.42 \times 10^{19}$ and $1.63 \times 10^{12}$ configurations,
respectively, while no model is found for four or more generations, even when
allowing for a massive hyper charge.

\fig{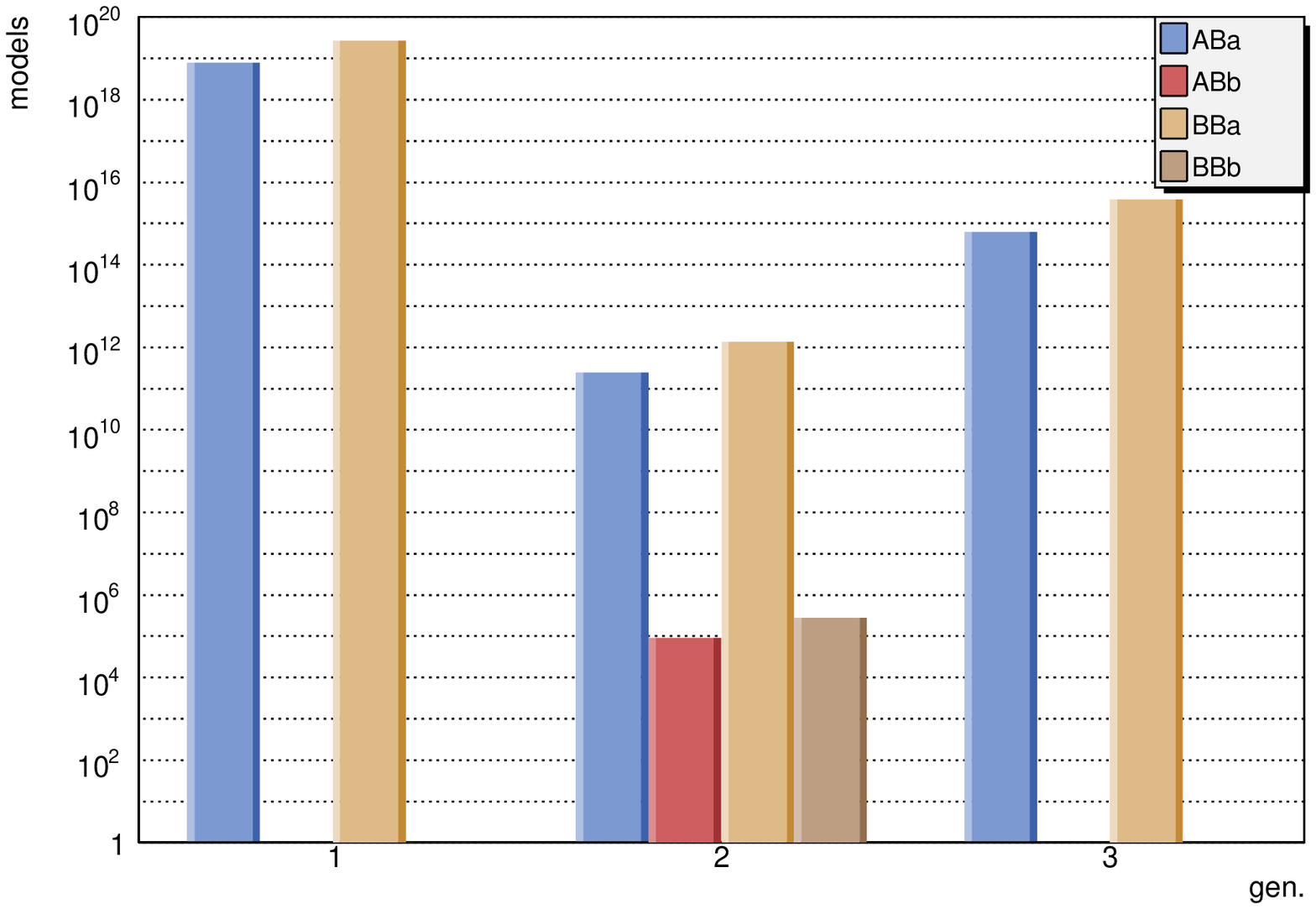}{%
Distribution of supersymmetric solutions to the RR tadpole equations with gauge
group and chiral matter content of the standard model for different numbers of
generations and the four geometries where such models occur (the other four
possibilities {\bf AAa/b} and {\bf BAa/b} do not lead to any standard model--like solutions).
The hidden sectors and number of chiral exotics have not been constrained.
Note that this plot includes all possibilities, more than three generations
do not occur.}

Out of the eight possible geometries only four, namely $\mathbf{ABa/b}$ and $\mathbf{BBa/b}$, allow
for standard model--like solutions, where the variants where the third torus is tilted contribute
only to models with two quark-lepton families. The situation is depicted in figure~\ref{fig:smdist}.

\subsection{Complex structure dependence}
In order to compare with the analysis of~\cite{balo08}, in figure~\ref{fig:rhodist} and table~\ref{Tab:Rho-Dist}
we plot the number of standard model--like spectra in dependence of the number $n$ of generations 
and complex structure parameter $\varrho=\frac{\sqrt{3}}{2} \frac{R_2}{R_1}$.

The results for three generation models differ from those found in~\cite{balo08}, 
which were $\varrho= \frac{1}{2},\frac{3}{2},1,\frac{1}{4}$ for {\bf AAa} and $\frac{1}{2},\frac{3}{2},3,\frac{3}{4}$
for {\bf BAa} and $3$ for {\bf BBa} and $\frac{1}{4}$ for {\bf ABa} in our notation, 
since the first two  lattices in~\cite{balo08}, 
{\bf AAa} and {\bf BAa}, are ruled out in our analysis by the RR tadpole cancellation conditions, 
which had not been imposed in~\cite{balo08}.
The {\bf ABa} and {\bf BBa} lattices have supersymmetric solutions to the RR tadpole
conditions with three chiral quark-lepton generations where our complex structure parameter $\varrho$ takes 
in total five different values per lattice orientation
opposed to one each reported in~\cite{balo08} which turns out to be the one with lowest frequency in our analysis.  
This might be related to the fact that the constraint on  three left handed 
quark generations $(\chi^{ab},\chi^{ab'})=(2,1)$ or $(1,2)$  in~\cite{balo08} 
is more restrictive than our ansatz $\chi^{ab}+\chi^{ab'} = \pm 3$ in table~\ref{Tab:SM-Confs}.
\footnote{As we will see in section~\protect\ref{sec:noex} there do not exist any supersymmetric solutions
to the RR tadpole conditions without chiral exotics that have $(\chi^{ab},\chi^{ab'})=(2,1)$ or $(1,2)$.}

\fig{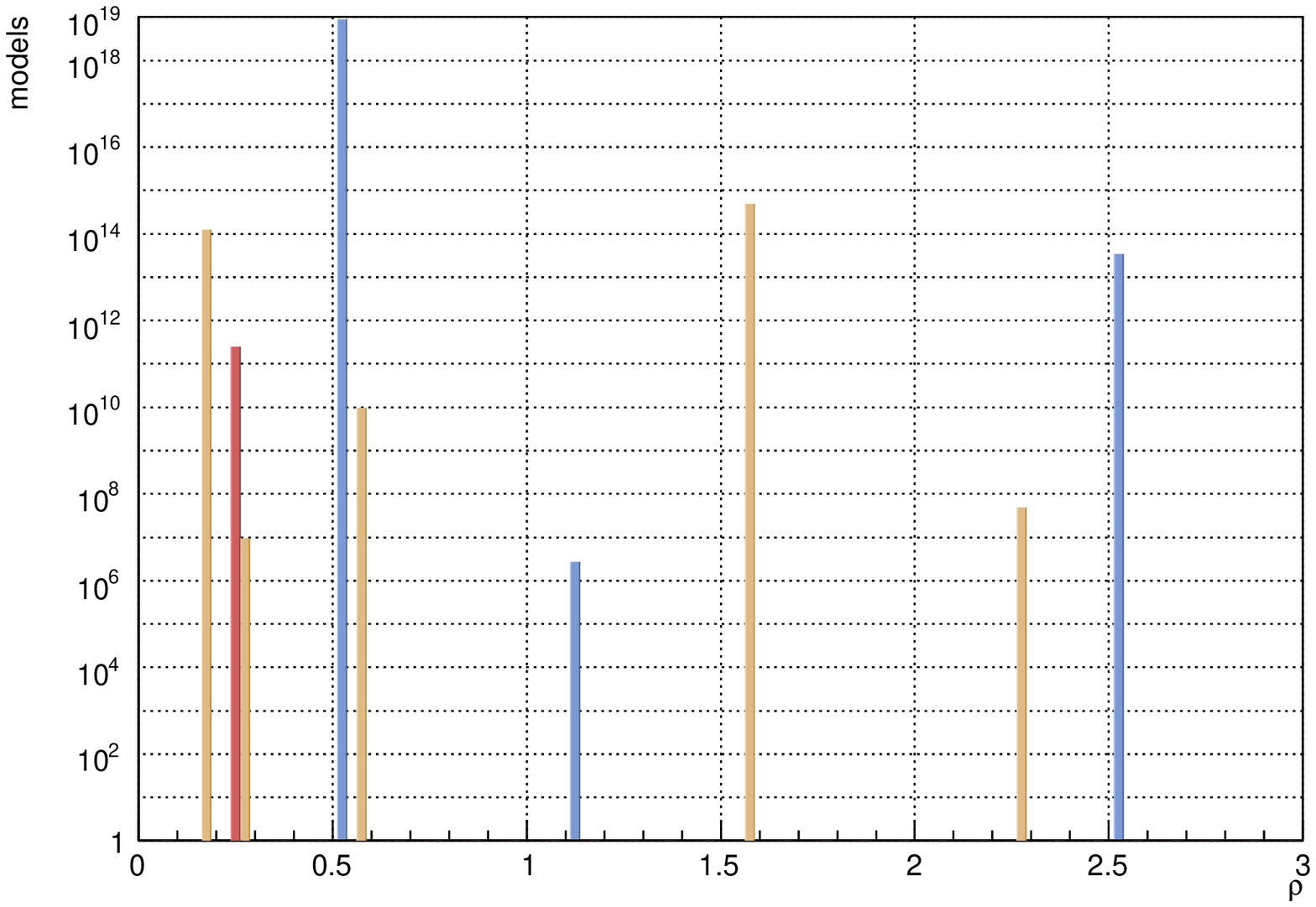}{%
Distribution of the complex structure parameters $\varrho$ for the {\bf ABa} geometry
depending on the 
number of generations: yellow, red and blue correspond to three, two and one generation,
respectively.}
\begin{table}[ht]
\begin{center}
\begin{equation*}
\begin{array}{|c|c||c|c||c|c|}
\hline
\multicolumn{6}{|c|}{\rule[-3mm]{0mm}{8mm}
\text{\bf Complex structure distribution}}
\\\hline\hline
\multicolumn{2}{|c||}{\text{1 generation}}
 & \multicolumn{2}{c||}{\text{2 generations}}
 & \multicolumn{2}{c|}{\text{3 generations}}
\\\hline\hline
\varrho & \# {\rm models}
 & \varrho & \# {\rm models}
 & \varrho & \# {\rm models}
\\\hline
\frac{1}{2} & 8.7 \cdot 10^{18}
 & \frac{1}{5} & 2.5 \cdot 10^{11}
 & \frac{1}{2} & 9.7 \cdot 10^{9}
\\
\frac{5}{2} & 3.4 \cdot 10^{13}
 & &
 & \frac{1}{4} & 9.6 \cdot 10^{6}
\\
\frac{7}{6} & 2.7 \cdot 10^{6}
 & &
 & \frac{1}{6} & 1.2 \cdot 10^{14}
\\
 &
 & &
 & \frac{3}{2} & 4.9 \cdot 10^{14}
\\
 &
 & &
 & \frac{9}{4} & 4.9 \cdot 10^{7}
\\\hline
\end{array}
\end{equation*}
\end{center}
\caption{Possible values of complex structure parameters $\varrho$ on $T^2_3$ 
and their abundance for $n$ generation models from the {\bf ABa} geometry.
The allowed complex structures on the {\bf BBa} lattice are obtained by 
identifying $\varrho_{\bf ABa} = \frac{3}{4\varrho_{\bf BBa}}$ with multiplicities 
which are by a factor $\mathcal{O}(1-10)$ bigger for each matching $\varrho$ value
than the ones displayed here.}
\label{Tab:Rho-Dist}
\end{table}

Due to the vast amount of data, in the following we will restrict our attention to one of the two
geometries, $\mathbf{ABa}$, where three generation models exist. 
For the {\bf BBa} geometry, we expect essentially the same 
statistical behaviour as confirmed by random samples, the equivalence of bulk solutions
proved in appendix~\ref{App_bulkrelation} as well as the check that the relations 
and abundances of complex structure parameters for standard model--like configurations are as discussed above.

\subsection{Gauge couplings}\label{Subsec:GC-SM}
The ratio $\alpha_s/\alpha_w=V_{U(2)_b}/V_{U(3)_a}$ for fine structure constants takes several different values
for different numbers of generations as depicted in figure~\ref{fig:gaugec}.
Also the values of the weak mixing angle~\eqref{Eq:sin2theta} at the string
scale are scattered broadly as shown in figure~\ref{fig:genasawsintheta}.

\fig{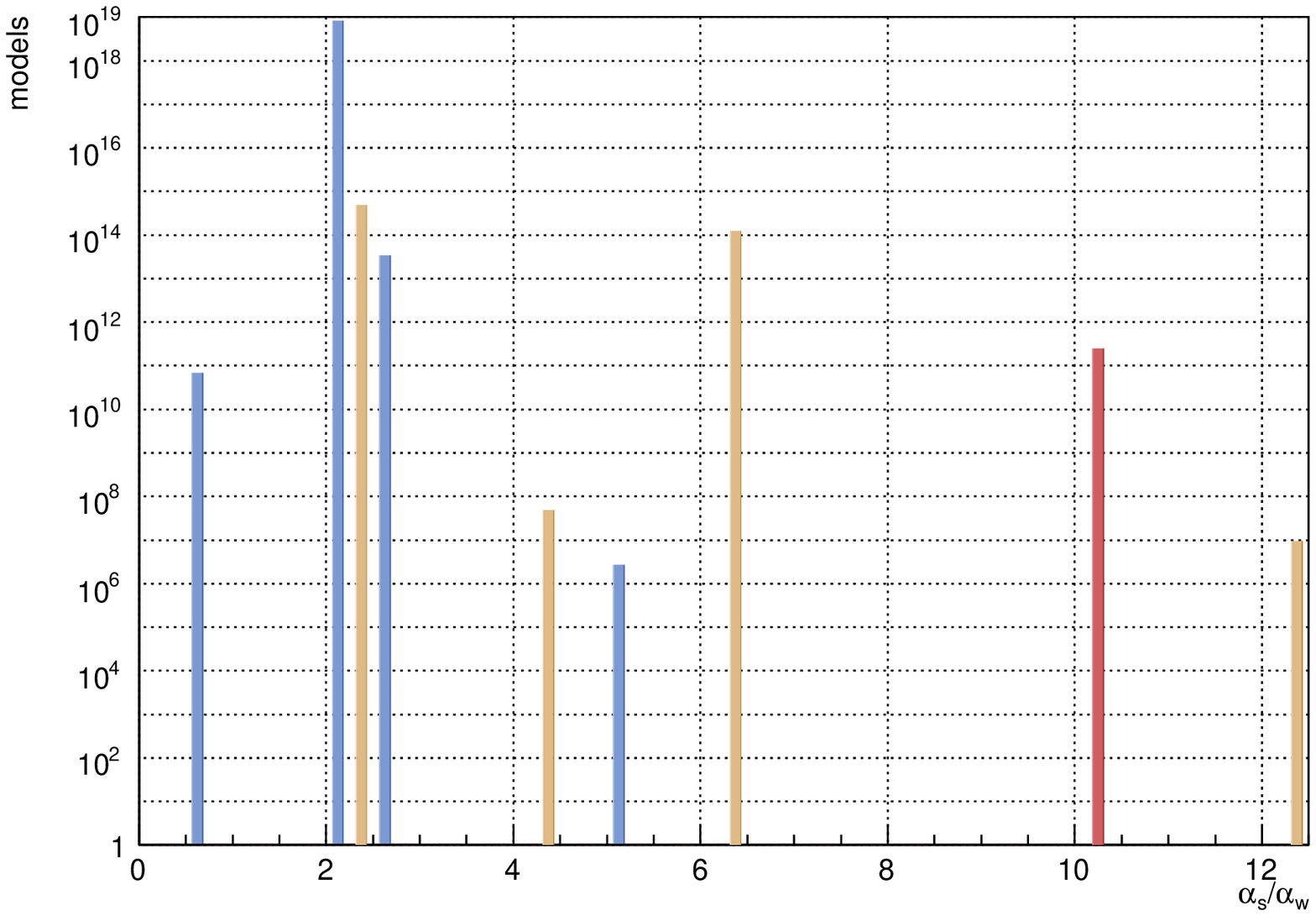}{%
The tree level ratio of fine structure constants $\alpha_s/\alpha_w$ for standard-like
models on the {\bf ABa} lattice (yellow, red and blue correspond to three, two
and one generation,respectively).}
\fig{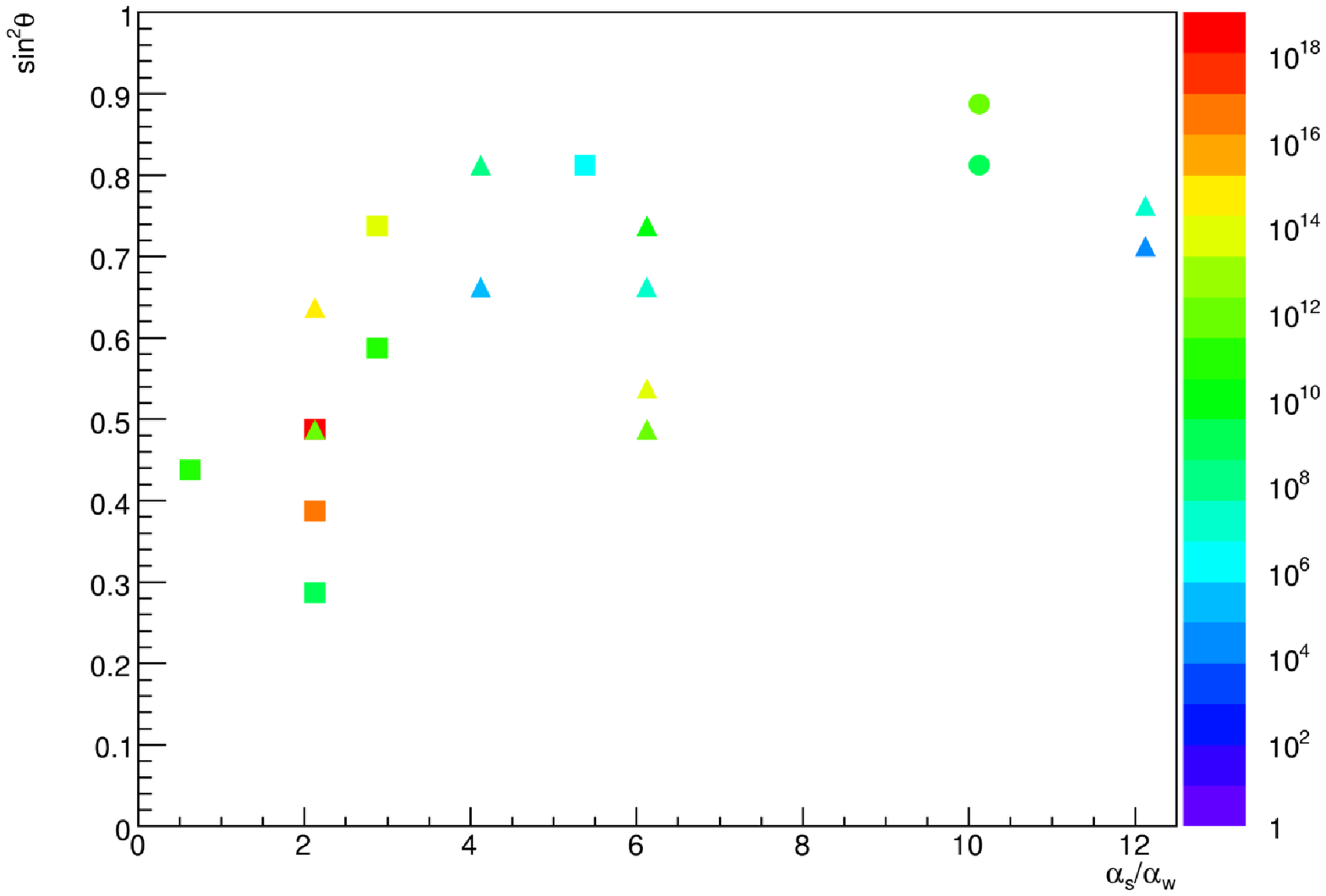}{%
Combined distribution of the tree level ratios of $\alpha_s/\alpha_w$ and the
weak mixing angle $\theta_w$ on the {\bf ABa} lattice. A triangle, circle and square
correspond to three, two and one generation models, respectively. Note that the
colour scale on the right, which encodes the number of models with given values
of $(\alpha_s/\alpha_w,{\rm sin}^2 \theta_w)$, is logarithmic.}

This distribution of values can be compared to the measured quantities at the electro-weak scale (\mbox{$M_Z \approx$ 91 GeV}), 
\begin{equation}\label{Eq:Mz-values}
\alpha_s(M_Z)  \approx 0.1,
\qquad
\alpha_{\rm electro-magnetic}( M_Z) \approx \frac{1}{128},
\qquad
{\rm sin}^2 \theta_w ( M_Z) \approx 0.23,
\end{equation} 
as well as results from intersecting D6-branes on other orbifolds.
On $T^6/\bZ_2 \times \bZ_2$~\cite{gbhlw05}, the Pati-Salam relation~\eqref{Eq:PS-fine-relations} on gauge couplings 
was fulfilled in 88\% of the models, but less than 3\% were compatible with the $SU(5)$ GUT relation~\eqref{Eq:SU5-relations}
on the non-Abelian gauge groups.\footnote{The analysis included only one and two generation models, since no explicit three
generation model was found due to the cut-off method used for the statistical treatment.}
In the $T^6/\bZ_6$ background with three generations~\cite{hoot04,gls07}, the situation was the reverse: since all 
D6-branes were wrapping the same bulk cycle, $\alpha_s = \alpha_w$ was fulfilled but the hyper charge as defined in 
table~\ref{Tab:SM-Confs} was not compatible with~\eqref{Eq:PS-fine-relations}. On the other hand,
redefining $Q_Y=(Q_a/3 + Q_c+Q_d+Q_e)/2$ admitted to interpret the additional chiral particles as 
three generations of Higgs multiplets with 
non-standard Yukawa couplings and~\eqref{Eq:PS-fine-relations} and~\eqref{Eq:SU5-relations} were both met.
The situation here differs from the other orbifolds in that  $\alpha_s \neq \alpha_w$ for any number of generations, 
and there is no obvious relation~\eqref{Eq:PS-fine-relations}. The values at the string scale also deviate considerably from those in equation~\eqref{Eq:Mz-values} 
at the electro-weak scale, which is, however, not too surprising in view of a high string scale $M_{\rm string} 
\sim 10^{16} - 10^{19}$ GeV and non-vanishing beta functions as well as threshold corrections for each gauge coupling.

\subsection{Number of Higgs multiplets}
The distribution of models with different numbers $h$ of ``chiral" Higgs generations
(plus some vector like lepton pairs in case of a massless $B-L$ symmetry), 
computed as given in~\eqref{eq:chh}, is shown in figure~\ref{fig:chhiggs}.
In this plot we consider only models with three generations of standard model--like
matter in the $\mathbf{ABa}$ geometry that have a massless
hyper charge.\footnote{As already mentioned earlier, models originating from the
$\mathbf{BBa}$ geometry are related by a rotational symmetry, so no new results can be
expected there.}

\fig{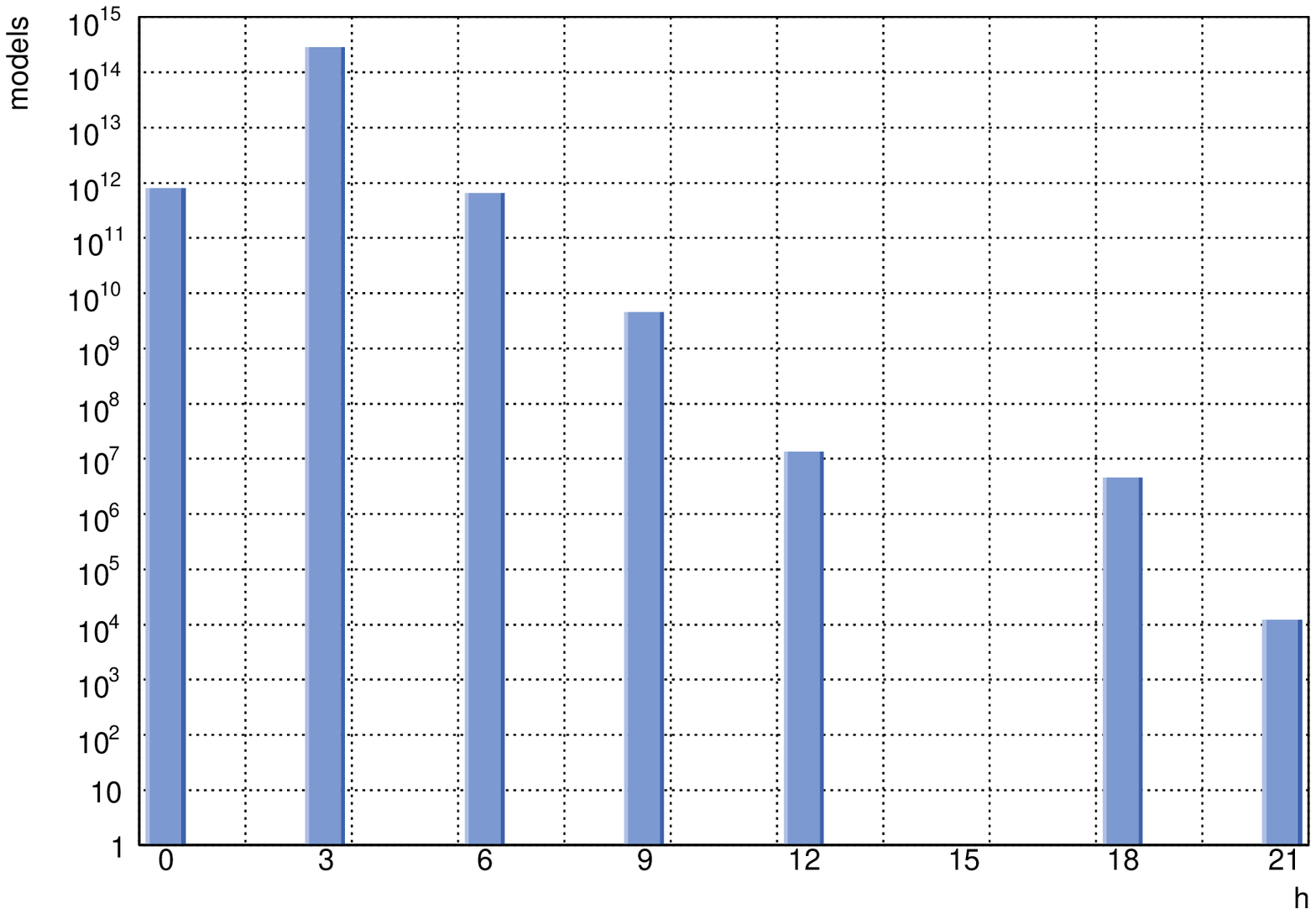}{%
Number of ``chiral" Higgs generation candidates (plus vector like lepton pairs in the presence of a $B-L$ symmetry)
as defined in equation~\protect\eqref{eq:chh} for
three-generation standard model--like configurations on {\bf ABa}.}

It is interesting to notice that ``chiral" Higgs generations (plus some vector like lepton pairs)
can only appear as multiples
of three. Up to the maximum value of 21 all possibilities with the exception of 15
appear. Statistically, a value of three chiral multiplets is clearly favoured, with
a total number of models two orders of magnitude above all other solutions.
This distribution suggests that the number of ``chiral" Higgs multiplets is correlated
with the constraint to obtain standard model matter with three families of quarks
and leptons on this particular orbifold background and is not expected to be of any
generality.

\subsection{Chiral exotics}
Within the three generation models, a wide variety of hidden sector configurations
is found and almost all of the models have a large number of chiral exotics.
To quantify their amount we use the total value of chiral matter multiplets between
the visible and hidden sector, defined as~\cite{gmho07}
\begin{equation}\label{eq:totex}
\xi := \sum_{\substack{v\in V\\h\in H}}\left|\chi^{vh}-\chi^{v'h}\right|,
\end{equation}
where $V=\{a,b,c,d\}$ is the set of branes in the standard model sector and $H$ contains all other stacks of 
(hidden) branes.
\wfig{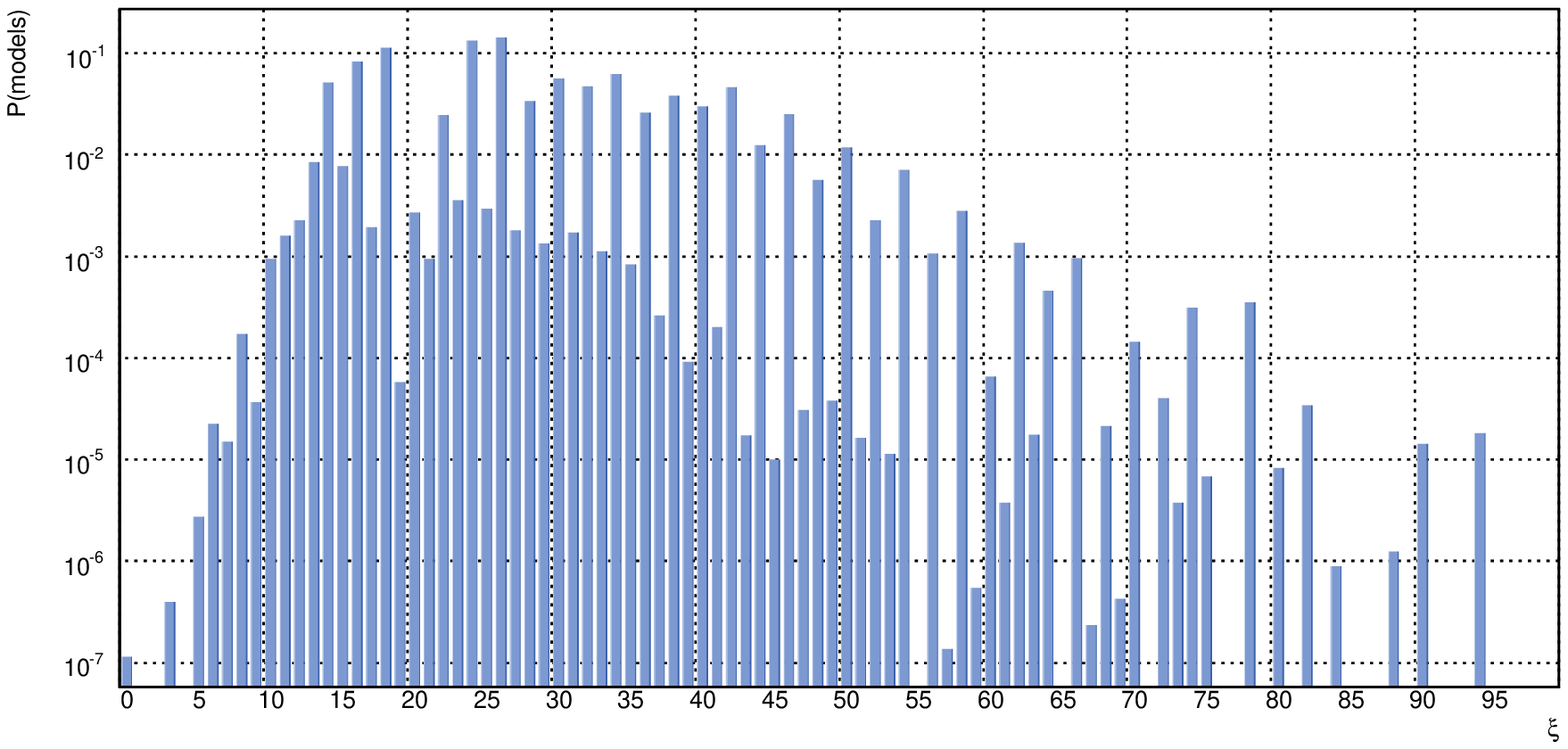}{%
Normalised distribution of the total number of chiral exotics $\xi$, as defined
in equation~\protect\eqref{eq:totex} for three-generation standard model--like
configurations on {\bf ABa}.}

For models on the $\mathbf{ABa}$ torus, the statistical result is shown in figure~\ref{fig:exotics}.
While interpreting this plot, one should keep in mind that the total amount of solutions
is presented on a logarithmic scale. In particular, there is a maximum of the distribution at $\xi=26$
and an exponential falloff towards larger numbers of exotic matter. This is to be expected, since
the appearance of ``long'' branes, admitting big intersection numbers, is exponentially suppressed
compared to shorter branes.
Additionally one observes that even numbers of chiral exotics are clearly preferred. This is an
artefact of the geometric set-up, much in the same way as an enhancement of even numbers for the
total rank of the gauge group was observed in~\cite{gbhlw05}.

Although statistically largely suppressed, there do exist models without any chiral exotics. A more
detailed analysis of these solutions will be the subject of section~\ref{sec:noex}.

\subsubsection{Correlation between ``chiral" Higgs multiplets and chiral exotics}

In table~\ref{Tab:Exotics-Higgses} and figure~\ref{fig:exotichiggs} the relation between the number $h$ of
``chiral" Higgs generations (plus some vector-like lepton pairs)
and the total number $\xi$ of chiral exotics is shown.
The two quantities are obviously not independent of each other. In particular one notes that it is
impossible to obtain models with an a priori small number of ``chiral" Higgs multiplets and a small number of chiral exotic
matter at the same time.

This result implies that even if one allows for a small number of chiral exotics, the number of Higgs multiplets
will always be larger than what is desired from a phenomenological perspective unless there exist terms in the 
low energy effective field theory which render all but one Higgs generation sufficiently massive.
A larger Higgs sector might eventually also be useful to address the mass hierarchy of standard model
particles~\cite{poze07}.

\begin{table}[ht]
\begin{center}
\resizebox{\linewidth}{!}{
\begin{tabular}{|r|r|r@{$\times$}l*{4}{||r|r|r@{$\times$}l}|}\hline
\multicolumn{20}{|c|}{\rule[-1.5mm]{0mm}{4mm}{\bf Counting of chiral exotics and chiral Higgs families}}\\\hline\hline
$\mathbf{\xi}$&$\mathbf{h}$&\multicolumn{2}{|c||}{$\mathbf{P}$}&$\mathbf{\xi}$&$\mathbf{h}$&\multicolumn{2}{|c||}{$\mathbf{P}$}&$\mathbf{\xi}$&$\mathbf{h}$&\multicolumn{2}{|c||}{$\mathbf{P}$}&$\mathbf{\xi}$&$\mathbf{h}$&\multicolumn{2}{|c||}{$\mathbf{P}$}&$\mathbf{\xi}$&$\mathbf{h}$&\multicolumn{2}{|c|}{$\mathbf{P}$}\\\hline\hline
$   0$&$  12$&$1.13$&$10^{-7}$&$  21$&$   3$&$9.36$&$10^{-4}$&$  34$&$   6$&$1.53$&$10^{-5}$&$  47$&$   3$&$3.08$&$10^{-5}$&$  62$&$   0$&$5.96$&$10^{-5}$\\\hline
$   0$&$  18$&$7.87$&$10^{-10}$&$  22$&$   3$&$2.44$&$10^{-2}$&$  34$&$   9$&$3.78$&$10^{-8}$&$  47$&$   6$&$2.48$&$10^{-8}$&$  62$&$   3$&$1.31$&$10^{-3}$\\\hline
$   0$&$  21$&$1.97$&$10^{-10}$&$  22$&$   6$&$9.79$&$10^{-5}$&$  35$&$   3$&$8.38$&$10^{-4}$&$  48$&$   0$&$2.67$&$10^{-4}$&$  63$&$   3$&$1.77$&$10^{-5}$\\\hline
$   3$&$   9$&$3.97$&$10^{-7}$&$  22$&$   9$&$5.44$&$10^{-6}$&$  36$&$   0$&$2.04$&$10^{-4}$&$  48$&$   3$&$4.45$&$10^{-3}$&$  64$&$   0$&$5.71$&$10^{-5}$\\\hline
$   3$&$  18$&$6.89$&$10^{-10}$&$  22$&$  18$&$9.44$&$10^{-9}$&$  36$&$   3$&$2.59$&$10^{-2}$&$  48$&$   6$&$9.09$&$10^{-4}$&$  64$&$   3$&$3.97$&$10^{-4}$\\\hline
$   5$&$   6$&$2.72$&$10^{-6}$&$  23$&$   3$&$3.60$&$10^{-3}$&$  36$&$   6$&$2.90$&$10^{-5}$&$  49$&$   0$&$9.92$&$10^{-8}$&$  64$&$   6$&$2.86$&$10^{-7}$\\\hline
$   6$&$   9$&$2.25$&$10^{-5}$&$  24$&$   3$&$1.32$&$10^{-1}$&$  37$&$   3$&$2.62$&$10^{-4}$&$  49$&$   3$&$3.78$&$10^{-5}$&$  66$&$   3$&$9.53$&$10^{-4}$\\\hline
$   6$&$  18$&$4.56$&$10^{-8}$&$  24$&$   6$&$4.51$&$10^{-4}$&$  38$&$   3$&$3.79$&$10^{-2}$&$  49$&$   6$&$3.14$&$10^{-7}$&$  66$&$   6$&$1.26$&$10^{-6}$\\\hline
$   7$&$   6$&$1.50$&$10^{-5}$&$  25$&$   3$&$2.95$&$10^{-3}$&$  38$&$   6$&$6.19$&$10^{-7}$&$  50$&$   0$&$1.45$&$10^{-6}$&$  67$&$   6$&$2.36$&$10^{-7}$\\\hline
$   8$&$   3$&$1.60$&$10^{-5}$&$  25$&$   6$&$2.20$&$10^{-8}$&$  39$&$   3$&$9.04$&$10^{-5}$&$  50$&$   3$&$1.18$&$10^{-2}$&$  68$&$   0$&$1.14$&$10^{-6}$\\\hline
$   8$&$   6$&$1.56$&$10^{-4}$&$  26$&$   3$&$1.43$&$10^{-1}$&$  39$&$   6$&$5.78$&$10^{-7}$&$  50$&$   6$&$3.33$&$10^{-5}$&$  68$&$   6$&$2.01$&$10^{-5}$\\\hline
$   9$&$   3$&$1.09$&$10^{-5}$&$  26$&$   6$&$5.04$&$10^{-5}$&$  40$&$   0$&$1.51$&$10^{-7}$&$  51$&$   0$&$1.26$&$10^{-6}$&$  69$&$   0$&$4.23$&$10^{-7}$\\\hline
$   9$&$   6$&$2.58$&$10^{-5}$&$  27$&$   0$&$3.97$&$10^{-7}$&$  40$&$   3$&$2.99$&$10^{-2}$&$  51$&$   3$&$1.51$&$10^{-5}$&$  70$&$   0$&$1.64$&$10^{-6}$\\\hline
$   9$&$   9$&$2.75$&$10^{-9}$&$  27$&$   3$&$1.79$&$10^{-3}$&$  40$&$   6$&$5.79$&$10^{-5}$&$  52$&$   0$&$3.49$&$10^{-4}$&$  70$&$   3$&$1.41$&$10^{-4}$\\\hline
$  10$&$   3$&$8.95$&$10^{-5}$&$  27$&$   6$&$3.15$&$10^{-7}$&$  41$&$   0$&$2.31$&$10^{-6}$&$  52$&$   3$&$1.91$&$10^{-3}$&$  72$&$   0$&$1.99$&$10^{-5}$\\\hline
$  10$&$   6$&$8.56$&$10^{-4}$&$  28$&$   3$&$3.33$&$10^{-2}$&$  41$&$   3$&$1.96$&$10^{-4}$&$  52$&$   6$&$2.39$&$10^{-7}$&$  72$&$   3$&$1.56$&$10^{-5}$\\\hline
$  11$&$   3$&$1.59$&$10^{-3}$&$  28$&$   6$&$2.24$&$10^{-4}$&$  41$&$   6$&$2.18$&$10^{-6}$&$  53$&$   3$&$1.13$&$10^{-5}$&$  72$&$   6$&$4.31$&$10^{-6}$\\\hline
$  12$&$   3$&$7.91$&$10^{-4}$&$  29$&$   3$&$1.34$&$10^{-3}$&$  42$&$   0$&$9.48$&$10^{-4}$&$  54$&$   0$&$2.85$&$10^{-5}$&$  73$&$   3$&$3.77$&$10^{-6}$\\\hline
$  12$&$   6$&$1.48$&$10^{-3}$&$  29$&$   6$&$3.31$&$10^{-8}$&$  42$&$   3$&$4.48$&$10^{-2}$&$  54$&$   3$&$7.11$&$10^{-3}$&$  74$&$   3$&$3.12$&$10^{-4}$\\\hline
$  13$&$   3$&$8.46$&$10^{-3}$&$  30$&$   0$&$2.63$&$10^{-5}$&$  42$&$   6$&$1.91$&$10^{-4}$&$  54$&$   6$&$2.83$&$10^{-8}$&$  74$&$   6$&$5.19$&$10^{-7}$\\\hline
$  14$&$   3$&$5.10$&$10^{-2}$&$  30$&$   3$&$5.60$&$10^{-2}$&$  43$&$   0$&$8.73$&$10^{-6}$&$  56$&$   3$&$7.68$&$10^{-4}$&$  75$&$   3$&$6.76$&$10^{-6}$\\\hline
$  15$&$   3$&$7.70$&$10^{-3}$&$  30$&$   6$&$5.22$&$10^{-5}$&$  43$&$   3$&$4.76$&$10^{-6}$&$  56$&$   6$&$3.05$&$10^{-4}$&$  78$&$   0$&$9.09$&$10^{-7}$\\\hline
$  16$&$   3$&$8.32$&$10^{-2}$&$  31$&$   0$&$1.32$&$10^{-7}$&$  43$&$   6$&$3.67$&$10^{-6}$&$  57$&$   6$&$1.36$&$10^{-7}$&$  78$&$   3$&$3.50$&$10^{-4}$\\\hline
$  16$&$   6$&$1.63$&$10^{-5}$&$  31$&$   3$&$1.70$&$10^{-3}$&$  44$&$   0$&$1.08$&$10^{-3}$&$  58$&$   0$&$1.13$&$10^{-7}$&$  80$&$   3$&$8.31$&$10^{-6}$\\\hline
$  17$&$   3$&$1.93$&$10^{-3}$&$  31$&$   6$&$1.65$&$10^{-8}$&$  44$&$   3$&$1.11$&$10^{-2}$&$  58$&$   3$&$2.76$&$10^{-3}$&$  82$&$   3$&$3.43$&$10^{-5}$\\\hline
$  18$&$   3$&$1.13$&$10^{-1}$&$  32$&$   3$&$4.66$&$10^{-2}$&$  44$&$   6$&$2.12$&$10^{-4}$&$  58$&$   6$&$4.33$&$10^{-5}$&$  84$&$   6$&$8.92$&$10^{-7}$\\\hline
$  18$&$   6$&$1.88$&$10^{-4}$&$  32$&$   6$&$2.28$&$10^{-4}$&$  45$&$   0$&$6.16$&$10^{-6}$&$  59$&$   0$&$5.45$&$10^{-7}$&$  88$&$   0$&$1.23$&$10^{-6}$\\\hline
$  18$&$   9$&$2.26$&$10^{-7}$&$  33$&$   0$&$3.60$&$10^{-6}$&$  45$&$   3$&$3.88$&$10^{-6}$&$  60$&$   0$&$2.45$&$10^{-5}$&$  90$&$   3$&$1.43$&$10^{-5}$\\\hline
$  19$&$   3$&$5.77$&$10^{-5}$&$  33$&$   3$&$1.12$&$10^{-3}$&$  46$&$   0$&$2.11$&$10^{-4}$&$  60$&$   3$&$1.53$&$10^{-6}$&$  94$&$   3$&$1.82$&$10^{-5}$\\\hline
$  20$&$   3$&$2.10$&$10^{-3}$&$  34$&$   0$&$1.09$&$10^{-5}$&$  46$&$   3$&$2.47$&$10^{-2}$&$  60$&$   6$&$3.98$&$10^{-5}$&\multicolumn{4}{c|}{}\\\cline{1-16}
$  20$&$   6$&$6.04$&$10^{-4}$&$  34$&$   3$&$6.21$&$10^{-2}$&$  46$&$   6$&$2.61$&$10^{-4}$&$  61$&$   3$&$3.77$&$10^{-6}$&\multicolumn{4}{c|}{}\\\hline
\end{tabular}}
\caption{Normalised frequencies $P$ of three generation standard-like models for a
given number $\xi$ of chiral exotics and $h$ of ``chiral" Higgs families
(or vector-like lepton pairs).}
\label{Tab:Exotics-Higgses}
\end{center}
\end{table}

\clearpage

\wfig{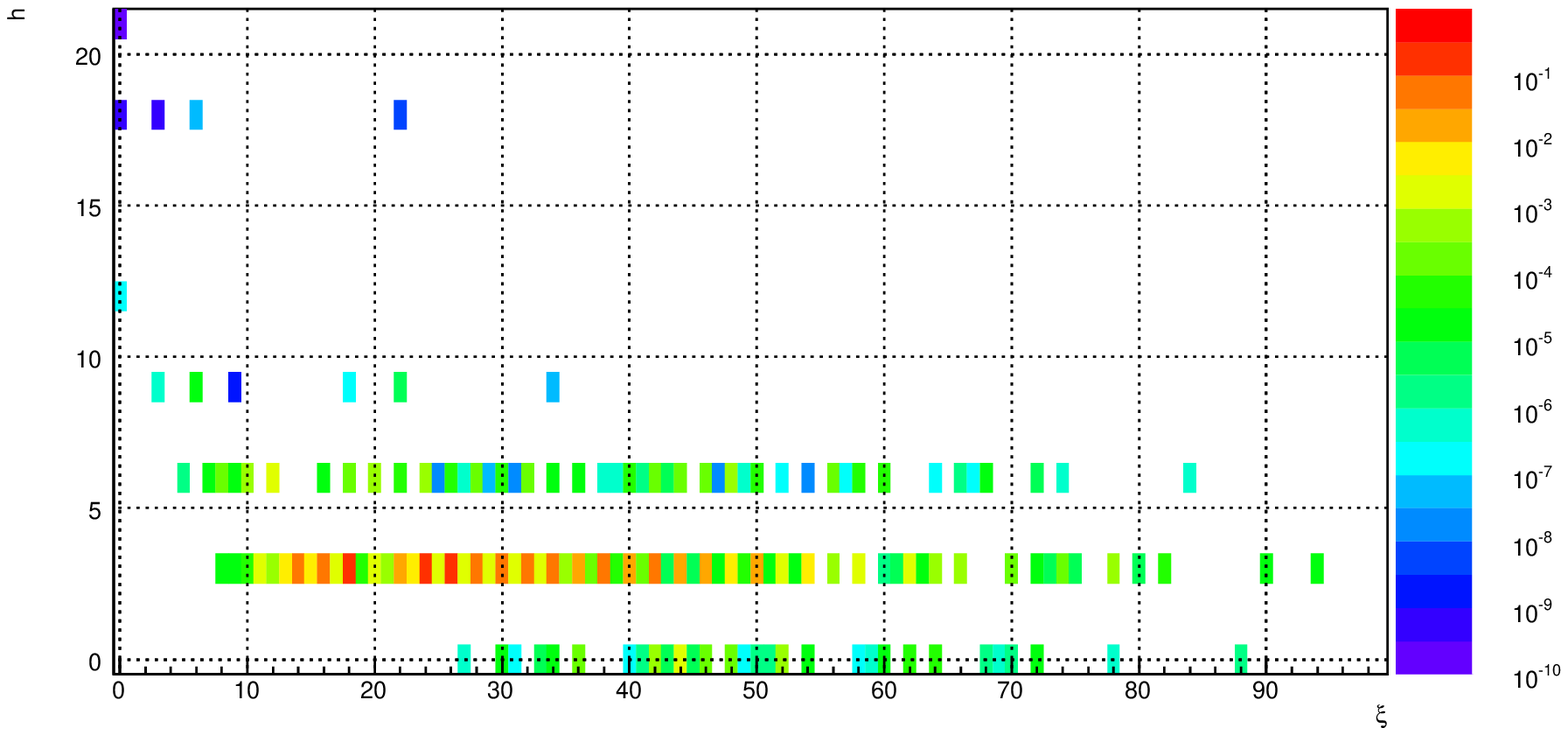}{Relation between the number $h$ of Higgs families  (plus some vector-like lepton pairs)
and total number $\xi$ of chiral exotics. The normalised frequency of a given pair $(\xi,h)$is encoded in the logarithmic
colour scale on the right.}

\subsection{Models without chiral exotics}\label{sec:noex}
In total, 7,139,328 models without chiral exotics as defined
in~\eqref{eq:totex} exist on the {\bf ABa} lattice.
The vast majority of these models has the same ratios of tree level gauge
couplings, namely $\alpha_s/\alpha_w=6.0$ and $\sin^2\theta_w=0.654$, as well
as twelve families of ``chiral" Higgses $(H_u,H_d)$ candidates (or vector-like lepton pairs)
which due to the massless $B-L$ symmetry discussed below actually split into nine  $(H_u,H_d)$
generations plus three vector-like lepton pairs.
Besides that there are two smaller groups of models with different ratios of gauge couplings
and more Higgs candidates as displayed in table~\ref{tab:zeroexstat}.

\begin{table}[ht]
\begin{center}
\begin{tabular}{|rrr|r|}\hline
{\bf h} & $\mathbf{\alpha_s/\alpha_w}$ & $\mathbf{\sin^2\theta_w}$
& {\bf \# models}\\\hline\hline
12 &  6 & 0.654 & 7,077,888 \\
18 &  4 & 0.667 &    49,152\\
21 & 12 & 0.720 &    12,288\\\hline
\end{tabular}
\caption{Number of ``chiral" Higgs candidates $h$, ratio of $\alpha_s/\alpha_w$
and weak mixing angle for models without chiral exotic matter.}
\label{tab:zeroexstat}
\end{center}
\end{table}

\subsubsection{Visible sector}\label{sec:noexvisible}
The standard model sectors are all very similar in the models without chiral
exotics. The chiral matter spectrum, that is responsible for the standard model
particles as shown in table~\ref{Tab:SM-Confs}, is identical for all models
with the same number of Higgs candidates:
\begin{equation}\label{eq:vsmatter}
\begin{array}{r@{\,=\,}lr@{\,=\,}lr@{\,=\,}lr@{\,=\,}lr@{\,=\,}lr@{\,=\,}lr@{\,=\,}l}
  h&12:\;&\chi^{ab}& 0,&\chi^{ab'}&3,&\chi^{ac}& -3,&\chi^{ac'}& -3,&\chi^{ad}& 0,&\chi^{ad'}& 0,\\
                         \muc{6}{c}{}&\chi^{bc}& -9,&\chi^{bc'}& -9,&\chi^{bd}& 6,&\chi^{bd'}& 3,\\
                                                       \muc{10}{c}{}&\chi^{cd}&-3,&\chi^{cd'}& 3;\\
  h&18:\;&\chi^{ab}&-3,&\chi^{ab'}&6,&\chi^{ac}& -3,&\chi^{ac'}& -3,&\chi^{ad}& 0,&\chi^{ad'}& 0,\\
                         \muc{6}{c}{}&\chi^{bc}& -6,&\chi^{bc'}& -6,&\chi^{bd}&15,&\chi^{bd'}&12,\\
                                                       \muc{10}{c}{}&\chi^{cd}&-3,&\chi^{cd'}& 3;\\
  h&21:\;&\chi^{ab}& 0,&\chi^{ab'}&3,&\chi^{ac}& -3,&\chi^{ac'}& -3,&\chi^{ad}& 0,&\chi^{ad'}& 0,\\
                         \muc{6}{c}{}&\chi^{bc}&-18,&\chi^{bc'}&-18,&\chi^{bd}& 6,&\chi^{bd'}& 3,\\
                                                       \muc{10}{c}{}&\chi^{cd}&-3,&\chi^{cd'}& 3.
\end{array}
\end{equation}
Furthermore it turns out that in the absence of chiral exotics
brane $c$ is always parallel to some $\OR\theta^{-k}$ invariant 
plane with  $\vec{Y}_c=0$ (as defined in table~\ref{Tab:Def-XYL} and
equation~\eqref{Eq:App-X3X6}) and therefore carries an $U(1)_c$ gauge group~\footnote{Brane 
$c$ in all these models belongs to type (1) in the classification of invariant branes in~\cite{gmho07}.}
which arises from the breaking of an $SO(2)_c$ or $Sp(2)_c$ gauge group by parrallel
displacement on $T^2_2$ away from the orientifold plane. Together with the
condition that the hyper charge is massless (see table~\ref{Tab:SM-Confs} and~\eqref{Eq:Q_Y-cycle})
this leads to the conclusion that there exists always a massless $U(1)_{B-L}$ symmetry.
A more detailed discussion of this issue using explicit examples can be found
in section~\ref{App:Realisations}.

\subsubsection{Hidden sector}\label{sec:noexhidden}
The hidden sectors of these models are in general very small, with a maximum of
three additional branes. This is a direct consequence of the orbifold geometry
leading to small maximal values in the RR tadpole equations, which distinguishes
$T^6/\bZ'_6$ from other orbifolds that allow for larger hidden sectors.

The distribution of the numbers of hidden sector stacks and the ranks of their
gauge groups is shown in table~\ref{tab:hstacks} and figure~\ref{fig:hstacks}.

\begin{table}[ht]
\begin{center}
\begin{tabular}{|r|r|r|}\hline
$\mathbf{s}$ & $\mathbf{\{N_i\}}$ & {\bf \# models}\\\hline\hline
0 &       &    61,440\\\hline
1 & 1     &   147,456\\
  & 3     &   442,368\\\hline
2 & 2,1   & 2,433,024\\\hline
3 & 1,1,1 & 4,055,040\\\hline
\end{tabular}
\caption{Number of hidden sector branes $s$ and rank distribution $\{N_i\}$ of the gauge
groups for standard-like models without chiral exotic matter.}
\label{tab:hstacks}
\end{center}
\end{table}

\twofig{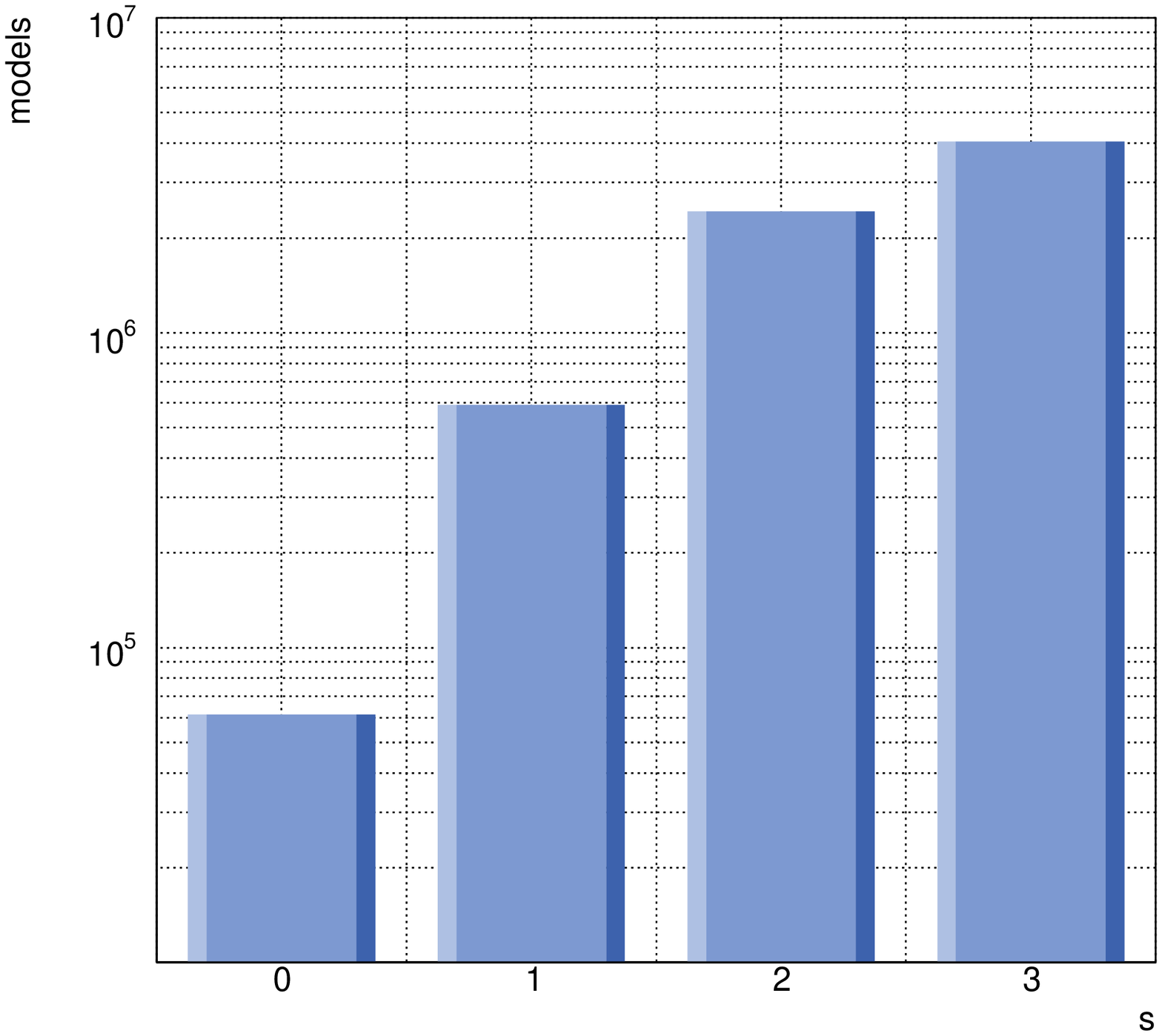}{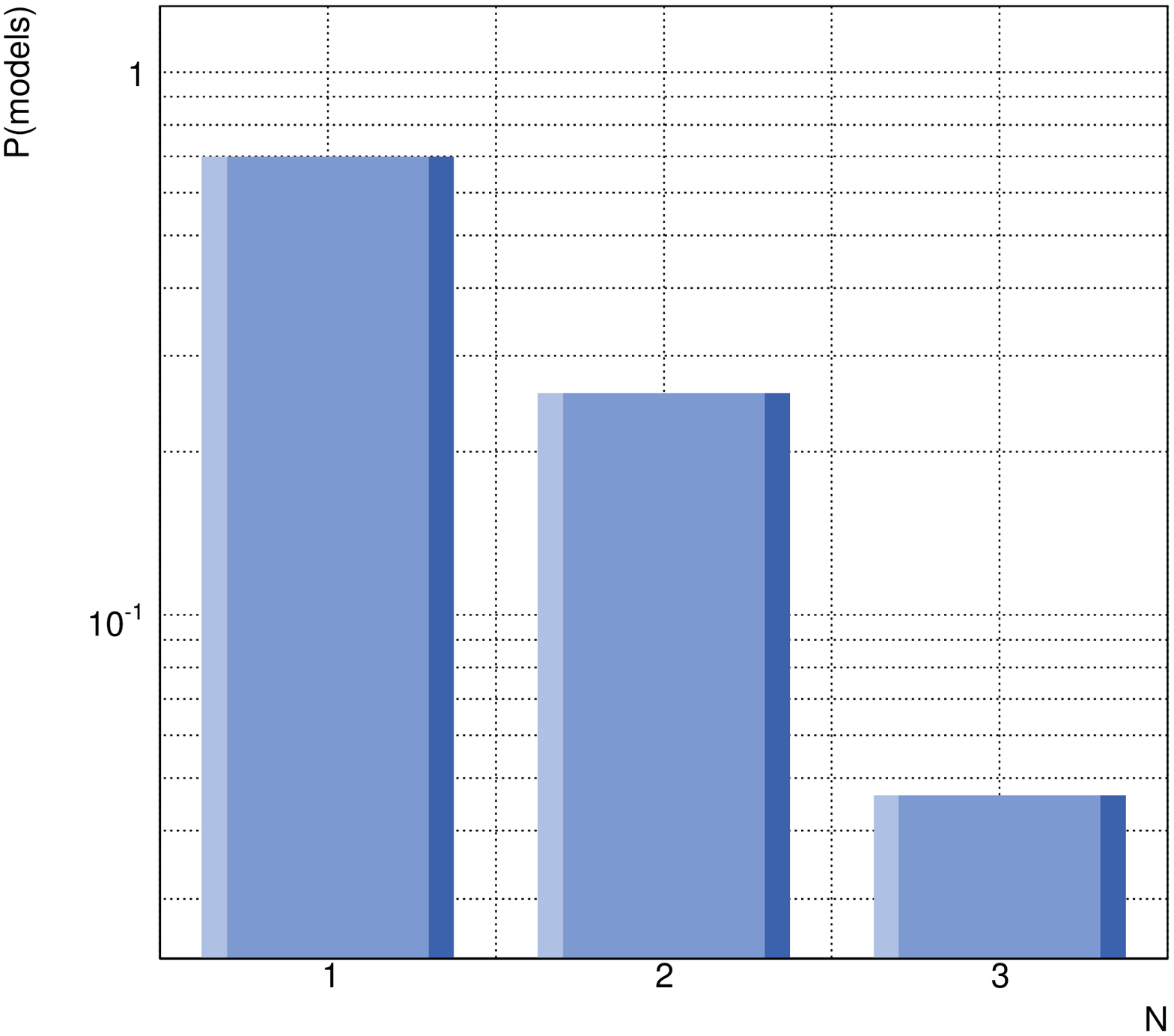}{%
Distributions of properties of the hidden sector of standard model--like
configurations without chiral exotic matter.
(a) the number of hidden sector branes $s$;
(b) probability to find a hidden sector gauge group factor of rank $N$.}
{fig:hiddenstat}

The models without any hidden sector correspond to those with $h=18$ or 21
(decomposing into 6 and 18 ``chiral" Higgs multiplet pairs $(H_u,H_d)$ plus 12 and 3
vector-like lepton pairs, respectively, due to the $U(1)_{B-L}$ symmetry)
in table~\ref{tab:zeroexstat}.
This implies that all models with a hidden sector have the same amount
of 9 ``chiral" Higgs multiplets plus 3 vector-like lepton pairs arising from
non-vanishing intersection numbers as well as the same tree level values for
$\alpha_s/\alpha_w$ and $\sin^2\theta_w$ at the string scale.

\begin{table}[ht]
\begin{center}
\begin{tabular}{|r|r|r|}\hline
$\mathbf{N}$ & {\bf \# models} & $\mathbf P(N)$\\\hline\hline
1 & 6,635,520 & 0.929\\
2 & 2,433,024 & 0.341\\
3 &   442,368 & 0.062\\\hline
\end{tabular}
\caption{Distribution of hidden sector gauge groups of rank $N$ and probability
to find a gauge group of specific rank for standard-like models without chiral exotic
matter.}
\label{tab:hrank}
\end{center}
\end{table}

In table~\ref{tab:hrank} and figure~\ref{fig:hrank} the probability distribution to
find a gauge group of rank $N$ within the hidden sector is shown.
Besides the fact that it is cut off at a maximal rank of three, the
distribution does not show differences to the more general case investigated
in~\cite{gmho07}. This is in good agreement with the results obtained for
different orbifold backgrounds~\cite{gbhlw05,gls07}, where it has also been
found that the distribution of properties in the hidden sector does not vary
significantly with respect to the distribution in the full set of solutions.

\subsubsection{Adjoint representations}\label{sec:adjstat}
The number of adjoint representations for the different gauge groups in the
visible and hidden sectors are very similar for the different realisations
of standard-like models without chiral exotics.
The result of a statistical analysis is shown in table~\ref{tab:adjstat},
where for hidden $SO/Sp(2N)$ gauge groups instead of adjoints
the number of symmetric plus antisymmetric representations
is listed, i.e. $\varphi^{\Adj_{U(N)}} \simeq \varphi^{\Sym_{SO/Sp2N)}} +\varphi^{\Anti_{SO/Sp2N)}}$
if the toroidal wrapping numbers $(n_i,m_i)$ are identical.

This result was to be expected, since the exceptional part of the brane
configuration does not play a role in determining the number of adjoints, as
can be seen from the discussion in section~\ref{sec:ospec}.

\begin{table}[ht]
\begin{center}
\begin{tabular}{|r|r|rrrr|rrr|r|}\hline
& & \muc{4}{|c|}{\it visible sector} & \muc{3}{|c|}{\it hidden sector} & \\\hline
$\mathbf{s}$ & $\mathbf{h}$                & $\mathbf{\varphi}^{\Adj_\mathbf{a}}$
  & $\mathbf{\varphi}^{\Adj_\mathbf{b}}$   & $\mathbf{\varphi}^{\Adj_\mathbf{c}}$
  & $\mathbf{\varphi}^{\Adj_\mathbf{d}}$   & $\mathbf{\varphi}^{\Adj_\mathbf{h_1}}$
  & $\mathbf{\varphi}^{\Adj_\mathbf{h_2}}$ & $\mathbf{\varphi}^{\Adj_\mathbf{h_3}}$
  & {\bf \# models}\\\hline\hline
0 & 18 &  2 &  4 &  4 & 10 & \muc{3}{|c|}{}      &    49,152\\
  & 21 &  2 & 10 &  4 & 10 & \muc{3}{|c|}{}      &    12,288\\\hline
1 & 12 &  2 & 10 &  4 & 10 &  2 & \muc{2}{c|}{} &   147,456\\
  & 12 &  2 & 10 &  4 & 10 & 10 & \muc{2}{c|}{} &   442,368\\\hline
2 & 12 &  2 & 10 &  4 & 10 &  2 &  2 &           & 2,433,024\\\hline
3 & 12 &  2 & 10 &  4 & 10 &  2 &  2 &  2        & 4,055,040\\\hline
\end{tabular}
\caption{Number of hidden sector branes $s$, number $h$ of ``chiral" Higgs candidates 
plus vector-like lepton pairs from non-vanishing 3-cycle intersection numbers
and number of adjoint representations for standard model-like spectra without chiral exotic
matter. In case the hidden brane $h_i$ is of $SO/Sp(2N)$ type, the counting of 
adjoints is replaced by counting symmetric plus antisymmetric representations,
$\varphi^{\Adj_{h_i}} \rightarrow  \varphi^{\Sym_{h_i}} +\varphi^{\Anti_{h_i}}$.}
\label{tab:adjstat}
\end{center}
\end{table}

\section{Explicit standard model--like realisations}\label{App:Realisations}

Three examples with the chiral spectrum of the standard model, nine ``chiral'' Higgs generations, 
a massless $B-L$ symmetry
and three different types of hidden sectors are discussed in section~\ref{Subsec:Examples-with-hidden},
and subsequently a model without hidden sector, 18  ``chiral'' Higgs generations and a massless $U(1)_{B-L}$ 
is presented in section~\ref{Subsec:Example2-nohidden}.

\subsection{Three generation models with hidden sectors}\label{Subsec:Examples-with-hidden}

In the following, we present three explicit realisations of three generation models without 
chiral exotics with hidden sector gauge groups $SO/Sp(6)$, $SO/Sp(4) \times SO/Sp(2)$ and $\widehat{SO/Sp(2)}$,
 respectively,~\footnote{The gauge group for a stack of $N$ orientifold invariant branes can be either $SO(2N)$ or 
$Sp(2N)$. We abbreviate this by $SO/Sp(2N)$ since the correct assignment is unknown.} where
the bulk part of the hidden stacks $h_3$ and $(h_2,h_1)$ are identical and the complex structure is $\varrho=1/2$ 
for all three choices of hidden sectors. 
The bulk configuration is given in table~\ref{tab:sm_bulk_branes} and the exceptional part 
with displacements $\sigma_i$ and Wilson lines $\tau^j$ in table~\ref{tab:sm_ex_branes}.
The hidden sector brane $h_1$ differs from $h_3$ and $h_2$ only in the assignment of $(\tau^0,\tau^1)$.
\begin{table}[ht]
\begin{center}%
\begin{tabular}{|c||c|c|rrrr|}\hline
\muc{7}{|c|}{\textbf{Standard model ex.: bulk brane configuration}}\\\hline\hline
\textbf{brane}& $(n_1,m_1;n_2,m_2;n_3,m_3)$ & $N$&$\tilde{a}_1\equiv P,$&$\tilde{a}_2\equiv Q,$&$\tilde{a}_3 \equiv U,$&$\tilde{a}_4 \equiv V$
\\\hline\hline
$a$ & $(1,-1;1,0;0,1)$ & 3 & 0, & 0, & 1, & -1
\\\hline
$b$ & $(1,1;2,-1;1,1)$ & 2 & 3, & 0, & 3, & 0
\\\hline
$c$ & $(1,-1;-1,2;1,0)$ & 1 & 1, & 1, & 0, & 0
\\\hline
$d$ &  $(1,1;1,-2;0,1)$ & 1 & 0, & 0, & 3, & -3
\\\hline\hline
$h_3$ or $h_2+h_1$ &  $(1,-1;1,0;0,1)$ & 3 or 2+1 & 0, & 0, & 1, & -1
\\\hline\hline
$\hat{h}_1$ & $(1,1;1,-2;0,1)$ & 1 & 0, & 0, & 3, & -3
\\\hline
\end{tabular}%
\caption{Bulk brane configuration for standard model--like examples with  hidden sectors 
$SO/Sp(6)_{h_3}$ or $SO/Sp(4)_{h_2} \times SO/So(2)_{h_1}$
or $\widehat{SO/Sp(2)}_{\hat{h}_1}$.}
\label{tab:sm_bulk_branes}
\end{center}
\end{table}
\setlength{\tabcolsep}{0.5\tabcolsep}
\begin{table}[ht]%
\begin{center}%
\begin{tabular}{|c||rrrr|rrrr||rrrr|rrr|}\hline
\muc{16}{|c|}{\textbf{Standard model ex.: exceptional brane configuration}}\\\hline\hline
\textbf{brane}&$d_1,$&$d_2,$&$d_3,$&$d_4$&$e_1,$&$e_2,$&$e_3,$&$e_4$&$\sigma_1,$&$\sigma_2,$&$\sigma_5,$&$\sigma_6$&$\tau^0,$&$\tau^1,$&$\tau^3$\\\hline\hline
$a$ & 0, & 1, & -1, & 0 &  0, & 1, & -1, & 0 &  $\frac{1}{2}$, & 0, & $\frac{1}{2}$, & 0 & 0, & 1, & 1
\\\hline
$b$ & -3, & 0, & -3, & 0 & 0, & 0, & 0, & 0 & $\frac{1}{2}$, & 0, & 0, & 0 & 1, & 1, & 0
\\\hline
$c$ & 0, & 0, & 3, & -3 &  0, & 0, & -3, & 3 & $\frac{1}{2}$, & 0, & 0, & $\frac{1}{2}$ & 0, & 1, & 1
\\\hline
$d$ & 0, & -1, & 1, & 0 &  0, & -1, & 1, & 0 & 0, & 0, & $\frac{1}{2}$, & 0 & 0, & 1, & 1
\\\hline\hline
$h_3$ or $h_2$ & 1, & 0, & 0, & 1 & -1, & 0, & 0, & -1 & 0, & 0, & 0, & 0 & 0, & 1, & 0
\\
$h_1$ & & & & & & & &
& & & &
& 1, & 0, & 0
\\\hline\hline
$\hat{h}_1$& 3, & 0, & 0, & 3 & -3, & 0, & 0, & -3
& $\frac{1}{2}$, & 0, & 0, & 0  & 0, & 1, & 0
\\\hline
\end{tabular}%
\caption{Exceptional brane configuration for standard model--like examples with  different hidden sectors.}
\label{tab:sm_ex_branes}
\end{center}
\end{table}
Although brane $c$ is parallel to the $\OR\theta^{-4}$ plane with also the exceptional contributions 
orientifold invariant and in the classification of~\cite{gmho07} corresponds to brane (1c) with $SO(2)_c$ or $Sp(2)_c$ gauge group,
we assume here that brane $c$ is displaced on the $\bZ_2$ invariant torus $T^2_2$ and the gauge group broken to $U(1)_c$.
We list here {\it all} states which arise in the decomposition $SO/Sp(2)_c \rightarrow U(1)_c$, namely 
$\Anti_c \rightarrow \1_0$ and $\Sym_c \rightarrow \1_0 + \1_2 + \1_{-2}$, however, one should keep in mind that 
by parallely separating brane $c$ and $(\theta c')$ on $T^2_2$, at least one of the final states acquires a mass  proportional to the distance of $c$ and $(\theta c')$.
The intersection numbers are then given in table~\ref{tab:sm_chiral_spec},
\begin{table}[ht]%
\begin{center}%
\begin{tabular}{|c||r|r||r|r|r||r|r|r||r|r|}\hline
\muc{11}{|c|}{\textbf{Standard model ex.: chiral matter spectrum}}\\\hline\hline
\textbf{brane} &
$\mathbf{\chi^{\Anti}}$ & $\mathbf{\chi^{\Sym}}$ &
$\mathbf{\chi^{\cdot b}}$ & $\mathbf{\chi^{\cdot c}}$ & $\mathbf{\chi^{\cdot d}}$
& $\mathbf{\chi^{\cdot b'}}$ & $\mathbf{\chi^{\cdot c'}}$ & $\mathbf{\chi^{\cdot d'}}$
&  $\mathbf{\chi^{\cdot h}}$ & $\mathbf{\chi^{\cdot h'}}$
\\\hline\hline
$a$ &  0 & 0 & 0 & -3 $(d_R)$ & 0 & 3 $(Q_L)$ & -3 $(u_R)$ & 0 & 0 & 0
\\\hline
$b$ &   0 & 0 &  &-9 $(H_u)$ & 6 $(L)$ & & -9 $(H_d)$ & 3 $(\ov{L})$ &  0 & 0
\\\hline
$c$ &   0 & 0 & \muc{2}{c|}{} & -3 $(\nu_R)$  &  \muc{2}{c|}{} & 3 $(e_R)$ & 0 & 0
\\\hline
$d$ &   0 & 0 & \muc{3}{c|}{} &  \muc{3}{c|}{} &  0 & 0
\\\hline
\end{tabular}
\caption{The chiral spectrum $[C]$ derived from non-vanishing 3-cycle intersection numbers
is identical for all three configurations with hidden branes $h_3$
or $h_2 + h_1$ or $\hat{h}_1$. The identifications with standard model--like particles are given in
parenthesis, see also equation~\protect\eqref{Eq:Ex1-Chiral}.}
\label{tab:sm_chiral_spec}
\end{center}%
\end{table}
and also the chiral plus non-chiral part of the spectrum only charged under the observable sector involving 
branes $a,b,c$ and $d$
in table~\ref{tab:sm__full_observable_spec} is universal.
\begin{table}[ht]%
\begin{center}%
\begin{tabular}{|c||c|c|c||c|c|c||c|c|c|}\hline
\muc{10}{|c|}{\textbf{Standard model ex.: complete observable matter spectrum}}\\\hline\hline
\textbf{brane}
& $\mathbf{\varphi^{\Adj}}$ &
$\mathbf{\varphi^{\Anti}}$ & $\mathbf{\varphi^{\Sym}}$ &
$\mathbf{\varphi^{\cdot b}}$ & $\mathbf{\varphi^{\cdot c}}$ & $\mathbf{\varphi^{\cdot d}}$
& $\mathbf{\varphi^{\cdot b'}}$ & $\mathbf{\varphi^{\cdot c'}}$ & $\mathbf{\varphi^{\cdot d'}}$
\\\hline\hline
$a$ &  2
& 6 & 0
& 0 & 3 & 6 & 5 & 3 & 6
\\\hline
$b$ &  10
& 8 & 12
& & 11 & 8 & & 11 & 7
\\\hline
$c$ &   4
& $(2x)$ & $8-2x$
&  \muc{2}{c|}{} & 5 & \muc{2}{c|}{} & 5
\\\hline
$d$ &   10
& (22) & 0
& \muc{3}{c|}{} &  \muc{3}{c|}{}
\\\hline
\end{tabular}
\caption{The universal chiral plus non-chiral matter spectrum of the standard model--like examples with only
charges under branes $a, b, c, d$. The models differ in their hidden sector gauge groups and non-chiral
matter charged under them as listed in table~\protect\ref{tab:sm_hidden_spec}. $0\leq x\leq4$ depends on the 
unknown $\OR\theta^{-k}$ eigenvalues of the massless states at $c(\theta^{k-1} c)$ intersections.}
\label{tab:sm__full_observable_spec}
\end{center}%
\end{table}

The models differ in the purely non-chiral part of the spectrum with hidden sector charges. 
Their multiplicities are given in table~\ref{tab:sm_hidden_spec}. Branes $h_i$ are orthogonal to the 
$\OR\theta^{-1+3k}$ invariant planes on $T^2_2$ and
carry $SO/Sp(2i)$ 
gauge factors~\footnote{The classification (1) and (2) of D6-branes in table 20 of~\protect\cite{gmho07} with $SO(2N)$ or 
$Sp(2N)$ gauge groups comprised only those parallel to some $\OR\theta^{-k}$ planes along all two-tori. Branes $h_i$ are 
instead orthogonal to some  $\OR\theta^{-k}$ invariant plane on $T^2_2$ and also orthogonal on either $T^2_1$ or $T^2_3$
while being parallel to the same orientifold plane on the remaining two-torus, a case overlooked before. 
One might worry about a new K-theory constraint from the two new kinds of invariant branes (3) and (4), 
but it turns out that it is trivially fulfilled also in
these cases since the line of argument in~\protect\cite{gmho07} involved combinatorics of 
odd and even numbers only, and, e.g., $\Pi_{\hat{h}_1}=3 \Pi_{h_i}$ as well as $(n_j^{\hat{h}_1},m_j^{\hat{h}_1})
= (n_j^{h_i},m_j^{h_i}) \,{\rm mod} \, 2$ do not alter this reasoning.
We had furthermore verified explicitly in~\protect\cite{gmho07} that the invariant branes (1) and (2) do not lead to any
standard model-like solutions with an $Sp(2)_b$ gauge factor. There might potentially be such configurations with the new invariant 
branes of type (3) or (4), but since it is at present not clear if symplectic or orthogonal gauge factors arise, this option is 
not taken into account in the present analysis.
}
whereas $\hat{h}_1$ is parallel to the $\OR\theta^{-1}$ plane on all three tori corresponding to 
brane (2a) in the classification of~\cite{gmho07} providing a hidden $\widehat{SO(2)}_{\hat{h}_1}$ or $\widehat{Sp(2)}_{\hat{h}_1}$ gauge factor.
\begin{table}[ht]
\begin{center}%
\begin{tabular}{|c||c|c||c||c||c|c|}\hline
\muc{7}{|c|}{\textbf{Standard model ex.: non-chiral exotic matter}}\\\hline\hline
\textbf{brane}
& $\mathbf{\varphi^{\cdot h_3}} =\mathbf{\varphi^{\cdot h_2}} $
& $\mathbf{\varphi^{\cdot h_1}}$
& $\mathbf{\varphi^{\cdot \hat{h}_1}}$
& \textbf{brane}
& $\mathbf{\varphi^{\Anti}}$ & $\mathbf{\varphi^{\Sym}}$
 \\\hline\hline
$a$ &  0 & 0 & 0 
& $h_3,h_2,h_1$ &  
$z$ & $2-z$ 
\\\hline
$b$ &  2 & 2 & 8
& $\hat{h}_1$ & 
$y$ & $10-y$
\\\hline
$c$ & 2 & 2 & 10
&\muc{3}{c|}{}
\\\hline
$d$ & 0 & 0 & 0
&\muc{3}{c|}{}
\\\hline
$h_2$ & & 2 
& & \muc{3}{c|}{}
\\\hline
\end{tabular}
\caption{Matter at intersection with hidden branes for the three different models with
hidden sector gauge groups $SO/Sp(6)$, $SO/Sp(4) \times SO/Sp(2)$ and $\widehat{SO/Sp(2)}$ corresponding to
branes $h_3$, $h_2+h_1$ and $\hat{h}_1$, respectively. $0 \leq z \leq 2$ and 
$0\leq y \leq 10$ depend on the undetermined  $\OR\theta^{-k}$ eigenvalues of massless states at 
$h_i(\theta^{k-1} h_i)$ and  $\hat{h}_1 (\theta^{k-1} \hat{h}_1)$ intersections, respectively.}
\label{tab:sm_hidden_spec}
\end{center}%
\end{table}
In the observable sector, out of the four $U(1)$ factors, two stay massless. We choose
the hyper charge $Y$ and $B-L$ symmetry as basis,
\begin{equation}\label{Eq:ex-massless-U1}
Q_Y=\frac{1}{6}Q_a + \frac{1}{2} Q_c +\frac{1}{2} Q_d,
\qquad
Q_{B-L} = \frac{1}{3} Q_a + Q_d.
\end{equation}
The hidden sector gauge groups are either of type $SO(2N)$ or $Sp(2N)$. When broken to $U(N)$, 
the Abelian part will stay massless. 
Remember that the transverse degrees of freedom of the two massive $U(1)$ factors, i.e. $Q_b$ and $3 Q_a - Q_d$, 
are provided by uncharged {\it closed} string RR states and that the supersymmetry conditions~\eqref{Eq:SUSY}
imply that {\it no Fayet-Iliopoulos term} appears. 

The (open string) matter spectrum thus consists of three sectors, $[C]+[V_U] + [V_H]$, 
where $[C]$ contains all (chiral) matter derived from non-vanishing 3-cycle intersection numbers,
$[V_U]$ the vector like matter occurring in all three models (these are the adjoint representations and
${\cal N}=2$ supersymmetric sectors in  complex representations ${\bf R} + \cc$)
and $[V_H]$ the vector like matter which differs for the three choices of hidden sectors
(it contains symmetric and antisymmetric representations of the hidden $SO/Sp(2N)$ gauge factors 
as well as ${\cal N}=2$ sectors in complex representations ${\bf R} + \cc$):
\begin{enumerate}
\item
The chiral spectrum of $\left( (S)U(3)_a \times (S)U(2)_b \right)_{Q_Y \times Q_{B-L}}$ 
which is identical in all models. In order to make the perturbatively allowed pattern of Yukawa couplings
more obvious, we also list the charges under the unphysical $U(1)s$ as upper indices ${}^{(Q_c,Q_d)}$.
All states in this sector form ${\cal N}=1$ multiplets at intersections of two different stacks of branes.
For shortness, the (trivial) representation under the hidden gauge group is not listed,
\begin{equation}\label{Eq:Ex1-Chiral}
\begin{aligned}
&[C] =3\times\bigg[ 
  \left(\3,\2\right)_{\bf 1/6, 1/3}^{(0,0)}
  + \left(\ov{\3},\1\right)_{\bf 1/3,-1/3}^{(1,0)}
  + \left(\ov{\3},\1\right)_{\bf -2/3, -1/3}^{(-1,0)}
\\
&\qquad\qquad\quad
  + \left(\1,\1\right)_{\bf 1,1}^{(1,1)}
  + \left(\1,\1\right)_{\bf 0,1}^{(-1,1)}
  + 2 \times \left(\1,\2\right)_{\bf -1/2,-1}^{(0,-1)}
\\
&\qquad\qquad\quad
  + \left(\1,\2\right)_{\bf 1/2,1}^{(0,1)}
  + 3 \times  \left(\1,\ov{\2}\right)_{\bf -1/2,0}^{(-1,0)}
  + 3 \times  \left(\1,\ov{\2}\right)_{\bf 1/2,0}^{(1,0)}
\bigg] 
\\
&\quad\;\equiv 3\times\bigg[
  Q_L + d_R + u_R + e_R + \nu_R + 2 \times L + \ov{L}\bigg] + 9 \times \bigg[  H_d + H_u  \bigg],
\end{aligned}
\end{equation}
with an net-number of three chiral lepton and nine $(H_u,H_d)$ families.
The $B-L$  charge reveals that there are furthermore three lepton-anti-lepton generations.

One can read off that the standard combinations for Yukawa couplings 
involving 
\begin{equation}\label{Eq:Ex1-W-allowed}
u_R \cdot  Q_L\cdot  H_u, \quad
 d_R \cdot Q_L\cdot  H_d,\quad
e_R\cdot  L \cdot H_d \quad {\rm and} \quad 
\nu_R \cdot L\cdot  H_u 
\end{equation}
are perturbatively allowed, whereas 
a $\mu$ term type combination \mbox{$H_u\cdot  H_d$} is perturbatively forbidden by the $U(1)$ symmetry inside $U(2)_b$.
Since global (anomalous) symmetries are generally broken by non-perturbative effects, we expect
an exponentially suppressed $\mu$-term to be generated by instanton corrections. 
Contrarily, a right-handed  neutrino Majorana mass term and quartic coupling appearing in the see-saw mechanism,
\begin{equation}\label{Ex1-W-forbidden}
\nu_R \cdot \nu_R
\quad {\rm and} \quad 
L \cdot L \cdot H_u \cdot H_u ,
\end{equation}
are forbidden by the massless  $B-L$ symmetry.
%
\item
The universal part of the non-chiral spectrum containing matter only charged
under stacks $a,b,c,d$ consists of adjoints in the first line 
and ${\cal N}=2$ supersymmetric sectors in the second to fifth line
consisting of bifundamental, symmetric
 and antisymmetric representations, the latter two being denoted by lower indices $S$ and $A$,
respectively. We list the ${\cal N}=2$ part of the spectrum in complex representations  
in square brackets in the form of chiral representations $+\cc$ (complex conjugate)
where the ``$+\cc$'' applies to {\it all} representations inside the square bracket.
\begin{equation}
\begin{aligned}
& [V_U]=
2 \times \left(\bf{8},\1\right)_{\bf 0,0}^{(0,0)}
  + 10 \times \left(\1,\3\right)_{\bf 0,0}^{(0,0)}
  + 26  \times \left(\1,\1\right)_{\bf 0,0}^{(0,0)}
\\
& \qquad\quad + \bigg[
    \left(\3,\2\right)_{\bf 1/6,1/3}^{(0,0)}
  + 3 \times \left(\ov{\3},\1\right)_{\bf 1/3,2/3}^{(0,1)}
  + 3 \times \left(\ov{\3},\1\right)_{\bf -2/3,-4/3}^{(0,-1)}
\\
&\qquad\qquad\quad
  + \,3 \times  \left(\ov{\3}_A,\1\right)_{\bf 1/3,2/3}^{(0,0)}
  + 4 \times \left(\1,\1_A\right)_{\bf 0,0}^{(0,0)}
  + 6 \times  \left(\1,\3_S\right)_{\bf 0,0}^{(0,0)}
\\
&\qquad\qquad\quad
  + \left(\1,\ov{\2}\right)_{\bf -1/2,0}^{(-1,0)}
  + \left(\1,\2\right)_{\bf -1/2,-1}^{(0,-1)}
  + \left(\1,\ov{\2}\right)_{\bf 1/2,0}^{(1,0)}
  + 2 \times \left(\1,\2\right)_{\bf 1/2,1}^{(0,1)}
\\
&\qquad\qquad\quad
  + \left(\1,\1\right)_{\bf 1,1}^{(1,1)}
  +  \left(\1,\1\right)_{\bf 0,1}^{(-1,1)}
  + (4-x) \times \left(\1,\1 \right)_{\bf 1,0}^{(2,0)}
  + \;  \cc \; \bigg] .
\end{aligned}
\end{equation}
The parameter $0 \leq x \leq 4$ depends on the undetermined $\OR\theta^{-k}$ eigenvalues 
of the massless states on brane $c$.
\\
The vector-like pairs $\left[\left(\1,\1\right)_{\bf 0,1}^{(-1,1)} +\,\cc\right]$ might potentially trigger
the breaking of $U(1)_{B-L}$ without a simultaneous supersymmetry breaking.
\item
The non-universal part containing matter charged under the hidden sector gauge group
where hidden sector representations are the last one or two entries separated from the observable
sector charges by a semicolon. Remarkably, no matter arises at intersections of brane $a$ and $d$ 
with the hidden branes leading to $Q_{B-L}=0$ in this sector. The notation for the non-chiral
matter is  analogous to the one for the universal part of the spectrum with antisymmetric and 
symmetric representations of $SO/Sp(2N)$ replacing the adjoints of $U(N)$.
\begin{enumerate}
\item
For the hidden gauge group $SO/Sp(6)_{h_3}$
\begin{equation}
\begin{aligned}
& [V_{H_1}]=
z \times (\1,\1;{\bf 15})^{(0,0)}_{\bf 0,0} +
(2-z) \times  (\1,\1;{\bf 21})^{(0,0)}_{\bf 0,0} 
\\
& \qquad\quad + \bigg[ 
(\1,\2;\6)^{(0,0)}_{\bf 0,0} + (\1,\1;\6)^{(1,0)}_{\bf 1/2,0}  
 \; + \cc \; \bigg] .
\end{aligned}
\end{equation}
\item
Or for hidden sector gauge factors $SO/Sp(4)_{h_2} \times SO/Sp(2)_{h_1}$
\begin{equation}
\begin{aligned}
& [V_{H_2}]=
z \times(\1,\1;\6,\1)_{\bf 0,0}^{(0,0)}
+(2-z) \times(\1,\1;{\bf 10},\1)_{\bf 0,0}^{(0,0)}
 \\
 & \qquad\quad
+z \times(\1,\1;\1,\1)_{\bf 0,0}^{(0,0)}
+(2-z) \times(\1,\1;\1,\3)_{\bf 0,0}^{(0,0)}
\\
&\qquad\quad
+ 2 \times (\1,\1;\4,\2)_{\bf 0,0}^{(0,0)}
 \\
 & \qquad\quad
+\bigg[ 
(\1,\2;\4,\1)_{\bf 0,0}^{(0,0)}
+(\1,\2;\1,\2)_{\bf 0,0}^{(0,0)}
\\
&\qquad\qquad\quad
+(\1,\1;\4,\1)_{\bf 1/2,0;0}^{(1,0)}
+(\1,\1;\1,\2)_{\bf 1/2,0}^{(1,0)}
 \; + \cc \; \bigg].
\end{aligned}
\end{equation}
\item
Or for a hidden $\widehat{SO/Sp(2)}_{\hat{h}_1}$
\begin{equation}
\begin{aligned}
&  [V_{H_3}]=
y \times (\1,\1;\1_A)^{(0,0)}_{\bf 0,0}
+ (10-y) \times (\1,\1;\3_S)^{(0,0)}_{\bf 0,0}
\\
& \qquad\quad
+\bigg[ 4 \times (\1,\2;\2)^{(0,0)}_{\bf 0,0}  + 5 \times (\1,\1;\2)^{(1,0)}_{\bf 1/2,0} 
 \; + \cc \; \bigg],
\end{aligned}
\end{equation}
\end{enumerate}
where the parameters $z$ and $y$ depend on the undetermined $\OR\theta^{-k}$ eigenvalues of the massless states
at intersections of $h_i$ and $\hat{h}_1$ with their orbifold images $(\theta^{k-1} h_i)$ and $(\theta^{k-1} \hat{h}_1)$,
respectively.

These models belong to the class of spectra with 9+3=12 ``chiral" Higgs plus lepton-anti-lepton 
families and no chiral exotics whose tree level ratio of gauge couplings are given in table~\ref{tab:zeroexstat}.
The beta function coefficients for these models can be computed explicitly,
\begin{equation}
\begin{aligned}
b_{SU(3)_a} &=14, \qquad
&
b_{SU(2)_b} &=\left\{\begin{array}{cc} 70 & h_3, h_2+h_1
\\ 72 & \hat{h}_1
\end{array}\right. ,
\\
b_{U(1)_{B-L}} &= \frac{280}{3},
&
b_{U(1)_Y} &= \left\{\begin{array}{cc} \frac{305}{6} - 2x & h_3, h_2+h_1
\\\frac{157}{3} - 2x & \hat{h}_1
\end{array}\right..
\end{aligned}
\end{equation}
\end{enumerate}
Although interactions have to our knowledge not yet been computed for the type of fractional D6-branes 
used in the construction, some of the non-chiral matter in the $[V_U]+[V_H]$ sectors can be made massive by simply
parallely displacing the one-cycles along $T^2_2$ away from the origin and each other. The parallel transport
preserves the supersymmetry conditions~\eqref{Eq:SUSY}, and the situation is depicted in figure~\ref{fig:symmetry-breaking} 
for branes $b$ and $c$. 

\twofigvar{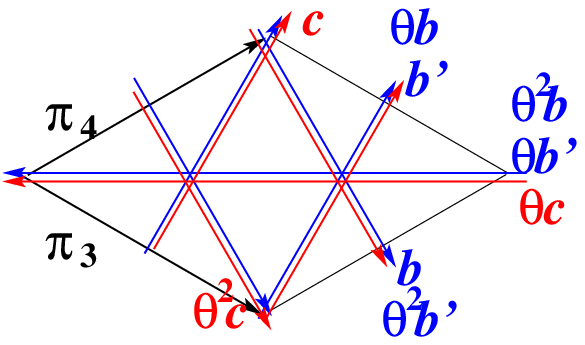}{.32}{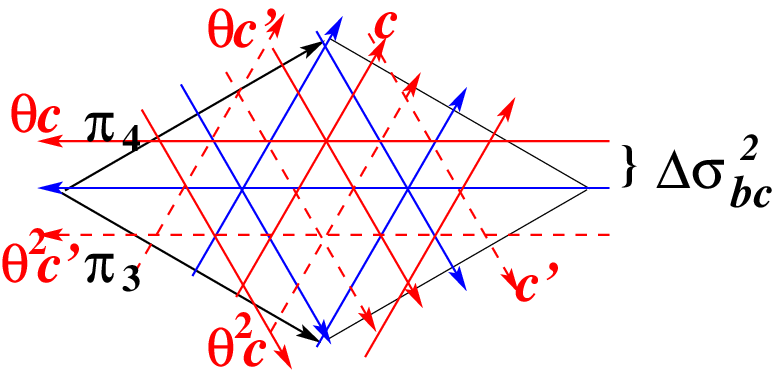}{.4}{%
Gauge symmetry breaking and masses through parallel displacement on $T^2_2$.
(a) branes $b$ and $c$ pass through the origin and carry gauge groups $U(2)_b \times SO/Sp(2)_{c}$. 
For brane $c$, the orientifold images are related by  $(\theta^k c)=(\theta^{k+1} c')$. 
(b) The displacement of brane $c$ breaks $SO/Sp(2)_{c}\rightarrow U(1)_c$ and renders 
states at $b(\theta^2 c)$ and $bc'$ intersections massive.}{fig:symmetry-breaking}

The mass scale is again set by the continuously variable
distance $\Delta\sigma^2$ according to equation~\eqref{Eq:masses}.

After displacing all brane stacks relative to each other along $T^2_2$, the chiral sector $[C]$ remains unchanged, 
but the vector like matter sector $[V_U]$ is reduced  (with $0 \leq \tilde{x} \leq 3$),
\begin{equation}
\begin{aligned}
{}\!
[V_U]^{(\Delta\sigma^2)} &=2 \times \left(\bf{8},\1\right)_{\bf 0,0}^{(0,0)}
  + 10 \times \left(\1,\3\right)_{\bf 0,0}^{(0,0)}
  + 25  \times \left(\1,\1\right)_{\bf 0,0}^{(0,0)}
\\
& \qquad + \bigg[
    \left(\3,\2\right)_{\bf 1/6,1/3}^{(0,0)}
  + 3 \times \left(\ov{\3},\1\right)_{\bf 1/3,2/3}^{(0,1)}
  + 3 \times \left(\ov{\3},\1\right)_{\bf -2/3,-4/3}^{(0,-1)}
\\
&\qquad\qquad
  + \left(\ov{\3}_A,\1\right)_{\bf 1/3,2/3}^{(0,0)}
  + 6 \times  \left(\1,\3_S\right)_{\bf 0,0}^{(0,0)}
  + (3-\tilde{x}) \times \left(\1,\1 \right)_{\bf 1,0}^{(2,0)}
  + \;  \cc \; \bigg] ,
\end{aligned}
\end{equation}
whereas $ [V_{H_1}]$ and  $ [V_{H_2}]$ are not affected.
$\widehat{SO/Sp(2)}_{\hat{h}_1}$ can be broken to $U(1)_{\hat{h}_1}$ via parallel displacements
leading to 
\begin{equation}
\begin{aligned}
{}
 [V_{H_3}]^{(\Delta\sigma^2)} &=
9 \times (\1,\1)^{(0,0)}_{\bf 0,0;0}
+ (9-\tilde{y}) \times \bigg[(\1,\1)^{(0,0)}_{\bf 0,0,1} \; + \cc \; \bigg]
\\
& 
+3 \times \bigg[ (\1,\2)^{(0,0)}_{\bf 0,0;1} + (\1,\2)^{(0,0)}_{\bf 0,0;-1}  
 +  (\1,\1)^{(1,0)}_{\bf 1/2,0;1}  +  (\1,\1)^{(1,0)}_{\bf 1/2,0;-1}
 \; + \cc \; \bigg],
\end{aligned}
\end{equation}
with $0 \leq \tilde{y} \leq 9$.
 
The beta function coefficients are then lowered to 
\begin{equation}
b_{SU(3)_a}^{(\Delta\sigma^2)} =12,\quad
b_{SU(2)_b}^{(\Delta\sigma^2)} = 65,\quad
b_{U(1)_Y}^{(\Delta\sigma^2)} =
42 - 2 \tilde{x},\quad
b_{U(1)_{B-L}}^{(\Delta\sigma^2)} = 72  .
\end{equation}
These string-scale values are clearly not realistic, in particular $b_{SU(3)_a}^{(\Delta\sigma^2)}$ has the wrong sign.

\subsection{A three generation model without hidden sector}\label{Subsec:Example2-nohidden}
As mentioned in section~\ref{sec:noex}, we found 61,440 models without any
hidden sector. There are two types of models which can be classified by the
number of $h$ Higgs-up and $h$ Higgs-down multiplets (and vector-like lepton pairs) that occur.
As shown in equation~\eqref{eq:vsmatter}, there are two groups with $h=18$ and 21, respectively. 
In the following we analyse one example of a model with $h=21$ Higgs candidates,
or more precisely 18 Higgs generations and three vector-like lepton pairs distinguished 
by their $B-L$ charge, in more detail.

The model is realised with a value of $\varrho=1/4$ for the complex
structure parameter on $T^2_3$. The configuration of bulk branes is given in
table~\ref{tab:sm_bulk_branes2} and the exceptional part containing the
displacements $\sigma_i$ and Wilson lines $\tau^j$ in
table~\ref{tab:sm_ex_branes2}.

\begin{table}[ht]
\begin{center}%
\begin{tabular}{|c||c|c|rrrr|}\hline
\muc{7}{|c|}{\textbf{Ex. w/o hidden sector: bulk brane configuration}}\\\hline
\textbf{brane}& $(n_1,m_1;n_2,m_2;n_3,m_3)$ & $N$&$\tilde{a}_1\equiv P,$&$\tilde{a}_2\equiv Q,$&$\tilde{a}_3\equiv U,$&$\tilde{a}_4\equiv V$
\\\hline\hline
$a$ &  $(1,-1;1,0;0,1)$ &  3 & 0, & 0, & 1, & -1 
\\\hline
$b$ &  $(1,1;2,-1;1,2)$ & 2 & 3, & 0, & 6, & 0
\\\hline
$c$ & $(1,-1;-1,2;1,0)$ & 1 & 1, & 1, & 0, & 0  
\\\hline
$d$ &  $(1,1;1,-2;0,1)$ & 1 & 0, & 0, & 3, & -3  
\\\hline
\end{tabular}
\caption{Bulk brane configuration for a standard model example without hidden sector. Only 
$m^b_3$ differs from the bulk configurations with hidden branes in section~\protect\ref{Subsec:Examples-with-hidden}.}
\label{tab:sm_bulk_branes2}
\end{center}
\end{table}

\begin{table}[ht]%
\begin{center}%
\begin{tabular}{|c||rrrr|rrrr||rrrr|rrr|}\hline
\muc{16}{|c|}{\textbf{Ex. w/o hidden sector: exceptional brane configuration}}\\\hline
\textbf{brane}&$d_1,$&$d_2,$&$d_3,$&$d_4$&$e_1,$&$e_2,$&$e_3,$&$e_4$&$\sigma_1,$&$\sigma_2,$&$\sigma_5,$&$\sigma_6$&$\tau^0,$&$\tau^1,$&$\tau^3$\\\hline\hline
$a$ & 0, & -1, & -1, & 0 & 0, & -1, & -1, & 0 
& $\frac{1}{2}$ & 0 & $\frac{1}{2}$ & 0
& 1 & 1 & 0
\\\hline
$b$ & -3, & -3, & 0, & 0 & 0, & 0, & 0, & 0 
& $\frac{1}{2}$ & 0 & 0 & 0
& 1 & 1 & 0
\\\hline
$c$ & 3, & 3, & 0, & 0 & -3, & -3, & 0, & 0 
& $\frac{1}{2}$ & 0 & 0 & 0
& 1 & 1 & 0
\\\hline
$d$ & 0, & 1, & 1, & 0 & 0, & 1, & 1, & 0 
& 0  & 0 & $\frac{1}{2}$ & 0
& 0 & 0 & 0
\\\hline
\end{tabular}%
\caption{Exceptional brane configuration for a standard model example without hidden sector.}
\label{tab:sm_ex_branes2}
\end{center}
\end{table}

In tables~\ref{tab:nohchi} and~\ref{tab:smO_ex2_fullspec} we list the
intersection numbers leading to the chiral and full matter spectrum,
respectively.

Concerning the $U(1)$ factors, we again find that $U(1)_b$ and the combination
$3U(1)_a-U(1)_d$ are massive, but we have a massless $U(1)_{B-L}$ symmetry
given by $\frac{1}{3}U(1)_a+U(1)_d$. Furthermore brane $c$ leads to an
additional massless $U(1)$, or provides a (purposefully broken) $SO/Sp(2)$ symmetry (brane $c$ corresponds to (1a)
in the classification of orientifold invariant branes in~\cite{gmho07}). 
We can thus take $U(1)_{B-L}$ and $U(1)_Y$ to span the space of massless 
Abelian gauge factors as in equation~\eqref{Eq:ex-massless-U1}.

\begin{table}[ht]
\begin{center}
\begin{tabular}{|c||r|r||r|r|r||r|r|r|}\hline
\muc{9}{|c|}{\textbf{Ex. w/o hidden sector: chiral matter spectrum}}\\\hline\hline
\textbf{brane} &
$\mathbf{\chi^{\Anti}}$ & $\mathbf{\chi^{\Sym}}$ &
$\mathbf{\chi^{\cdot b}}$ & $\mathbf{\chi^{\cdot c}}$ & $\mathbf{\chi^{\cdot d}}$
& $\mathbf{\chi^{\cdot b'}}$ & $\mathbf{\chi^{\cdot c'}}$ & $\mathbf{\chi^{\cdot d'}}$
\\\hline\hline
$a$ &  0 & 0 & 0 & -3 $(d_R)$ & 0 & 3 $(Q_L)$ & -3 $(u_R)$ & 0
\\\hline
$b$ &   -9 & 0 &  &-18 $(H_u)$ & 6 $(L)$ & & -18 $(H_d)$ & 3 $(\ov{L})$
\\\hline
$c$ &   0 & 0 & \muc{2}{c|}{} & -3 $(\nu_R)$  &  \muc{2}{c|}{} & 3 $(e_R)$
\\\hline
$d$ &   0 & 0 & \muc{3}{c|}{} &  \muc{3}{c|}{}
\\\hline
\end{tabular}
\caption{Chiral spectrum of a model without a hidden sector.
The identifications with standard model--like particles are given in
parenthesis, see also equation~\protect\eqref{Eq:ex2-explicitspectrum-chiral}.}
\label{tab:nohchi}
\end{center}
\end{table}

\begin{table}[ht]
\begin{center}
\begin{tabular}{|c||r|r|r||r|r|r||r|r|r|}\hline
\muc{10}{|c|}{\textbf{Ex. w/o hidden sector: complete matter spectrum}}\\\hline
\textbf{brane} 
& $\mathbf{\varphi^{\Adj}}$ &
$\mathbf{\varphi^{\Anti}}$ & $\mathbf{\varphi^{\Sym}}$ &
$\mathbf{\varphi^{\cdot b}}$ & $\mathbf{\varphi^{\cdot c}}$ & $\mathbf{\varphi^{\cdot d}}$ 
& $\mathbf{\varphi^{\cdot b'}}$ & $\mathbf{\varphi^{\cdot c'}}$ & $\mathbf{\varphi^{\cdot d'}}$ 
\\\hline\hline
$a$ & 2 & 6 & 0
& 0 & 3 & 6 
& 5 & 3 & 6
\\\hline
$b$ & 10 & 11 & 20
& & 22 & 10 
& & 22 & 5
\\\hline
$c$ & 4 & $(2x)$ & $8-2x$
 &  \muc{2}{c|}{} &  5  & \muc{2}{c|}{} & 5
\\\hline
$d$ &  10 & (22) & 0
& \muc{3}{c|}{} &  \muc{3}{c|}{}
\\\hline
\end{tabular}
\caption{The complete non-chiral matter spectrum of the standard model example without hidden sector.  }
\label{tab:smO_ex2_fullspec}
\end{center}%
\end{table}

The complete matter spectrum can be divided into a chiral part  $[C]$
derived from non-vanishing 3-cycle intersection numbers,
\begin{equation}\label{Eq:ex2-explicitspectrum-chiral}
\begin{aligned}
{}
[C] &=3\times\bigg[ 
    \left(\3,\2\right)_{\bf 1/6, 1/3}^{(0,0)}
  + \left(\ov{\3},\1\right)_{\bf 1/3,-1/3}^{(1,0)}
  + \left(\ov{\3},\1\right)_{\bf -2/3, -1/3}^{(-1,0)}
  + \left(\1,\1\right)_{\bf 1,1}^{(1,1)}
  +  \left(\1,\1\right)_{\bf 0,1}^{(-1,1)}
\\
&\qquad\qquad
  + 2 \times \left(\1,\2\right)_{\bf -1/2,-1}^{(0,-1)}
  + \left(\1,\2\right)_{\bf 1/2,1}^{(0,1)}
  + 6 \times  \left(\1,\ov{\2}\right)_{\bf -1/2,0}^{(-1,0)}
  + 6 \times  \left(\1,\ov{\2}\right)_{\bf 1/2,0}^{(1,0)}
\\
&\qquad\qquad
  + 3 \times \left(\1,\1_{\ov{A}}\right)_{\bf 0,0}^{(0,0)}
\bigg]
\\
& \equiv 3 \times \bigg[Q_L +d_R + u_R + e_R +\nu_R + 2\times L + \ov{L}  \bigg]
+ 18 \times \bigg[ H_d + H_u\bigg]
+ 9 \times S ,
\end{aligned}
\end{equation}
and a non-chiral part $[V]$, which consists of matter in the adjoint representation
of the gauge group and matter from the ${\cal N}=2$ sectors, listed here as chiral
representations $+\cc$ (complex conjugate) inside square brackets.
\begin{equation}\label{Eq:ex2-explicitspectrum-non-chiral}
\begin{aligned}
{}
[V] &= 2 \times \left(\bf{8},\1\right)_{\bf 0,0}^{(0,0)}
  + 10 \times \left(\1,\3\right)_{\bf 0,0}^{(0,0)}
  + 26  \times \left(\1,\1\right)_{\bf 0,0}^{(0,0)}
\\
&\quad + \bigg[   
     \left(\3,\2\right)_{\bf 1/6,1/3}^{(0,0)}
  + 3\times \left(\ov{\3},\1\right)_{\bf 1/3,2/3}^{(0,1)}
  + 3 \times \left(\ov{\3},\1\right)_{\bf -2/3,-4/3}^{(0,-1)}
\\
&\qquad\quad
  + (3-x+1_m) \times \left(\1,\1 \right)_{\bf 1,0}^{(2,0)}
  + \left(1+2_m  \right) \times (\ov{\3}_A,\1)_{\bf 1/3,2/3}^{(0,0)}
\\
&\qquad\quad
  + \left(9+1_m  \right) \times (\1,\3_S)_{\bf 0,0}^{(0,0)}
  + 2_m \times  \left(\1,\ov{\2}\right)_{\bf -1/2,0}^{(-1,0)}
\\
&\qquad\quad
  + 2_m \times \left(\1,\ov{\2}\right)_{\bf 1/2,0}^{(1,0)}
  + 2_m \times  \left(\1,\2\right)_{\bf -1/2,-1}^{(0,-1)}
  + 1_m \times  \left(\1,\2\right)_{\bf 1/2,1}^{(0,1)}
\\
&\qquad\quad
  + 1_m \times (\1,\1_A)_{\bf 0,0}^{(0,0)}
  + 1_m \times   (\1,\1)_{\bf 0,-1}^{(1,-1)} 
  + 1_m \times (\1,\1)_{\bf 1,1}^{(1,1)}  
  + \;\cc\;\bigg].
\end{aligned}
\end{equation}
To keep the notation compact, the multiplicities of non-chiral states which arise from sectors with D6-branes parallel
along $T^2_2$ and thus massive when branes are parallely displaced in a continuous manner are labelled by a lower index $m$. The 
last two lines of $[V]$ disappear completely from the massless spectrum if suitable distances $\Delta\sigma^2$ are chosen, and the 
multiplicities of states in the third and fourth line of $[V]$ are reduced. 

All states in the ``chiral'' sector $[C]$ derive from non-vanishing intersection numbers, but as in the previous examples, 
since $U(1)_b$ acquires a mass through the Green Schwarz mechanism by absorbing a {\it neutral} closed string field, 
the pairings of Higgs fields, $H_d+H_u$, and leptons, $L+\ov{L}$, and the states $S\equiv \left(\1,\1_{\ov{A}}\right)_{\bf 0,0}^{(0,0)}$ are vector-like with respect to  the massless gauge group 
$SU(3)_a \times SU(2)_b \times U(1)_Y \times U(1)_{B-L}$ of this model. 
The selection rules for perturbative superpotential couplings of standard model particles
are identical to the previous examples, see equations~\eqref{Eq:Ex1-W-allowed} and~\eqref{Ex1-W-forbidden}, since all quarks, leptons and Higgses arise from the same intersection 
pattern as displayed in equation~\eqref{eq:vsmatter} or by comparison of tables~\ref{tab:sm_chiral_spec} 
and~\ref{tab:nohchi}.

The beta function coefficients for this model are 
\begin{equation}
\begin{array}{ccc}
& \text{$[C]+[V]$,} 
& \text{$[C]+[V]$ minus massive states in~\protect\eqref{Eq:ex2-explicitspectrum-non-chiral},}
\\
b_{SU(3)_a}= & 14, & 12,
\\
b_{SU(2)_b}= & 91, & 80,
\\
b_{U(1)_Y}= & \frac{181}{3}-2x, & 48 - 2x,
\\
b_{U(1)_{B-L}}= & \frac{280}{3}, & 72.
\end{array}
\end{equation}
Again, in order to obtain realistic values, more non-chiral fields have to acquire a mass which necessitates
the exact computation of couplings or $n$-point functions and clearly goes beyond the scope of this paper.

\section{Conclusions}\label{Sec:Concl}

In this article, we have continued the statistical analysis of supersymmetric intersecting D6-brane models on
$T^6/\bZ_6'$ with the result that the seemingly large variety of spectra and gauge couplings 
observed in~\cite{gmho07} is subject to very strong correlations as displayed for the 
number of chiral exotics versus Higgses in table~\ref{Tab:Exotics-Higgses}. Most of the models 
without chiral exotics contain nine generations of chiral up- and down-type Higgs multiplets plus three vector-like leptons 
pairs stemming from non-vanishing 3-cycle intersection numbers 
in the examples where a $B-L$ symmetry distinguishes Higgses and sleptons.
Considering the hidden sectors, only gauge groups of at most rank three occur. All of these models 
have the same weak mixing angle and ratio of strong and weak gauge couplings at the string scale.

We have thus seen that there exist correlations between certain properties of models with the gauge
and chiral matter content of the standard model and an unconstrained hidden sector.
Whether this is true in a bigger context or an artefact of the specific geometric set-up of these
models is an interesting open question.
The correlations between other properties of theses models certainly also deserve
further study and should be compared with different constructions.\footnote{For some results
concerning correlations on $T^6/(\bZ_2\!\times\!\bZ_2)$ and $T^6/\bZ_6$
see~\cite{gbhlw05,gm07,WIP}.} Unfortunately, a statistical analysis of the complete non-chiral
matter spectrum has turned out to be too ambitious due to the numerous distinctions of special cases 
discussed in section~\ref{sec:ospec}.

We found that in the examples discussed in detail in section~\ref{App:Realisations}
the one-loop beta function coefficient for the strong coupling has the wrong sign, unless
most of the non-chiral states acquire masses at a high enough scale. To see if this can occur
for states other than those parallel on $T_2^2$, it 
will be necessary to determine interaction terms for the fractional branes under consideration.\footnote{For
{\it toroidal} branes, some results have been reported in~\cite{lmrs04}, and for {\it toroidal} instanton computations
see e.g.~\cite{bcw06,ibur06}. Threshold corrections and K\"ahler metrics on $T^6/(\bZ_2\!\times\!\bZ_2)$
with {\it rigid} branes have been considered in~\cite{blsc07}.} We hope to come back to this point in future work.

The analyses of the  $T^6/\bZ_6$ and $T^6/\bZ_6'$ intersecting D6-brane orbifold models 
suggest that three standard model--like generations  are naturally realised in the presence of a $\bZ_3$ 
symmetry, which was not expected from the originally proposed configuration in~\cite{imr01}, but 
occurs also in heterotic orbifold models, see e.g.~\cite{le06} and references therein.
It seems therefore natural to expect phenomenologically interesting models also on, e.g.,  $T^6/\bZ_6 \times \bZ_2$
orbifold backgrounds with torsion and rigid cycles~\cite{WIP}.

\subsection*{Acknowledgements}
G.~H. thanks Oleg Lebedev, Fernando Marchesano and Angel Uranga for helpful discussions.
The work of F.~G. is supported by the Foundation for Fundamental Research of
Matter (FOM) and the National Organisation for Scientific Research (NWO).

\appendix
\section{Chan-Paton labels}\label{App:CPlabels}

The representations of four dimensional massless open string states in a $T^6/\bZ_{2N}$ background 
depend on the $\bZ_2\equiv \theta^N$ orbifold and $\OR\theta^{-k}$ orientifold projections on the Chan-Paton labels
$\lambda$,
\begin{equation}\label{Eq:CP-projections}
\begin{aligned}
&\lambda = a_{\bZ_2} \left( \gamma_{\bZ_2}\lambda \gamma_{\bZ_2}^{-1} \right),
\\ 
&\lambda = a_{\OR\theta^{-k}}\left(\gamma_{\OR\theta^{-k}} \lambda \gamma_{\OR\theta^{-k}}^{-1} \right)^T,
\\ 
&\lambda = a_{\OR\theta^{-k+N}}\left(\gamma_{\OR\theta^{-k+N}} \lambda \gamma_{\OR\theta^{-k+N}}^{-1} \right)^T,
\end{aligned}
\end{equation}
with the eigenvalues $a_{\bZ_2},a_{\OR\theta^{-k}} = \pm 1$ of massless states 
and the consistency conditions on 
the combined orbifold and orientifold action 
\begin{equation}
a_{\OR\theta^{-k+N}} = a_{\bZ_2} \cdot a_{\OR\theta^{-k}},
\quad\quad
\gamma_{\OR\theta^{-k+N}} = \gamma_{\OR\theta^{-k}} \cdot \gamma_{\bZ_2}.
\end{equation}

The transformation properties of the intersection points determine which projection conditions have to be imposed
on the Chan-Paton matrices $\lambda$.
\eqref{Eq:CP-fractional} applies to intersection points which are neither 
$\bZ_2$ nor $\OR\theta^{-k}$
or $\OR\theta^{-k+N}$ invariant, which corresponds to the notation
$(a_{\bZ_2}, a_{\OR\theta^{-k}}, a_{\OR\theta^{-k+N}})=(*,*,*)$. 
The remaining cases are listed in table~\ref{Tab:CP-proj}.
\begin{table}[ht]
  \begin{center}
\begin{equation*}
\begin{array}{|c|c|c|c|}
\hline
 \multicolumn{4}{|c|}{\rule[-3mm]{0mm}{8mm}
\text{\bf Projections of Chan-Paton matrices}}
\\\hline\hline
\bZ_2 & \OR\theta^{-k} & \OR\theta^{-k+N} & {\rm rep}
\\\hline\hline
+ & + &  + & \left(\begin{array}{cc} ({\bf N^1}_a,{\bf N^2}_a)& 0 \\ 0  &  * \end{array}\right)
\\
+ & - &  - & \left(\begin{array}{cc} ({\bf N^1}_a,{\bf N^2}_a)& 0 \\ 0  & * \end{array}\right)
\\
- & + & - & \left(\begin{array}{cc} 0 &  \Sym^1_a \\ \Sym^2_a   & 0 \end{array}\right)
\\
- & - & + & \left(\begin{array}{cc} 0 & \Anti^1_a  \\ \Anti^2_a  & 0 \end{array}\right)
\\\hline
+ & * & * &  \left(\begin{array}{cc} ({\bf N^1}_a,\ov{\bf N^1}_b) & 0 \\ 0  & ({\bf N^2}_a,\ov{\bf N^2}_b)\end{array}\right)
\\
- & * & * &  \left(\begin{array}{cc} 0 &  ({\bf N^1}_a,\ov{\bf N^2}_b) \\ ({\bf N^2}_a,\ov{\bf N^1}_b) & 0 \end{array}\right)
\\\hline
* & + & * & \left(\begin{array}{cc} ({\bf N^1}_a,{\bf N^2}_a)  & \Sym^1_a \\ \Sym^2_a & * \end{array}\right)
\\
* & - & * & \left(\begin{array}{cc} ({\bf N^1}_a,{\bf N^2}_a)  &  \Anti^1_a\\ \Anti^2_a & *  \end{array}\right)
\\\hline
* & * & + &  \left(\begin{array}{cc} ({\bf N^1}_a,{\bf N^2}_a)  & \Anti^1_a\\ \Anti^2_a & * \end{array}\right)
\\
* & * & - &  \left(\begin{array}{cc} ({\bf N^1}_a,{\bf N^2}_a)  & \Sym^1_a \\ \Sym^2_a & * \end{array}\right)
\\\hline
\end{array}
\end{equation*}
 \end{center}
\caption{Projections of Chan-Paton labels for various combinations of orbifold and orientifold
eigenvalues $(a_{\bZ_2},a_{\OR\theta^{-k}},a_{\OR\theta^{-k+N}})$. The entry  $*$ for an eigenvalue
denotes no projection condition. A star in the lower right entry of the Chan-Paton label appears whenever
this entry is by the orientifold projection related to the upper left entry thereby not providing any new 
degrees of freedom (d.o.f.).}
\label{Tab:CP-proj}
\end{table}
\clearpage

\section{Massless open string states}\label{A:MasslessStates}

The general mass formula for states localised at brane intersections reads
\begin{equation}
\frac{\alpha'}{4}m^2=\frac{1}{2}p^2 + E_T - \frac{1}{2} \qquad
\text{with }
\qquad
E_T=\frac{1}{2}\sum_{i=1}^3 \phi_i(1-\phi_i) \quad \text{
  and } \quad \phi_i \in [0,1),
\end{equation}
where $\pi \phi_i$ is the angle between the branes on $T^2_i$ up to a possible shift of $\pi$.

Setting $\phi_0 \equiv 0$, in the NS sector the states at the brane intersection take the form
\begin{equation}
p= |\vec{n}-\vec{\phi}\rangle
\end{equation}
with $n_i \in \bZ$, and the raising and lowering oscillators  are 
$
\psi^{\mu}_{-1/2+m},
\psi^i_{-1/2+\phi_i+m},
\psi^{\bar{i}}_{-1/2-\phi_i+m},
$
with $i=1,2,3$ and $m \in \bZ$. GSO invariant states comply with $\sum_i n_i = $ odd.

The R sector states at brane intersections are given by
\begin{equation}
p= |\vec{n}+\vec{\frac{1}{2}}-\vec{\phi}\rangle
\end{equation}
with corresponding oscillator states
$
\psi^{0}_{m},
\psi^i_{\phi_i +m},
\psi^{\bar{i}}_{-\phi_i+m}
$, where $\psi^{0}_0$ flips the four dimensional chirality of a given state.
We use the convention that the GSO projection  enforces $\sum_i n_i =$ even.

The resulting massless and GSO projected open string states for supersymmetric configurations of D6-branes 
at angles are given in table~\ref{Tab:Massless-States}.
Generically, the state $a(\theta^k b)$ at a supersymmetric brane intersection with angles $\pi(\phi_1,\phi_2,\phi_3)$
and $\sum_{i=1}^3 \phi_i = 0$ mod 2
provides one massless NS and R sector d.o.f.
with a given $\bZ_2$ parity. Its inverse sector $(\theta^k b)a$ at angles $\pi(-\phi_1,-\phi_2,-\phi_3)$
furnishes the missing d.o.f. to form a full chiral multiplet.
The $\bZ_2$ parity is read off from the fact that the unique (tachyonic) NS ground-states $|0\rangle_{\rm NS}$ 
is $\bZ_2$ even,
while in the sector with one vanishing angle, e.g. $\pi(\phi, - \phi,0)$, the oscillator $\psi^{2}_{-1/2+\phi}$ 
is also $\bZ_2$ even, but $\psi^{\ov{1}}_{-1/2+\phi}$ is $\bZ_2$ odd.
\begin{table*}[ht]
\centering
\resizebox{\linewidth}{!}{%
$\begin{array}{|c||c|c||c|}
\hline
\multicolumn{4}{|c|}{\rule[-3mm]{0mm}{8mm}
\text{\bf Massless open string states}}
\\\hline\hline
(\phi_1,\phi_2,\phi_3) & \mbox{NS} & \mbox{R} & \tilde{a}_{\bZ_2} 
\\\hline\hline
(0,0,0) & |\pm 1,0,0,0\rangle_{\rm NS} =\psi^{\mu}_{-1/2} |0\rangle_{\rm NS}
& \pm |\half,\half,\half,\half\rangle_{\rm R}   
= |0\rangle_{\rm R}, \left(\prod_{i=0}^3 \psi^i_0  \right)|0\rangle_{\rm R}
& +
\\
&  |0,\pm 1,0,0\rangle_{\rm NS}  =\psi^{1,\ov{1}}_{-1/2} |0\rangle_{\rm NS} 
&   \psi^0_0 \psi^1_0  |0\rangle_{\rm R}, \psi^2_0 \psi^3_0  |0\rangle_{\rm R}
& -
\\
& |0,0,\pm 1,0\rangle_{\rm NS}  =\psi^{2,\ov{2}}_{-1/2} |0\rangle_{\rm NS} 
&  \psi^0_0 \psi^2_0  |0\rangle_{\rm R}, \psi^1_0 \psi^3_0  |0\rangle_{\rm R}
& +
\\
& |0,0,0,\pm 1\rangle_{\rm NS}  =\psi^{3,\ov{3}}_{-1/2} |0\rangle_{\rm NS} 
&  \psi^0_0 \psi^3_0  |0\rangle_{\rm R}, \psi^1_0 \psi^2_0  |0\rangle_{\rm R}
& -
\\\hline
 (\phi, - \phi,0) &  |0,- \phi,\phi-1,0\rangle_{\rm NS}  =\psi^{2}_{-1/2+\phi} |0\rangle_{\rm NS} 
&  |\half,\half-\phi,\phi-\half,-\half\rangle_{\rm R}=   |0\rangle_{\rm R}
& +
\\
&  |0,1- \phi,\phi,0\rangle_{\rm NS}  =\psi^{\ov{1}}_{-1/2+\phi} |0\rangle_{\rm NS} 
&  |-\half,\half-\phi,\phi-\half,\half\rangle_{\rm R}= \psi^{0}_0 \psi^{3}_0  |0\rangle_{\rm R}
& -
\\
 ( \phi,0, - \phi) &  |0,- \phi,0,\phi-1 \rangle_{\rm NS}  =\psi^{3}_{-1/2+\phi} |0\rangle_{\rm NS} 
&  |\half,\half-\phi,-\half,\phi-\half\rangle_{\rm R}=   |0\rangle_{\rm R}
& -
\\
&  |0,1- \phi,0,\phi \rangle_{\rm NS}  =\psi^{\ov{1}}_{-1/2+\phi} |0\rangle_{\rm NS} 
&  |-\half,\half-\phi,\half,\phi-\half\rangle_{\rm R}= \psi^{0}_0 \psi^{2}_0  |0\rangle_{\rm R}
& -
\\
 (0, \phi, - \phi) & |0,0,- \phi,\phi-1 \rangle_{\rm NS}  =\psi^{3}_{-1/2+\phi} |0\rangle_{\rm NS}  
&  |\half,-\half,\half-\phi,\phi-\half\rangle_{\rm R}=  |0\rangle_{\rm R}
& -
\\
& |0,0,1- \phi,\phi \rangle_{\rm NS}  =\psi^{\ov{2}}_{-1/2+\phi} |0\rangle_{\rm NS} 
&  |-\half,\half,\half-\phi,\phi-\half\rangle_{\rm R}= \psi^{0}_0 \psi^{1}_0  |0\rangle_{\rm R}
& +
\\\hline
 (\phi_1,\phi_2, -\phi_1-\phi_2) & & &
\\
{\rm for} \;  0<\phi_1+\phi_2 <1
&  |0,-\phi_1,-\phi_2,\phi_1+\phi_2-1 \rangle_{\rm NS} 
&  |-\half,\half-\phi_1,\half-\phi_2,\phi_1+\phi_2-\half \rangle_{\rm R}
& +
\\
{\rm for} \; 1<\phi_1+\phi_2 <2
&  |0,1-\phi_1,1-\phi_2,\phi_1+\phi_2-1 \rangle_{\rm NS}
&   |\half,\half-\phi_1,\half-\phi_2,\phi_1+\phi_2-\frac{3}{2} \rangle_{\rm R}
& +
\\\hline
\end{array}
$}
\caption{GSO projected massless NS and R states for supersymmetric configurations of D6-branes at angles. The $\bZ_2$
subgroup of $T^6/\bZ_{2N}$ is given by $N\vec{v}=(\half,0,-\half)$.  
For shortness, $\pm |\half,\half,\half,\half\rangle_{\rm R}$
denotes the two states $|\half,\half,\half,\half\rangle_{\rm R}$ and 
$|-\half,-\half,-\half,-\half\rangle_{\rm R}$. The parameter range is chosen to be $0< \phi,\phi_1,\phi_2 < 1$.
In the absence of Wilson lines, the sign $\tilde{a}_{\bZ_2}$ coincides with the one occurring in the projection 
of the Chan-Paton labels, $a_{\bZ_2}$.}
\label{Tab:Massless-States}
\end{table*}
\section[Matter on SO(2N)/Sp(2N) branes]{Matter on $SO(2N)/Sp(2N)$ branes}\label{App:SO-Sp}

In~\cite{gmho07}, we (partially, see footnote in section~\ref{Subsec:Examples-with-hidden}) classified 
branes which carry $SO(2N)$ or $Sp(2N)$ gauge factors.
Independently of deriving the correct gauge group assignment, it is possible to 
compute the number of symmetric plus antisymmetric matter states as follows.
The $\gamma_{\bZ_2}$ matrix is the same as in equation~\eqref{Eq:Gamma-Matrix-Reps},
but the $\OR\theta^{-k}$ projection and thereby the identification of representations
in the Chan-Paton labels for a $\OR$ invariant brane $c$ changes,
\begin{equation}
\gamma_{\OR\theta^{-k}} = \unity
,\quad\quad
\lambda_{c(\theta^k c)}  \simeq \left(\begin{array}{cc}
\Anti_{c_1} + \Sym_{c_1} & ({\bf 2N}_{c_1},{\bf 2N}_{c_2})
\\
({\bf 2N}_{c_1},{\bf 2N}_{c_2}) & \Anti_{c_2} + \Sym_{c_2}
\end{array}\right),
\end{equation}
where as before we include both fractional branes $c_1 + c_2 =c$ which are individually $\OR$ invariant
and have opposite $\bZ_2$ eigenvalue and vanishing relative Wilson lines.

The matter states and their $\bZ_2$ eigenvalues are read off from table~\ref{Tab:Massless-States}
leading to one multiplet in the adjoint representation plus two multiplets transforming as $({\bf 2N}_{c_1},{\bf 2N}_{c_2})$ 
in the $cc$ sector. The $c(\theta^k c)$ sectors for $k=1,2$ have angles $\pi(\pm \frac{1}{3}, \mp \frac{1}{3},0)$,  
and the corresponding representations depend on the invariance properties of the intersection points
and the (undetermined) $\OR\theta^{-k}$ eigenvalue of the massless state,
\begin{itemize}
\item $z_{\OR\theta^{-k}+\bZ_2}$:
$\OR\theta^{-k}$ and $\bZ_2$ invariant points on $T_1^2 \times T_2^2$:
\begin{equation}
({\bf 2N}_{c_1},{\bf 2N}_{c_2}) +
\left\{\begin{array}{cc}
\OR\theta^{-k}: + & \Sym_{c_1}+\Sym_{c_2}
\\
\OR\theta^{-k}:- & \Anti_{c_1} +\Anti_{c_2}
\end{array}\right.
,
\end{equation}
\item $z_{\OR\theta^{-k(+N)}}$:
$\OR\theta^{-k(+N)}$ but not $\bZ_2$ invariant orbits on $T_1^2 \times T_2^2$:
\begin{equation}
2 \times ({\bf 2N}_{c_1},{\bf 2N}_{c_2}) + 2 \times
\left\{\begin{array}{cc}
\OR\theta^{-k(+N)}: + & \Sym_{c_1}+\Sym_{c_2}
\\
\OR\theta^{-k(+N)}:- & \Anti_{c_1} +\Anti_{c_2}
\end{array}\right.
,
\end{equation}
\item $z_{\bZ_2}$:
$\bZ_2$ but not $\OR\theta^{-k}$ invariant orbits:
\begin{equation}
2 \times ({\bf 2N}_{c_1},{\bf 2N}_{c_2}) +
\Sym_{c_1}+\Sym_{c_2}  +
\Anti_{c_1} +\Anti_{c_2}
,
\end{equation}
\item $z_0$:
orbits of points without invariance properties:
\begin{equation}
4 \times ({\bf 2N}_{c_1},{\bf 2N}_{c_2}) +
2 \times \left[
\Sym_{c_1}+\Sym_{c_2}  +
\Anti_{c_1} +\Anti_{c_2}
\right]
.
\end{equation}
\end{itemize}
Independently of the correct $\OR\theta^{-k}$ assignments,
the matter spectrum contains ($i=1,2$)
\begin{equation}
\begin{aligned}
\# \left(\Sym_{c_i}\right)  +   \#\left(\Anti_{c_i}\right)
&= 1+ | I_{c_i(\theta c_i)} |
,\\
\# ({\bf 2N}_{c_1},{\bf 2N}_{c_2})
&= 2 + | I_{c_1(\theta c_2)} |
.
\end{aligned}
\end{equation}
All other bifundamental representations are counted in the usual way
(up to the caveat that $a(\theta^k c')$ sectors are identified with 
  $a(\theta^k c)$ sectors).

\section{Bulk relations among different lattices}\label{App_bulkrelation}

The bulk wrapping numbers $\tilde{a}_1 \ldots \tilde{a}_4$ (labelled $P\ldots V$ in~\cite{gmho07})
in table~\ref{Tab:Def-XYL} can be expressed in
terms of one-cycle wrapping numbers $(n_i,m_i)_{i=1\ldots 3}$ on $T^2_i$ as follows
\begin{equation}
\tilde{a}_1 \equiv P = A\, n_3, 
\qquad
\tilde{a}_2 \equiv Q = B \, n_3,
\qquad
\tilde{a}_3 \equiv U = A \, m_3,
\qquad
\tilde{a}_4 \equiv V = B \, m_3,
\end{equation}
with the abbreviations
\begin{equation}
A \equiv n_1 \, n_2 - m_1 \, m_2, 
\qquad
B \equiv n_1 \, m_2 + m_1 \, n_2 + m_1 \, m_2. 
\end{equation}
A rotation of $\pi(1/3,0,-1/2)$ for $b=0$ acts on the one-cycle wrapping numbers as
\begin{equation}\label{Eq:app-trafonm}
\left(\begin{array}{cc} n_1 & m_1 \\ n_2 & m_2 \\ n_3 & m_3 \end{array}\right)
\rightarrow
\left(\begin{array}{cc}-m_1 & n_1 + m_1 \\ n_2 & m_2 \\ m_3 & -n_3\end{array}\right),
\qquad
\left(\begin{array}{c} A \\ B \end{array}\right)
\rightarrow
\left(\begin{array}{c} -B \\ A+B \end{array}\right).
\end{equation}
The bulk wrapping numbers are thus transformed,
\begin{equation}\label{App:eq-trafo-atilde}
(\tilde{a}_1,\tilde{a}_2,\tilde{a}_3,\tilde{a}_4) 
\rightarrow 
(-\tilde{a}_4,\tilde{a}_3+\tilde{a}_4 ,\tilde{a}_2,-\tilde{a}_1-\tilde{a}_2)
\equiv (\hat{a}_1,\hat{a}_2,\hat{a}_3,\hat{a}_4).
\end{equation}
This transformation translates the bulk RR tadpole cancellation condition~\eqref{Eq:RRtcc}
of the {\bf AAa} geometry to the {\bf BAa} lattice and the {\bf ABa} to the {\bf BBa} 
constraint. 

If the complex structure parameter $\varrho$ on the original torus
is replaced by $\hat{\varrho} = \frac{3}{4\varrho}$, also the bulk supersymmetry conditions~\eqref{Eq:SUSY}
are reinterpreted in terms of the transformed geometry.

The {\bf b} type orientations are included by replacing $(n_3,m_3) \rightarrow (n_3,m_3+b \, n_3)$ 
in~\eqref{Eq:app-trafonm} and~\eqref{App:eq-trafo-atilde}. 

In~\cite{gmho07}, a double counting of solutions was avoided as follows.
\begin{enumerate}
\item[(i)]
$(n_1,m_1)=({\rm odd},{\rm odd})$ selects one of the orbifold image cycles $(\theta^k a)$,
\item[(ii)]
$(n_3,m_3+b \, n_3) > (0,0)$ avoids counting the orbit $(\theta^k a)$ and its orientifold image $(\theta^l a')$
as two independent configurations,
\item[(iii)]
$n_1 >0$ forbids a simultaneous flip of the remaining orientations of one-cycles along $T^2_1 \times T^2_2$.
\end{enumerate}
A careful analysis of the solutions related by a $\pi(1/3,0,-1/2)$ rotation reveals that conditions (i) and (ii) are 
still met if the $(\theta \hat{a}')$ image of $a$ is chosen, which has wrapping numbers
\begin{equation}\label{Eq:transformed-wrappings}
{\bf BAa:}
\;
\left(\hat{n}_i,\hat{m}_i\right)_{(\theta \hat{a}')}
=
\left(\begin{array}{cc} m_1 & n_1 \\ n_2 & -(n_2 + m_2) \\ m_3 & n_3 
\end{array}\right),
\qquad
{\bf BBa:}
\left(\begin{array}{cc}  m_1 & n_1 \\ n_2+m_2 & -m_2 \\ m_3 & n_3 
\end{array}\right).
\end{equation}
Condition (iii) is replaced by $\hat{m}_1 >0$. The coprime condition on $T^2_1$ and $T^2_3$ is
clearly preserved under the transformation, and on $T^2_2$, it follows from the 
fact that a $\bZ_6'$ rotation $\theta$ which acts by $(n_2,m_2)
\stackrel{\theta}{\rightarrow}(-n_2-m_2,n_2)\stackrel{\theta}{\rightarrow}(m_2,-n_2-m_2)$ 
also conserves relative primes.

This proves that the number of supersymmetric solutions to the bulk RR tadpole cancellation conditions
on {\bf AA} and {\bf BA} geometries along $T^2_1 \times T^2_2$ is identical, similarly for {\bf AB} and {\bf BB}. 

Due to condition (i), the torus cycle is either passing through the $\bZ_2$ fixed points $\{1,6\}$ or $\{4,5\}$ on $T^2_1$ in the notation of~\cite{gmho07}
from which one derives the coefficients $d_j, e_j$ of the exceptional cycles in~\eqref{Eq:Def-frac}.
For some fixed $j_1$ which the torus cycle on $T^2_3$ passes through, one obtains
\begin{equation}
\begin{aligned}
(d_{j_1}, e_{j_1})_{\{1,6\}} &=\left((-1)^{\tau^0+\tau^1+1}\,(n_2 + m_2),(-1)^{\tau^0+\tau^1}\, n_2  \right),
\\
(d_{j_1},e_{j_1})_{\{4,5\}} &=\left((-1)^{\tau^0} \left( n_2 + (-1)^{\tau^1} m_2 \right) ,
(-1)^{\tau^0} \left((-1)^{\tau^1+1} n_2 + (1 - (-1)^{\tau^1}) m_2 \right)  \right),
\end{aligned}
\end{equation}
where $(j_1,j_2) \in \{(1,2),(4,3)\}$
or $\{(1,3),(2,4)\}$ or $\{(1,4),(2,3)\}$ are the pairs of fixed points traversed by the cycle 
for $(n_3,m_3)=$ (odd,even), (odd,odd) and (even, odd), respectively, according to the fixed point labels in~\cite{gmho07}.
The first and second set of points corresponds to no or some displacement of the cycle from the origin on $T^2_3$, 
respectively.
The coefficients $(d_{j_2}, e_{j_2})$ carry and additional global factor $(-1)^{\tau^3}$ where 
$\tau^i \in \{0,1\}$ parameterise the overall $\bZ_2$ eigenvalue ($i=0$) and Wilson lines on $T^2_1 \times T^2_3$ ($i=1,3$).

Upon the $\pi(1/3,0,-1/2)$ rotation, the wrapping numbers on $T^2_3$  transform as
$(n_3,m_3) \rightarrow (m_3,n_3)$ and $\bZ_2$ fixed points on 
 the {\bf a} geometry for $T^2_3$ are permuted as follows,
\begin{equation}
(1,2,3,4) \rightarrow (1,4,3,2).
\end{equation}
Inserting the transformed wrapping numbers $ \left(\hat{n}_2,\hat{m}_2\right)_{(\theta \hat{a}')}$
on $T^2_2$ given in~\eqref{Eq:transformed-wrappings} leads to 
\begin{equation}\label{Eq:App-de-hat}
\begin{aligned}
{\bf BAa:} &
\\
(\hat{d}_{\hat{j}_1}, \hat{e}_{\hat{j}_1})_{\{1,6\}} &=\left((-1)^{\tau^0+\tau^1}\, m_2,(-1)^{\tau^0+\tau^1} \, n_2\right),
\\
(\hat{d}_{\hat{j}_1},\hat{e}_{\hat{j}_1})_{\{4,5\}} &=\left((-1)^{\tau^0}\,\left((1-(-1)^{\tau^1}) n_2 + (-1)^{\tau^1+1} m_2 \right),
(-1)^{\tau^0+1}\,\left(n_2 + (1 + (-1)^{\tau^1+1}) m_2 \right)\right),
\\
\\
{\bf BBa:} & 
\\
(\hat{d}_{\hat{j}_1}, \hat{e}_{\hat{j}_1})_{\{1,6\}} &=\left((-1)^{\tau^0+\tau^1+1}\, n_2 ,(-1)^{\tau^0+\tau^1}\,(n_2 + m_2) \right),
\\
(\hat{d}_{\hat{j}_1},\hat{e}_{\hat{j}_1})_{\{4,5\}} &=\left((-1)^{\tau^0}\,\left(n_2 + (1 + (-1)^{\tau^1+1}) m_2 \right),
(-1)^{\tau^0+1}\, \left((-1)^{\tau^1 } n_2 + m_2\right)\right),
\end{aligned}
\end{equation}
in terms of the original numbers $(n_2,m_2)$ on {\bf AAa} and {\bf ABa}, respectively, 
and the analogous expressions with
an overall prefactor of $(-1)^{\tau^3}$ for $(\hat{d}_{\hat{j}_2}, \hat{e}_{\hat{j}_2})$.

The RR tadpole cancellation~\eqref{Eq:RRtcc} and massless hyper charge~\eqref{Eq:Q_Y-cycle} conditions
due to exceptional cycles are conveniently parameterised
for the {\bf a} geometry on $T^2_3$ by $(X_3 \ldots X_6)$ and $(Y_3 \ldots Y_6)$ with 
\begin{equation}\label{Eq:App-X3X6}
X_{2+j} =\left\{\begin{array}{cc}
2 \, e_j & {\bf AAa} \\ e_j - d_j & {\bf ABa} \\ d_j + e_j & {\bf BAa} \\ d_j + 2 \, e_j & {\bf BBa} 
\end{array}\right.
,\quad
Y_{2+j}  =\left\{\begin{array}{c}
2\, d_j + e_j \\ d_j + e_j \\ d_j - e_j \\ 2 \, e_j 
\end{array}\right.
\quad {\rm for}
\quad j=1\ldots 4.
\end{equation}
In case of a {\bf b} torus, $X_3$ and $X_6$ remain unchanged while the remaining two entries are subject to a 
shift
\begin{equation}\label{Eq:App-X4X5}
\left(\begin{array}{c} X_4 \\ X_5 \end{array}\right)
\rightarrow 
\left(\begin{array}{c} X_4 + 2b(e_3 - e_2) \\ X_5 - 2b(e_3 - e_2) \end{array}\right),
\end{equation}
and a similar expression for $(Y_4,Y_5)$.
This explicit form can be used to test the behaviour of the RR tadpole cancellation conditions from exceptional cycles 
under the $\pi(1/3,0,-1/2)$ rotation. The correspondence of the bulk relations does, however, not carry over immediately.
This is not too surprising since,e.g., the number of full solutions on {\bf BBa} is by one order of magnitude bigger 
than on {\bf ABa}~\cite{gmho07}.

\section{Trinification models}\label{App_Trinification}

Intersecting D-branes offer the possibility for trinification models with 
gauge group $U(3)_a \times U(3)_b \times U(3)_c$ and $n$ quark-lepton generations 
in 
\begin{equation}
n \times \left[
(\ov{\3}_a,\3_b,\1) + 
(\1,\ov{\3}_b,\3_c) +
(\3_a,\1,\ov{\3}_c)
\right].
\end{equation}
It turns out that on $T^6/\bZ_6'$, there are only supersymmetric solutions for $n=2$,
and all of them contain a large number of chiral exotics.
The chiral matter states involving only stacks $a$, $b$ and $c$ and the associated complex
structure value are displayed in table~\ref{Tab:Trinification-Confs}.
More chiral exotic states can arise at intersections with further brane stacks required 
for RR tadpole cancellation.

\begin{table}[ht!]
\begin{center}
\begin{equation*}
\begin{array}{|c|c|c|}
\hline
\multicolumn{3}{|c|}{\rule[-3mm]{0mm}{8mm}
\text{\bf Trinification models with } U(3)_a \times U(3)_b \times U(3)_c 
}
\\\hline\hline
\varrho & \text{ chiral states on $a,b,c$ only} & \# {\rm models}
\\\hline\hline
\frac{1}{3} &
\begin{array}{c}
2 \times \left[ (\ov{\3}_a,\3_b,\1) +  (\1,\ov{\3}_b,\3_c) + (\3_a,\1,\ov{\3}_c) \right] \\
+ 6 \, (\3_{\ov{A}_a},\1,\1) + 3 \, (\6_{\ov{S}_a},\1,\1)  \\
+ 4\,  (\3_a,\3_b,\1) + 2 \, (\3_a,\1,\3_c) + 2 \, (\1,\ov{\3}_b,\ov{\3}_c)
\end{array}
& 2864
\\\hline
1 & 
\begin{array}{c}
2 \times \left[ (\ov{\3}_a,\3_b,\1) +  (\1,\ov{\3}_b,\3_c) + (\3_a,\1,\ov{\3}_c) \right] \\
+  (\3_{\ov{A}_a},\1,\1) + 4 \, (\6_{\ov{S}_a},\1,\1)  \\
+ 4\,  (\3_a,\3_b,\1) + 2 \, (\3_a,\1,\3_c) + 2 \, (\1,\ov{\3}_b,\ov{\3}_c)
\end{array}
& 6.3 \times 10^{8}
\\\hline
\end{array}
\end{equation*}
\end{center}
\caption{Complex structure values $\varrho$ on $T^2_3$, frequency and chiral matter charged under $U(3)_a \times U(3)_b \times U(3)_c$
for trinification models. Further chiral states can arise from intersections with other branes needed for 
RR tadpole cancellation.}
\label{Tab:Trinification-Confs}
\end{table}

Due to the large number of chiral exotics these models are not very interesting from
a phenomenological point of view and we do not pursue a more detailed analysis.

\clearpage
\addcontentsline{toc}{section}{References}
\bibliographystyle{utphys}
\bibliography{refs_smz6p}

\end{document}